\documentclass{aa}  

\usepackage[colorlinks=true, linkcolor = blue,
            urlcolor  = blue,
            citecolor = blue,
            anchorcolor = blue]{hyperref}
\usepackage{graphicx}
\usepackage{txfonts}
\usepackage{natbib}
\usepackage{enumitem} 
\usepackage{color}
\usepackage{threeparttable}
\usepackage{tikz,amsmath}
\usetikzlibrary{shapes.geometric, arrows}
\usepackage{ulem}
\usepackage{wrapfig}
\usepackage{rotating}
\usepackage{lscape} 
\tikzstyle{process} = [rectangle, minimum width=1.5em, minimum height=3.5em, text centered, draw=blue, fill=gray!10]
\tikzstyle{process2} = [rectangle, minimum width=1.5em, minimum height=3.5em, text centered, draw=white, fill=white]

\tikzstyle{arrow} = [thick,->,>=stealth]
\usepackage{cancel}
\bibpunct{(}{)}{;}{f}{}{,} 

\newcommand{\teff}{$T_{\mathrm{eff}}$}

\newcommand{\logg}{\mbox{log \textit{g}}}
\newcommand{\loggiso}{\mbox{log \textit{g}}$_{\mathrm{iso}}$} 
\newcommand{\vsini}{$v\mathrm{sin}\,i$}
\newcommand{\titan}{\textsc{Titans}}

\begin{document}

  \title{Titans metal-poor reference stars. I.}

  \subtitle{Accurate effective temperatures and surface gravities for dwarfs and subgiants from 3D non-LTE H$\alpha$ profiles and Gaia parallaxes}
    
   \author{R. E. Giribaldi\inst{1}
         \and
          A. R. da Silva\inst{1}
         \and
          R. Smiljanic\inst{1}
         \and 
          D. Cornejo Espinoza\inst{2}
          }

   \institute{Nicolaus Copernicus Astronomical Center, Polish Academy of Sciences, ul. Bartycka 18, 00-716, Warsaw, Poland \\
             \email{riano@camk.edu.pl, rianoesc@gmail.com}
        \and
        Center of Radio Astronomy and Astrophysics at Mackenzie, Engineering School, Mackenzie Presbyterian University, S\~ao Paulo, Brazil
             }
             
    \date{Received March 8, 2021; Accepted April 26, 2021}

 
  \abstract
   {Several large stellar spectroscopic surveys are producing overwhelming amounts of data that can be used for determining stellar atmospheric parameters and chemical abundances. 
   Nonetheless, the accuracy achieved in the derived astrophysical parameters is still insufficient, mainly because of the paucity of adequate calibrators, particularly in the metal-poor regime ([Fe/H] $\leq -$1.0).}
   {Our aim is to increase the number of metal-poor stellar calibrators that have accurate parameters. Here, we introduce the \titan\ metal-poor reference stars: a sample of 41 dwarf and subgiant stars with accurate, but model-dependent, parameters.}
   {Effective temperatures (\teff) were derived by fitting observed H$\alpha$ profiles with synthetic lines computed using three-dimensional (3D) hydrodynamic model atmospheres that take into account departures from the local thermodynamic equilibrium (non-LTE effects). Surface gravities (\logg) were computed using evolutionary tracks and parallaxes from \textit{Gaia} early-data release 3.}
   {The same methods recover the \teff\ values of the \textit{Gaia} benchmark stars, which are mostly based on interferometric measurements, with a 1$\sigma$ dispersion of $\pm 50$ K. 
   We assume this to be the accuracy of the H$\alpha$ profiles computed from 3D non-LTE models for metal-poor dwarfs and subgiants,
   {although this is likely an upper-bound estimate dominated by the uncertainty of the standard \teff\ values.}
   We achieved an internal precision typically between 30-40~K, these errors dominated by instrumental effects. 
   The final {total uncertainty for the \teff\ values of the \titan\ are thus estimated to be of the order of $1\%$.} The typical error for \logg\ is $\leq$ 0.04 dex. 
   In addition, we identified a few members of \textit{Gaia}-Enceladus, of Sequoia, and of the Helmi stream in our sample. These stars can pave the way for the accurate chemical characterization of these Galactic substructures.}
   {Using the \titan\ as reference, large stellar surveys will be able to improve the internal calibration of their astrophysical parameters. Ultimately, this sample will help users of data from \textit{Gaia} and large surveys in reaching their goal of redefining our understanding of stars, stellar systems, and the Milky Way.}


   \maketitle
%

\section{Introduction}

The data obtained by the \textit{Gaia} mission \citep{Gai(a)} have triggered an on-going revolution in our understanding of stellar and Galactic astrophysics \citep[see e.g.][]{brown2021microarcsecond}. \textit{Gaia} is delivering astrometry and photometry for more than 10$^9$ stars. To complement \textit{Gaia}, with radial velocities and chemical abundances, several large stellar spectroscopic surveys have started (or are being planned). \par
Stellar characterization is essential, for example, for tracing the timeline of processes driving the formation and evolution of the Milky Way. Stellar chemical abundances and ages are powerful complements to kinematic and dynamic properties in studies of the stellar populations \citep[e.g.][]{2002ARA&A..40..487F}. Nevertheless, it is common to observe a large dispersion in the values of stellar parameters and abundances when comparing literature results or the performance of different analysis methods applied to the same spectra \citep{2012A&A...547A.108L,2014AJ....148...54H,2016ApJS..226....4H,2014A&A...570A.122S,2017A&A...601A..38J,2021arXiv210102242B}. This outcome is a consequence of the complexity of the task. Various astrophysical parameters must be derived from limited information carried by the, mostly incomplete, stellar data (e.g. spectra, photometric magnitudes, and parallax measurements). The data themselves are affected by random noise and systematic uncertainties. Moreover, the interpretation of the stellar data depends on the use of auxiliary models and data (e.g. for atomic transitions) which have their own limitations \citep[see, e.g.,][for different aspects of these problems]{amarsi2018,2016A&ARv..24....9B,2019MNRAS.486.2075B,2018A&A...611A..68B}.\par
Ideally, to characterize a star, one would directly measure its most fundamental parameters: luminosity ($L$) -- or, in practice, bolometric flux ($F_{\rm BOL}$) and distance; radius ($R$); and mass ($\mathcal{M}$). From these parameters, it is possible to compute values of effective temperature (\teff) and surface gravity (\logg). In reality, extracting fundamental parameters from direct measurements is only possible in a restricted number of cases \citep{2010A&ARv..18...67T,2014PASP..126..711B,2020FrASS...7....3H,2020arXiv200610868S}. Therefore, \teff\ and \logg\ values are mostly obtained from indirect, model-dependent methods \citep[see e.g.][]{2005MSAIS...8..130S}.\par
The spectroscopic method using Fe lines is a frequently used technique to simultaneously derive \teff, \logg, and metallicity ([Fe/H]\footnote{[A/B] = $ \log{\left( \frac{N(\text{A})}{N(\text{B})} \right )_\text{Star}} - \log{\left( \frac{N(\text{A})}{N(\text{B})} \right )_\text{Sun}} $, where $N$ denotes the number abundance of a given element.}). One approach, usually referred to as the equivalent width (EW) method, relies on measuring EWs to constrain the parameters using the excitation and ionization equilibrium of Fe lines \citep[e.g.][]{2013A&A...558A..38M,2014dapb.book..297S,2019A&A...628A.131T}. The equilibria of Fe lines can also be used via the direct fitting of the lines with synthetic spectra \citep[see, e.g., the codes described by][]{blanco-cuaresma2014,2014A&A...564A.109S}. An alternative is the so-called spectrum synthesis method, where large sections of the observed spectrum are compared to synthetic theoretical spectra in search for the best fitting one \citep[e.g.][]{2016A&A...591A..81W,2017A&A...597A..16P,2019A&A...629A..74G}.\par
Spectroscopic methods are often avoided for deriving \teff\ in metal-poor stars \mbox{([Fe/H] $\leq -1$~dex)} in order to prevent biased results. Instead, the Infrared Flux Method \citep[IRFM,][]{Blackwell1977, Blackwell1979, Blackwell1980} is preferred \citep[see][]{melendez2004,melendez2006}.\par
Effective temperatures derived with the IRFM are usually considered robust,  much less susceptible to the model limitations that affect spectroscopic methods, such as granulation effects or departures from the local thermodynamic equilibrium (so-called non-LTE effects). Nevertheless, the method is still affected by uncertainties in the reddening and inconsistencies in the absolute flux calibration \citep[see e.g.][and references therein]{gonzalez2009,Casagrande2010}.\par
Thanks to the increased sensitivity of instruments like the Precision Astronomical Visible Observations (PAVO) beam combiner \citep{ireland2008}, in the Center for High Angular Resolution Astronomy (CHARA) interferometric array \citep{tenBrummelaar2005}, angular diameters of nearby dwarfs and subgiants can now be measured, delivering data for a quasi-direct\footnote{Strictly speaking, \teff\ values based on interferometric measurements carry a marginal dependence on model atmospheres because of the need of limb-darkening corrections to estimate the angular diameters \citep{2000MNRAS.318..387D}.} \teff\ determination with accuracy within 100~K \citep[e.g.][]{bazot2011,white2013,karovicova2018,karovicova2020}. 
This enables the establishment of a standard \teff\ scale that can be used to evaluate the accuracy of model-dependent techniques. \par

On the base of this interferometric \teff\ scale, Balmer line profiles synthesised using three-dimensional (3D), non-LTE model atmospheres were recently found to accurately reproduce the observed profiles \citep{amarsi2018,giribaldi2019_Ha}. This important result restored confidence on a method frequently applied to avoid biases related to degeneracy among the atmospheric parameters, since the shape of the wings of the Balmer lines essentially depends only on \teff\ \citep[e.g.][among many others]{cayrel1985,1985IAUS..111..137C, fuhrmann1994,fuhrmann1998,mishenina2001}. Despite its potential for providing accurate \teff\ values, up to now only few works have made use of Balmer lines computed with 3D non-LTE models \citep{amarsi2019,nordlander2019,hanke2020}. Most likely, this is the case because of complications related to the normalization of wide profiles in modern echelle spectra.

The accurate calibration of stellar parameters has become imperative in the current era of large spectroscopic surveys \citep{2019ARA&A..57..571J}. On-going and future projects like Gaia-ESO \citep{2012Msngr.147...25G,2013Msngr.154...47R}, Galactic Archaeology with HERMES \citep[GALAH,][]{2015MNRAS.449.2604D}, Large Sky Area Multi-Object Fiber Spectroscopic Telescope \citep[LAMOST,][]{2012RAA....12.1197C}, Maunakea Spectroscopic Explorer \citep[MSE,][]{2019arXiv190404907T}, 4-metre Multi-Object Spectroscopic Telescope \citep[4MOST,][]{2019Msngr.175....3D}, Milky Way Mapper \citep[MWM,][]{2017arXiv171103234K}, will together obtain spectra for more than 10$^7$ stars. The analysis of this amount of data requires innovative methodologies such as given by codes like The Cannon \citep{2015ApJ...808...16N}, StarNet \citep{2018MNRAS.475.2978F}, and The Payne \citep{2019ApJ...879...69T}, or those explored by several other authors \citep[e.g.][]{2019OAst...28...68K,2019MNRAS.483.3255L,2020A&A...644A.168G,2020ApJ...891...23W}. Some of these methods depend on training samples of stars that have well-derived, accurate parameters. Lack of accuracy can result in systematic problems of the training sample propagating to the new analysis \citep[see e.g. the trend between \teff\ and {[Fe/H]} of stars in the Praesepe cluster displayed in Fig.\ 11 of][]{2020ApJ...898...58W}.

Stars with good-quality interferometric \teff\ have been compiled into a sample called the \textit{Gaia} benchmark stars \citep[GBS henceforth,][]{Heiter,jofre2014,2018RNAAS...2..152J}. This sample includes, in addition, a few stars for which \teff\ values were inferred by color calibrations based on interferometric measurements or, alternatively, by using the IRFM. The uncertainties of the interferometric \teff\ values is of 1\% or better. The uncertainties of the \teff\ values based on calibrations can be slightly higher, with typical values between 80 and 100~K; but still useful for accuracy diagnostics. The \logg\ values of the GBS are based on masses determined using stellar evolution models. For dwarfs and subgiants, typical uncertainties in \logg\ are also of the order of 1\% or better ($\leq$ 0.04 dex).

Apart from the GBS, several samples of calibrating stars, with accurate masses, have become available thanks to recent advances in asteroseismology \citep{2016AN....337..970V,2017A&A...600A..66V,2017ApJS..233...23S,2018ApJS..239...32P,2020A&A...643A..83W}. 
Stars in eclipsing binaries are an additional option for use as calibrators. Masses and radii of the components can be measured with very high accuracy \citep[e.g.][]{2019MNRAS.484..451H,2021MNRAS.500.4972R}, with the possibility of using these measurements to derive effective temperatures of high quality \citep{2020MNRAS.497.2899M}. There is, however, still strong need of larger samples of calibrators, in particular to improve the coverage of the parameter space towards low metallicities \citep[e.g.][]{2016A&A...592A..70H,2019A&A...627A.173V}.

In this work we introduce the  \titan\footnote{In Greek mythology, the Titans are the offsprings of Gaia.} reference stars, a selection of metal-poor stars for which we provide accurate and precise parameters and abundances, and that is intended to serve for the calibration of spectroscopic analysis methods. 
In this first publication, we focus on providing values of \teff\ and \logg\ for dwarf and subgiant stars with \teff\ within $5500$-$6600$~K and \logg\ within 3.7-4.6~dex. The metallicity range of this sample lies between $-1.0$ and $-3.1$~dex. We did not impose a lower limit in metallicity; this was rather set by the availability of the spectra required to apply our methodology. Effective temperatures were determined fitting  H$\alpha$ line profiles. H$\alpha$ becomes progressively weaker and less useful for our method as \teff\ decreases. This limit of usability sets at about $\sim4500$~K for giants and  $\sim5000$~K for dwarfs. However, once again, for the current sample of \titan, the lower limit in the \teff\ range was set by the availability of good-quality spectra.
As we discuss later in the paper, the H$\alpha$ profiles of metal-poor stars are significantly sensitive to \logg, which we inferred using \textit{Gaia} parallaxes from the early data release 3 \citep[EDR3,][]{2020arXiv201201533G}.  
Therefore, the atmospheric parameters in this sample owe their fine accuracy to the \textit{Gaia} mission. This is the reason why we call these stars \titan.

The paper is organised as follows. The sample and the observational data are described in Section \ref{sec:data}. The determination of atmospheric parameters and the accuracy of the results are discussed in Sections \ref{sec:parameters} and \ref{sec:accuracy}, respectively. In Section \ref{sec:substructures}, we separate the stars into different Galactic substructures. Finally, our results are summarised in Section \ref{sec:conclusions}.

\begin{figure}
    \centering
    \includegraphics[width=1\linewidth]{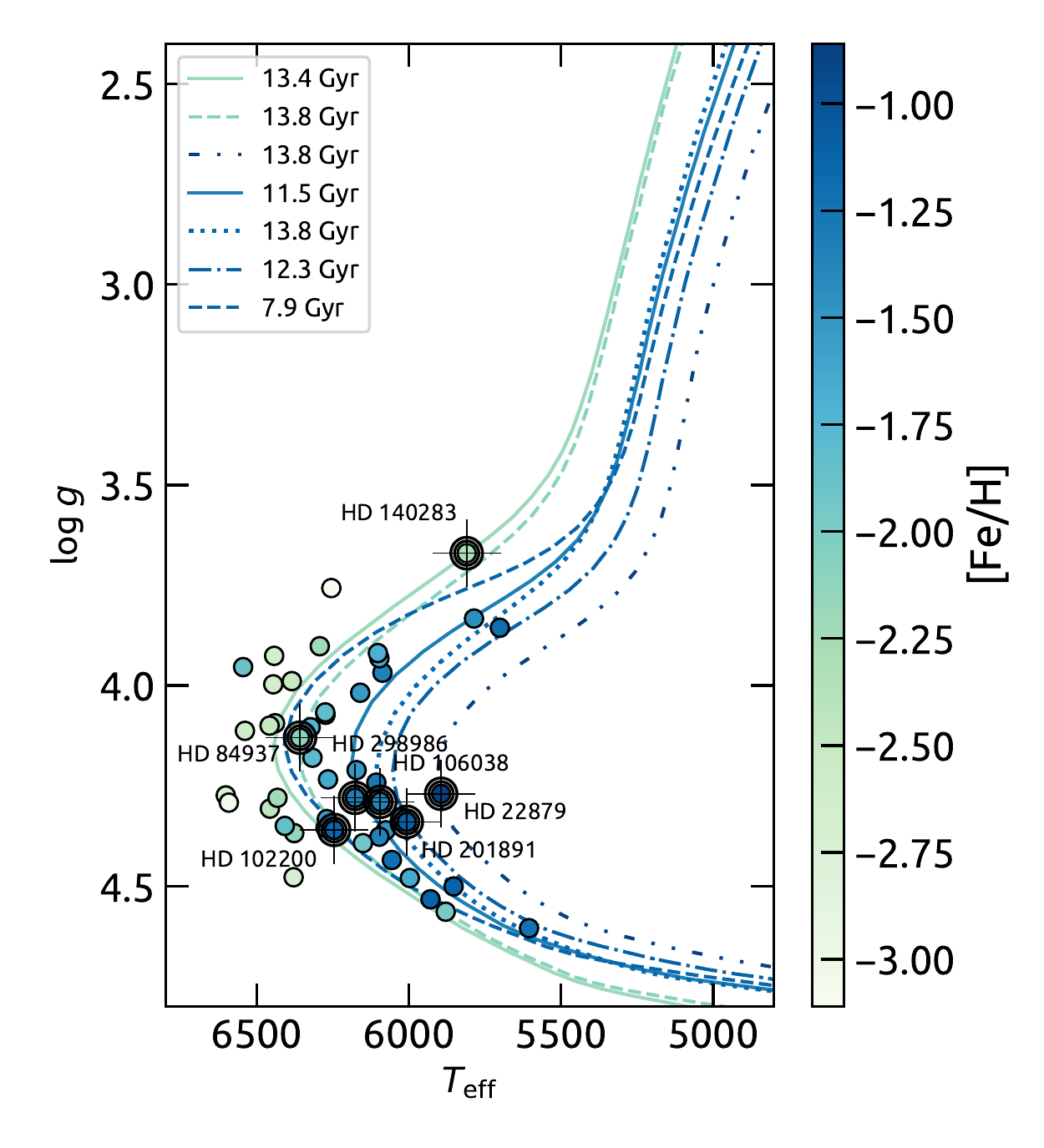}
    \caption{Distribution of the \titan\ in the Kiel diagram.
    The stellar parameters used in the plot are those displayed in Table~\ref{tab:CAMK_stars}. 
    The \textit{Gaia} benchmark stars with
    [Fe/H]~$<-0.8$~dex are identified with bold contours.  Isochrones corresponding to the age and [Fe/H] of each benchmark star are plotted.}
    \label{fig:Kiel_diagram}
\end{figure}

\section{Sample selection and observational data}\label{sec:data}

\subsection{Selection of the Titans sample}

The main source of sample stars was the work of \citet{Casagrande2010}. Additional candidates were assembled from \citet{1996ApJ...471..254R, 2006A&A...451.1065G, 2008MNRAS.391...95R, 2010A&A...511L..10N,2017ApJ...835...81B}. The stars were included in the preliminary sample if their previously determined atmospheric parameters were as follows: \teff\ $\geq 5000$~K, [Fe/H] $\leq -0.9$~dex, and \logg\ $\geq 3.5$~dex.

\begin{table}
\caption{Atmospheric parameters of the Titans in the literature.}
\label{tab:preliminar_parameters}
\centering
\scriptsize
\begin{threeparttable}
\begin{tabular}{lccccc}
\hline\hline
Star & \teff\ (K) & \logg & [Fe/H] &  Source & Spectra \\
\hline
BD+24 1676 & $6387 \pm 92$ & $3.84$ & $-2.54$ & Ca10 & 3\\
BD$-10$ 388 & $6260 \pm 61$ & $3.98$ & $-2.32$ & Ca10 & 3\\
BPS CS22166-030 & $6039 \pm 59$ & $4.00$ & $-3.36$ & Ca10 & 5\\
CD$-33$ 3337 & $6001 \pm 64$ & $4.03$ & $-1.32$ & Ca10 & 2\\
CD$-35$ 14849 & $6396 \pm 65$ & $4.22$ & $-2.35$ &  Ca10 & 3\\
HD 16031 & $6286 \pm 64$ & $4.17$ & $-1.74$ & Ca10 & 2\\
HD 166913 & $6267 \pm 62$ & $4.31$ & $-1.56$ & Ca10 & 4\\
HD 218502 & $6298 \pm 66$ & $3.96$ & $-1.76$ & Ca10 & 3\\
HD 34328 & $6056 \pm 67$ & $4.50$ & $-1.66$ & Ca10 & 2\\
HD 59392 & $6045 \pm 79$ & $3.87$ & $-1.62$ & Ca10 & 4\\
BD+02 4651 & $6349 \pm 67$ & $3.79$ & $-1.78$ & Ca10 & 4\\
BD+03 740 & $6419 \pm 83$ & $3.97$ & $-2.70$ & Ca10 & 3\\
BD$-13$ 3442 & $6434 \pm 80$ & $3.92$ & $-2.74$ & Ca10 & 6\\
BD+26 2621 & $6336 \pm 72$ & $4.00$ & $-2.54$ & Ca10 & 3\\
BD+26 4251 & $5950 \pm 100$ & $4.26$ & $-1.29$ & R\&L08 & 4\\
BD+29 2091 & $5974 \pm 44$ & $4.58$ & $-1.99$ & Ca10 & 3\\
CD$-30$ 18140 & $6373 \pm 74$ & $4.13$ & $-1.90$ & Ca10 & 4\\
CD$-33$ 1173 & $6685 \pm 68$ & $3.87$ & $-2.93$ & Ca10 & 3\\
CD$-48$ 2445 & $6453 \pm 66$ & $4.25$ & $-1.93$ & Ca10 & 3\\
CD$-71$ 1234 & $6408 \pm 69$ & $4.29$ & $-2.41$ & Ca10 & 4\\
G~24$-3$ & $6118 \pm 59$ & $4.27$ & $-1.59$ & Ca10 & 3\\
HD~108177 & $6333 \pm 68$ & $4.41$ & $-1.64$ & Ca10 & 3 \\
HD~116064 & $5895 \pm 58$ & $4.44$ & $-1.87$ & Ca10 & 4\\
HD~122196 & $5986 \pm 63$ & $3.71$ & $-1.82$ & Ca10 & 2\\
HD~126681 & $5638 \pm 50$ & $4.55$ & $-1.14$ & Ca10 & 2\\
HD~132475 & $5808 \pm 57$ & $3.87$ & $-1.51$ & Ca10 & 5\\
HD~160617* & $6048 \pm 65$ & $3.73$ & $-1.78$ & Ca10 & 4\\
HD~189558 & $5765 \pm 57$ & $3.77$ & $-1.12$ & Ca10 & 2\\
HD~193901 & $5920 \pm 52$ & $4.52$ & $-1.07$ & Ca10 & 4\\
HD~213657 & $6299 \pm 72$ & $3.90$ & $-1.96$ & Ca10 & 7\\
HD~241253 & $5953 \pm 100$ & $4.23$ & $-0.94$ &  R\&L08 & 2\\
HD~284248 & $6291 \pm 61$ & $4.42$ & $-1.56$ & Ca10 & 6\\
HD~74000 & $6361 \pm 98$ & $4.10$ & $-2.01$ & Ca10 & 2\\
HD~94028 & $6110 \pm 66$ & $4.36$ & $-1.44$ & Ca10 & 5\\
HE~0926$-0508$ & $6563$ & $4.14$ & $-2.98$ & Be17 & 1\\
LP~815$-43$ & $6535 \pm 68$ & $4.19$ & $-2.18$ & Ca10 & 2\\
LP~831$-70$ & $6408 \pm 67$ & $4.38$ & $-2.93$ & Ca10 & 3\\
Ross~453 & $6174$ & $3.96$ & $-1.95$ & Be17 & 1\\
Ross~892 & $5859 \pm 30$ & $4.27$ & $-1.20$ & N\&S10 & 1\\
UCAC2~20056019 & $6200 \pm 50$ & $4.00$ & $-2.74$ & Ry96 & 1\\
Wolf~1492 & $6407 \pm 96$ & $4.20$ & $-3.12$ & Ge06 & 1\\
\hline
\end{tabular}
\begin{tablenotes}
\item{} \textbf{Notes.} The fifth column displays the source of the parameters used to select the stars. The reference codes correspond to: \citet[][Ca10]{Casagrande2010}, \citet[][R\&L08]{2008MNRAS.391...95R}, \citet[][Be17]{2017ApJ...835...81B}, \citet[][N\&S10]{2010A&A...511L..10N}, \citet[][Ry96]{1996ApJ...471..254R}, \citet[][Ge06]{2006A&A...451.1065G}.
The last column displays the number of spectra used in the H$\alpha$ line fitting. Star HD~160617 is marked with the symbol (*) to indicate that it had HARPS spectra analyzed instead of UVES.
These parameters were used as initial guess values in our analysis.
\end{tablenotes}
\end{threeparttable}
\end{table}  

We then restricted the sample to stars with spectra obtained with the Ultraviolet and Visual Echelle Spectrograph \citep[UVES,][]{dekker2000} that were publicly available in the data archive\footnote{\url{http://archive.eso.org/wdb/wdb/adp/phase3_main/form}} of the European Southern Observatory (ESO). The spectra had to encompass the H$\alpha$ line at 6562.797~\AA, have resolution higher than $R = 40\,000$, and signal-to-noise ratio (SNR) higher than $100$. We made use of the processed, single-exposure UVES spectra. Data reduction was conducted with the ESO UVES pipeline \citep{2000Msngr.101...31B}.

During the analysis, several of the spectra that we collected were found to present distortions related to spectral order merging. 
For this first work, we did not attempt to re-reduce those spectra, but simply removed them from the working data set. 
This resulted in a substantial decrease of the original sample, as for many stars all their available spectra ended up being excluded. 
We plan to come back to these spectra in later stages of the project. 
For other stars, only one or two spectra remained. 
The final sample is given in Table \ref{tab:preliminar_parameters}, which includes \teff\ with its quoted uncertainties, \logg\ values, [Fe/H] values, and their corresponding sources in the literature.
The distribution of the stars in the Kiel diagram is presented in Fig.~\ref{fig:Kiel_diagram}, the parameters displayed are those derived in this work (Table~\ref{tab:CAMK_stars}). We note that for the star HD~160617, that had all UVES spectra excluded from the analysis, we instead analysed high-quality spectra obtained with the High Accuracy Radial velocity Planet Searcher \citep[HARPS,][]{Mayor2003} spectrograph, as we already had these spectra at hand for a different purpose.

\subsection{Comparison sample of Gaia benchmarks}

The GBS sample was used in our analysis to test the accuracy of the methods. Stars with [Fe/H] $> 1$~dex are also included for testing purposes. We included all GBS that have HARPS and/or UVES spectra with $R > 40\,000$ and SNR~$> 100$, and without signs of order merging defects. Most of the selected data have indeed SNR~$\sim 200$. For one star, $\tau$~Cet, we used also three UVES spectra that have SNR~$\sim50$.

The final list of selected Gaia benchmark stars was: $\beta$~Hyi, $\tau$~Cet, $\epsilon$~For, $\delta$~Eri, $\beta$~Vir, $\eta$~Boo, $\alpha$~Cen~A, 18~Sco, Sun\footnote{For the Sun we used the normalized HARPS spectra of \citet{giribaldi2019_Ha}}, $\beta$~Gem, $\xi$~Hya, HD~107328, $\epsilon$~Vir, HD~140283, HD~84937, HD~22879, HD~49933, HD~298986, HD~106038, HD~201891, and HD~102200.

\begin{figure*}
\centering
\hfill\\
{\scriptsize \begin{tikzpicture}[node distance=2em]

\node (T0) [process, minimum width=9em, minimum height=1.5em, yshift=0.0cm] {\scriptsize \textcolor{red}{Prior error}};
\node (T1) [process, right of=T0, minimum width=7.5em, minimum height=1.5em, xshift=2.5cm, yshift=0.0cm] {\scriptsize \textcolor{red}{H$\alpha$ fitting $\delta$(\teff)}};
\node (T2) [process, right of=T1, minimum width=11em, minimum height=0.5em, xshift=3.0cm, yshift=0.0cm] {\scriptsize \textcolor{red}{Fe lines $\delta$([Fe/H])~/~Isochrone $\delta$(\logg)}};
\node (T3) [process, right of=T2, minimum width=7.5em, minimum height=1.5em, xshift=3cm, yshift=0.0cm] {\scriptsize \textcolor{red}{H$\alpha$ fitting $\delta$(\teff)}};
\node (S0) [process, below of=T0, minimum width=9em, xshift=0.0cm, yshift=-0.3cm] {\parbox[t][][t]{2cm}{\centering $\delta$([Fe/H]) $\;\pm0.3$~dex\\ $\delta$(\logg) $\;\pm0.3$~dex}};
\node (Ss) [process2, left of=S0, xshift=-1.9cm, yshift=-0.0cm] {\rotatebox{0}{\scriptsize \textcolor{black}{{\parbox[t][][t]{2.3cm}{ \teff--[Fe/H] series --->\\ \teff--\logg\ series ~~~--->}}}}};
\node (S1) [process, below of=T1, minimum width=7.5em, xshift=0.0cm, yshift=-0.3cm] {\parbox[t][][t]{1.2cm}{\centering $\mp5\,\mathrm{to}\,\mp60$~K\\ $\mp105$~K}};
\node (S2) [process, below of=T2, minimum width=11em, xshift=0.0cm, yshift=-0.3cm] {\parbox[t][][t]{2.5cm}{\centering up to $\mp0.03$~dex / $\;\;\;\;$---\\ 
---$\;\;\;\;$ / $\mp0.05$~dex}};
\node (S3) [process, below of=T3, minimum width=7.5em, xshift=0.0cm, yshift=-0.3cm] {\parbox[t][][t]{1cm}{\centering $\;\pm0$~K\\ $\;\pm20$~K}};

\draw [arrow] (S0) -- (S1);
\draw [arrow] (S1) -- (S2);
\draw [arrow] (S2) -- (S3);

\end{tikzpicture}}
\caption{Error flow diagram up to the beginning of the second loop. Each box displays, for each technique, the magnitude of the changes in a certain parameter induced by the variations given for the parameters in the previous box.}
\label{fig:error_flow}
\end{figure*}
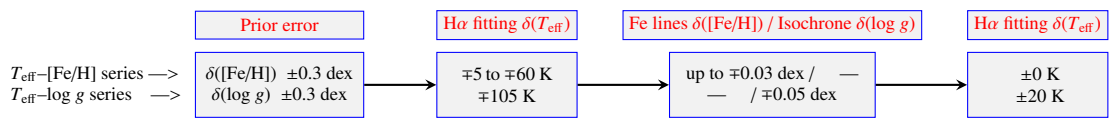

\subsection{Additional data processing}\label{sec:datareduction}

The spectra were Doppler corrected to rest-frame wavelength scale with the iSpec software framework \citep{blanco-cuaresma2014} using a  pre-computed synthetic solar spectrum available within iSpec as template (namely, Synth.Sun.300\_1100nm).
When the corrections are applied, iSpec also transforms the Doppler-shifts of the observational spectra from wavelength units to velocity units. These values are later used to compute heliocentric velocities as described in the next section. 
According to the software documentation, the template spectrum was generated by the code SPECTRUM \citep{1994AJ....107..742G} using a MARCS model \citep{gustafson2008}, the default VALD \citep{2011BaltA..20..503K} line list, and the parameters \teff = 5777 K, \logg = 4.44, [Fe/H] = 0.00, microturbulence velocity ($v_{mic}$) = 1.07 km s$^{-1}$, and macroturbulence ($v_{mac}$) = 4.19 km s$^{-1}$ with $R \sim 300\,000$. 

iSpec was also used to normalize the spectra for the metallicity determination. The routine used for the global normalization within iSpec was configured to minimize errors, avoiding for example flux counts above the unity that are clearly not related to the spectral noise. Every normalized spectrum was visually inspected in the regions around the Fe lines that were selected for the metallicity analysis.

Separately, we corrected a duplicated set of spectra using the IRAF\footnote{IRAF is distributed by the National Optical Astronomy Observatory, which is operated by the Association of Universities for Research in Astronomy (AURA) under a cooperative agreement with the National Science Foundation: \url{http://ast.noao.edu/data/software}.} task \textit{dopcor}. This set was used for the H$\alpha$ profile fitting, which was performed outside iSpec. Normalization and fitting of H$\alpha$ profiles is a specific procedure, described in detail in \citet[][]{giribaldi2019_Ha} and briefly explained in Section \ref{sec:teff} below.

\subsection{Kinematics and orbits}\label{sec:orbits}

We computed Galactic velocities and orbits for the sample stars, to be able to investigate their membership in different substructures of the Galaxy. For that, we used coordinates, proper motions and parallaxes from \textit{Gaia} EDR3 \citep{2020arXiv201203380L}. 
All parallax values had relative errors, $\sigma_\varpi/\varpi$, better than 20\%. We thus used a simple parallax inversion to compute the stellar distances \citep[see][]{BailerJones2015,Luri2018}. 
We applied corrections to parallax biases discussed in  \citet{2020arXiv201201742L} only to the stars for which the biases were found significant. 
These are BPS CS 22166-030 and UCAC2 20056019, and their corrections are $+0.05$ and $+0.06$~mas, respectively. However, they still induced negligible changes in the derived values of \logg\ and \teff.
These are approximately $-0.03$~dex and $+15$~K on each parameter, respectively. For simplicity, for the remaining stars the biases were ignored as their effect on the parameters would be even smaller.

Heliocentric radial velocities were derived by means of Astropy routines \citep{astropy:2013, astropy:2018}, using the Doppler corrections computed as described in Sec.~\ref{sec:datareduction}. 
The heliocentric radial velocities ($v_{helio}$) of the \titan\ are listed in Table \ref{tab:membership}. 
We also computed the orbits for the GBS using \textit{Gaia} EDR3 and their $v_{helio}$ values. The orbits of the stars were integrated for a total of 15 Gyr using the program \textit{galpy} \citep{Bovy2015} with the gravitational potential of \cite{McMillan2017}. A Monte Carlo method with 100 random samples was used to estimate the uncertainties in the orbital parameters. From these calculations, we extracted quantities like the orbital energy, the orbital angular momentum, and the orbital actions.

\section{Atmospheric parameters}\label{sec:parameters}

The atmospheric parameters were determined combining H$\alpha$ profile fitting, spectral synthesis of neutral and ionized iron lines, and the fitting of isochrones in two auto-consistent loop series. 
Each of these three techniques were used to derive \teff, [Fe/H], and \logg, respectively, while keeping the other atmospheric parameters fixed. 
As the starting values (priors) we used those in Table~\ref{tab:preliminar_parameters}. The series of steps in the analysis can be summarised as follows.

First, we run a \teff--[Fe/H] loop series that starts by determining \teff\ from the H$\alpha$ line fitting, keeping the prior values of \logg\ and [Fe/H] fixed. Then, [Fe/H] is derived through the fitting of the observed Fe lines with synthetic spectra computed with the new value of \teff\ together with the prior value of \logg. 
As the H$\alpha$ profile is not sensitive to [Fe/H] in the metal-poor regime, consistency between \teff\ and [Fe/H] was accomplished by the end of the first loop in all cases.
    
Second, we run a \teff--\logg\ loop series that starts by determining \logg\ using theoretical isochrones, adopting the \teff\ and [Fe/H] values computed in the first loop series. The newly derived \logg\ value substitutes the prior for the next iteration that rederives \teff. The loop series continues until consistency is reached between \teff\ and \logg. In most cases, it was enough to run this loop series twice.

Finally, as the last step, the \teff\ and \logg\ values derived in the second loop are fixed to derive the ultimate [Fe/H] value. 
The conclusive atmospheric parameters of the \titan\ are given in Table \ref{tab:CAMK_stars}.
The table also incorporates the metal-poor GBS (with [Fe/H] < $-0.8$~dex), as for many of them our new parameters are more accurate than the estimates previously available. Element abundances for the \titan\ and the metal-poor GBS will be discussed in subsequent works in this series.
Deriving accurate \teff\ values from H$\alpha$ fitting in metal-poor stars is more challenging than in metal-rich stars, because of the degeneracy between \teff\ and \logg\ that becomes relevant. In Fig.~\ref{fig:error_flow}, we show an error flow diagram that illustrates the impact of one given parameter on the other parameters calculated in the next steps of the loop series. This shows how each step of the analysis progressively removes parameter offsets. It can be seen that large variations in metallicity induce only small changes in \teff\ at the beginning of the loops. These changes become progressively negligible as the loops are completed. On the other hand, changes in \logg\ still induce small variations in \teff\ at the end of the first \teff--\logg\ loop. This explains why a second \teff--\logg\ loop was usually needed. The methods used to constrain each parameter are described in more detail below.

\subsection{Effective temperature}
\label{sec:teff}

The most important development with respect to the methodology used in \citet{giribaldi2019_Ha} is the adoption of a new grid of hydrodynamic 3D non-LTE H$\alpha$ profiles, synthesized with the codes provided by \citet{amarsi2018}. The grid covers the dimensions: \teff\ from $4000$ to $7000$~K (in steps of 100~K), \logg\ from 1.0 to 5.0~dex (in steps of 0.1~dex), and [Fe/H] from $-$4.0 to +1.0~dex (in steps of 0.5~dex). The projected rotational velocity (\vsini) and the Gaussian instrumental profile ($v\,broad$) were fixed to default values of 1.85 and 1.0 km s$^{-1}$, respectively. In any case, their contributions to the broadening of the H$\alpha$ wings are negligible.

The normalization-fitting procedure used to derive \teff\ from H$\alpha$ profiles is described in detail in Sections~3 and 4 of \citet{giribaldi2019_Ha}. Briefly, normalization and fitting are parts of one iterative procedure in which visual compatibility between the observed and synthetic profiles, along the wavelength axis, is required. In this way, we can adjust the procedure to correct for profile distortions induced by the continuum normalization.

\begin{figure*}
    \centering
    \includegraphics[width=0.85\linewidth]{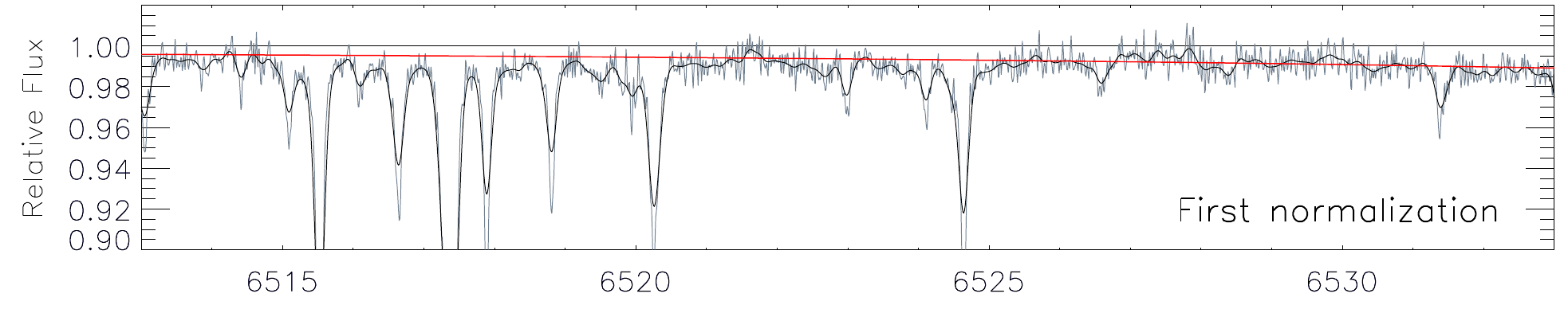}
    \includegraphics[width=0.85\linewidth]{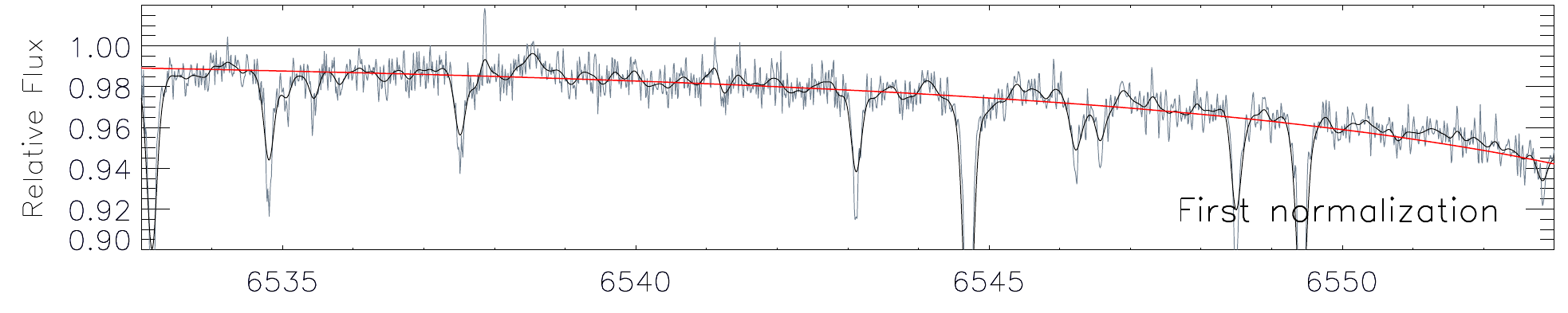}
    \includegraphics[width=0.85\linewidth]{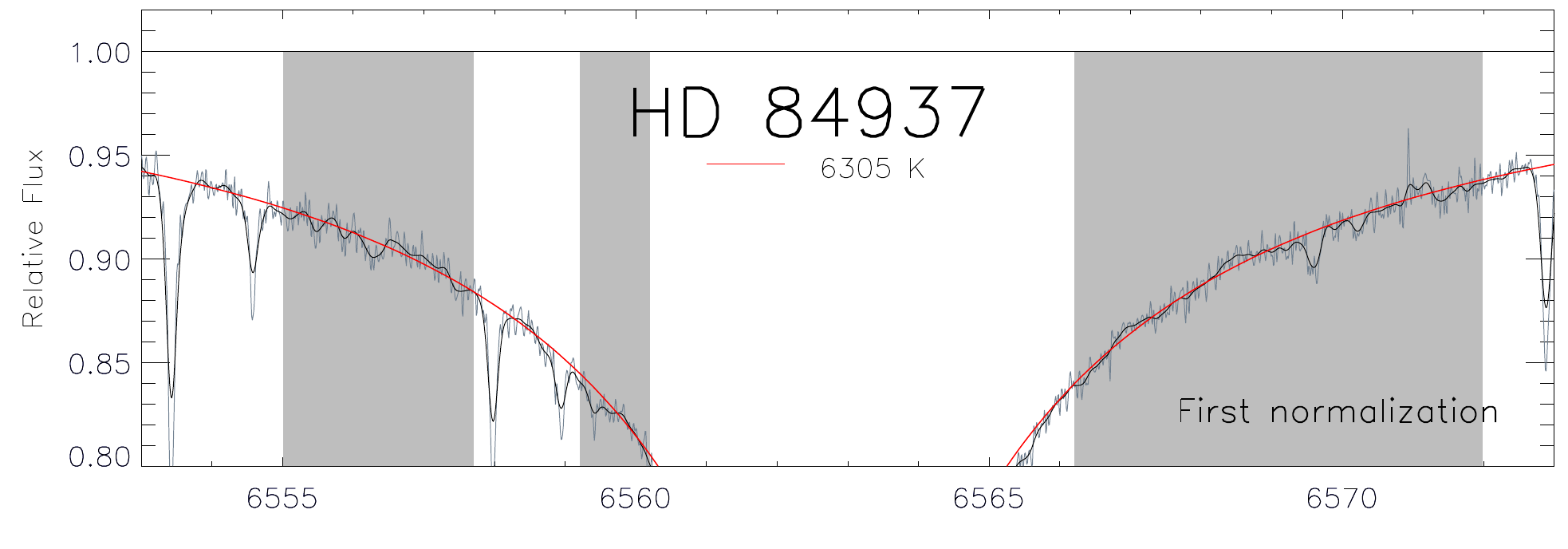}
    \includegraphics[width=0.85\linewidth]{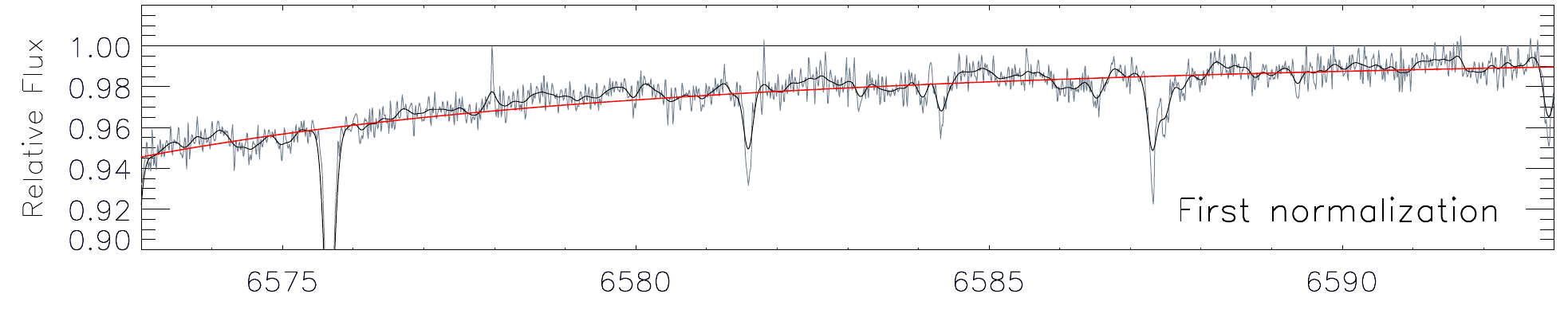}
    \includegraphics[width=0.85\linewidth]{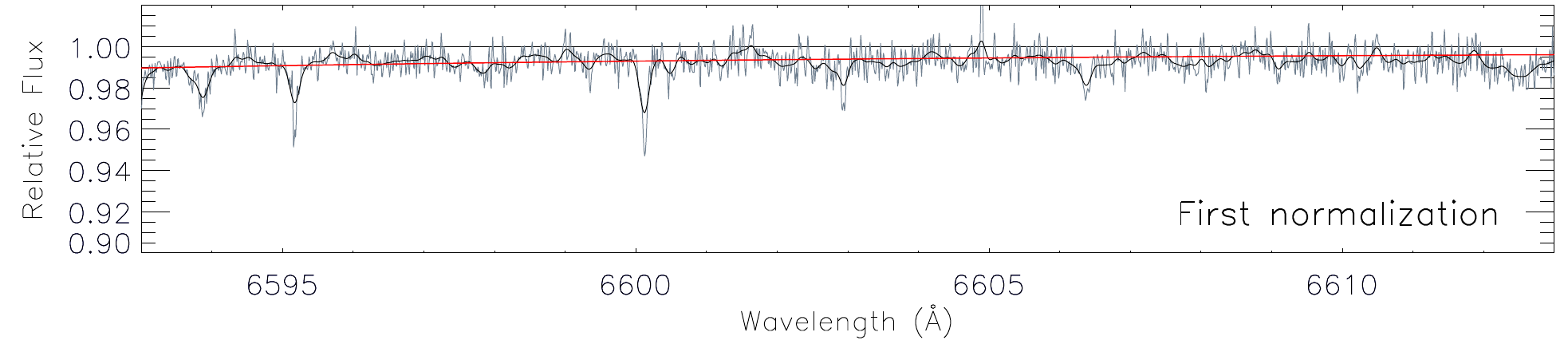}
    \includegraphics[width=0.5\linewidth]{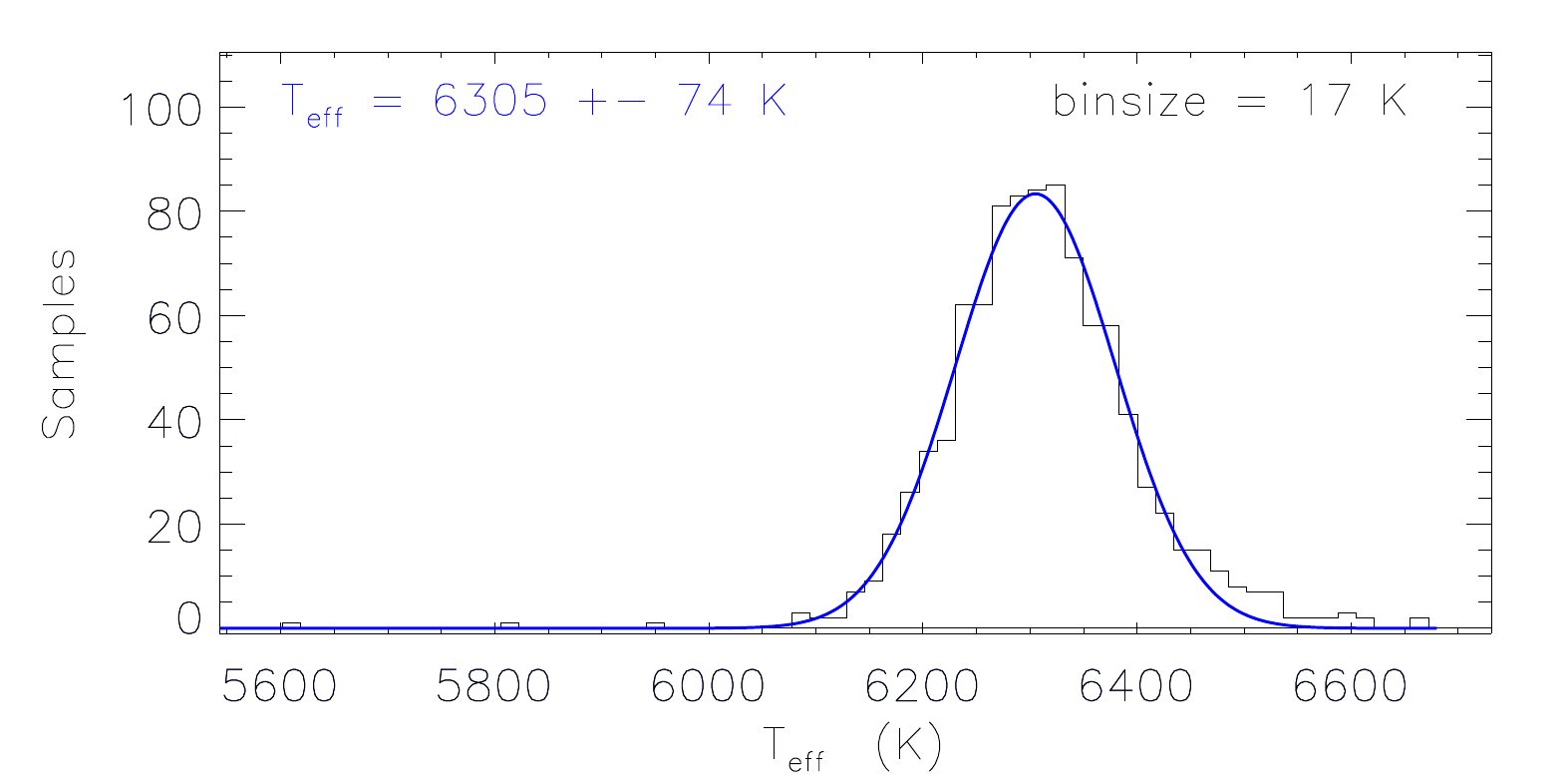}
    \caption{First normalization-fitting for HD~84937 with HARPS. The spectrum is segmented in peaces of 20~\AA\ to facilitate the visualization of the details.
    The original spectrum is plotted in gray, and the black line is the spectrum degraded to facilitate the comparison with the synthetic profile showed in red. The shades are the fitting masks.
    \textit{Bottom panel:} Gaussian fit to the histogram of temperatures related to the wavelength bins within the windows represented by the shades.}
    \label{fig:extension_error}
\end{figure*}

\begin{figure*}
    \centering
    \includegraphics[width=0.85\linewidth]{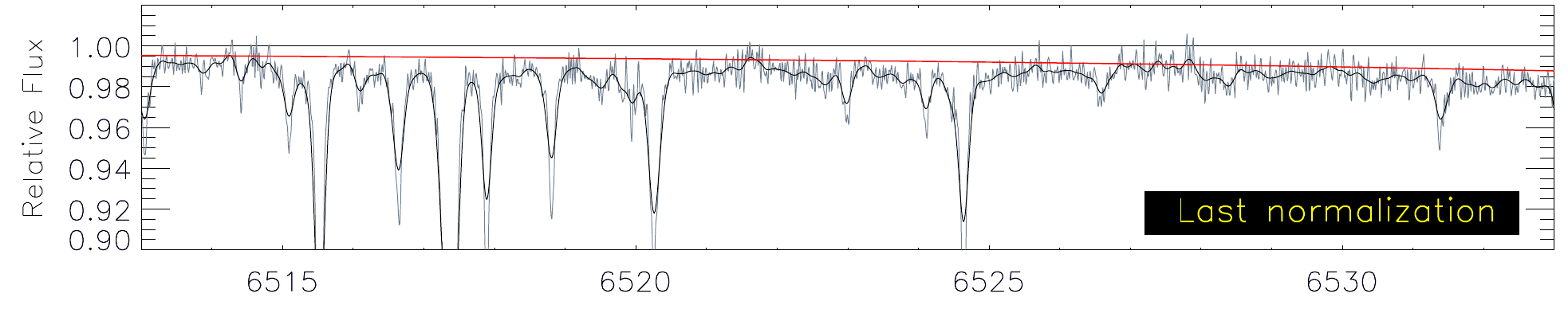}
    \includegraphics[width=0.85\linewidth]{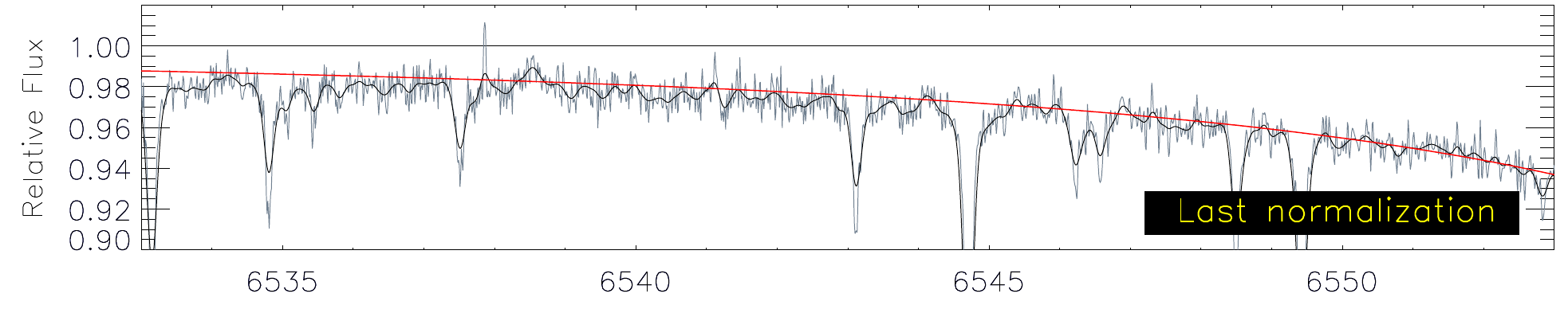}
    \includegraphics[width=0.85\linewidth]{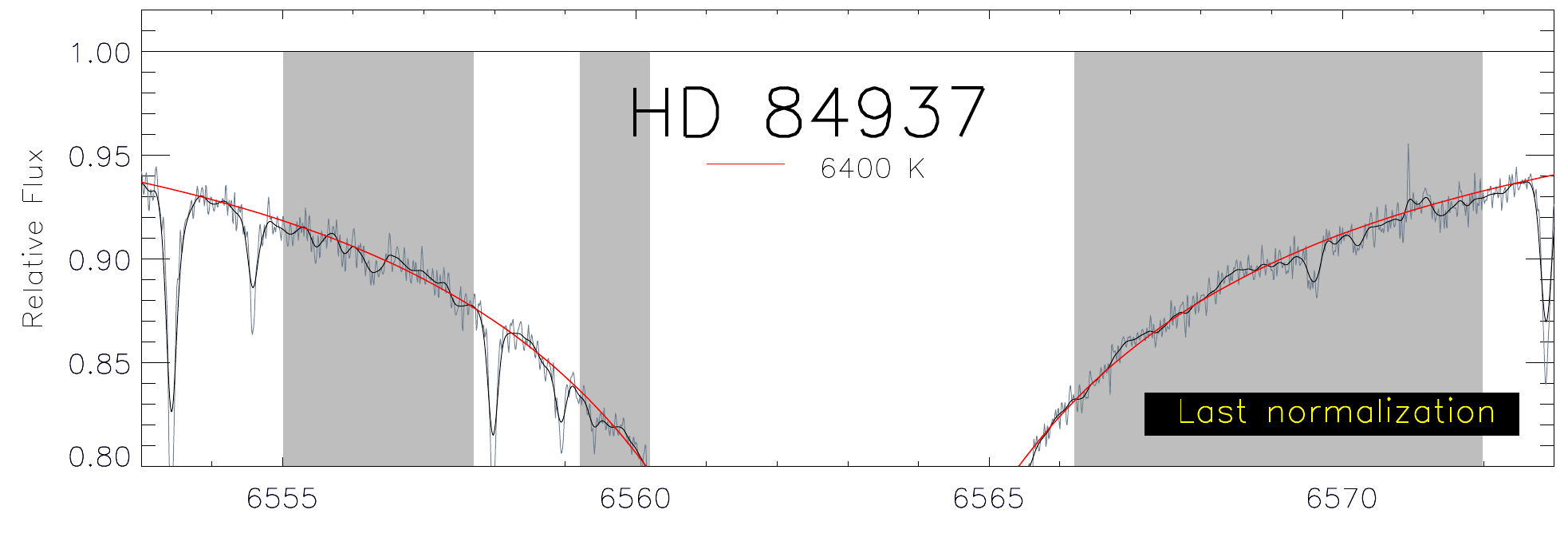}
    \includegraphics[width=0.85\linewidth]{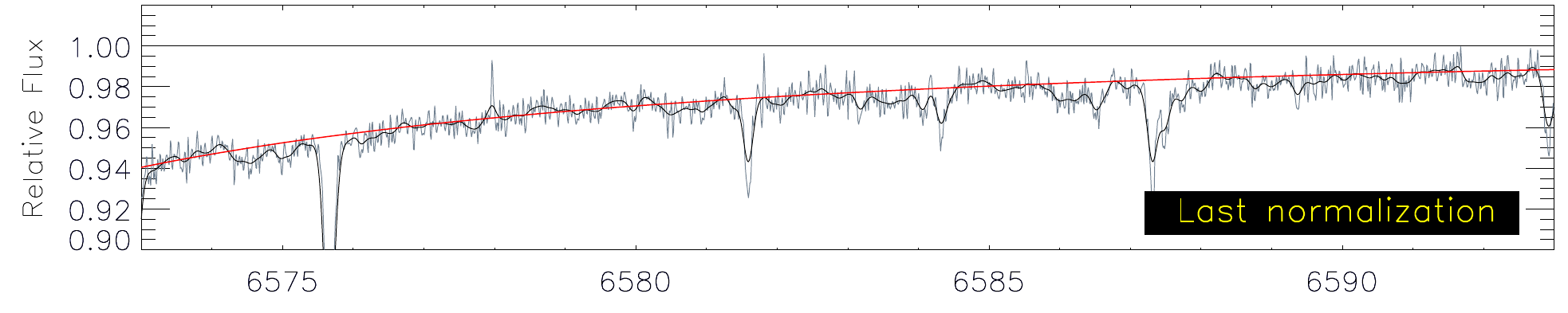}
    \includegraphics[width=0.85\linewidth]{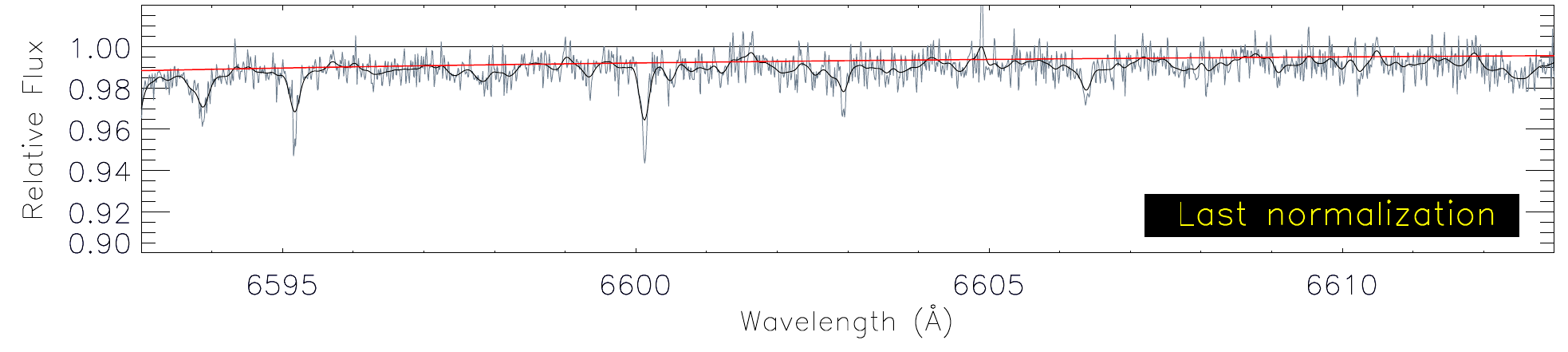}
    \includegraphics[width=0.5\linewidth]{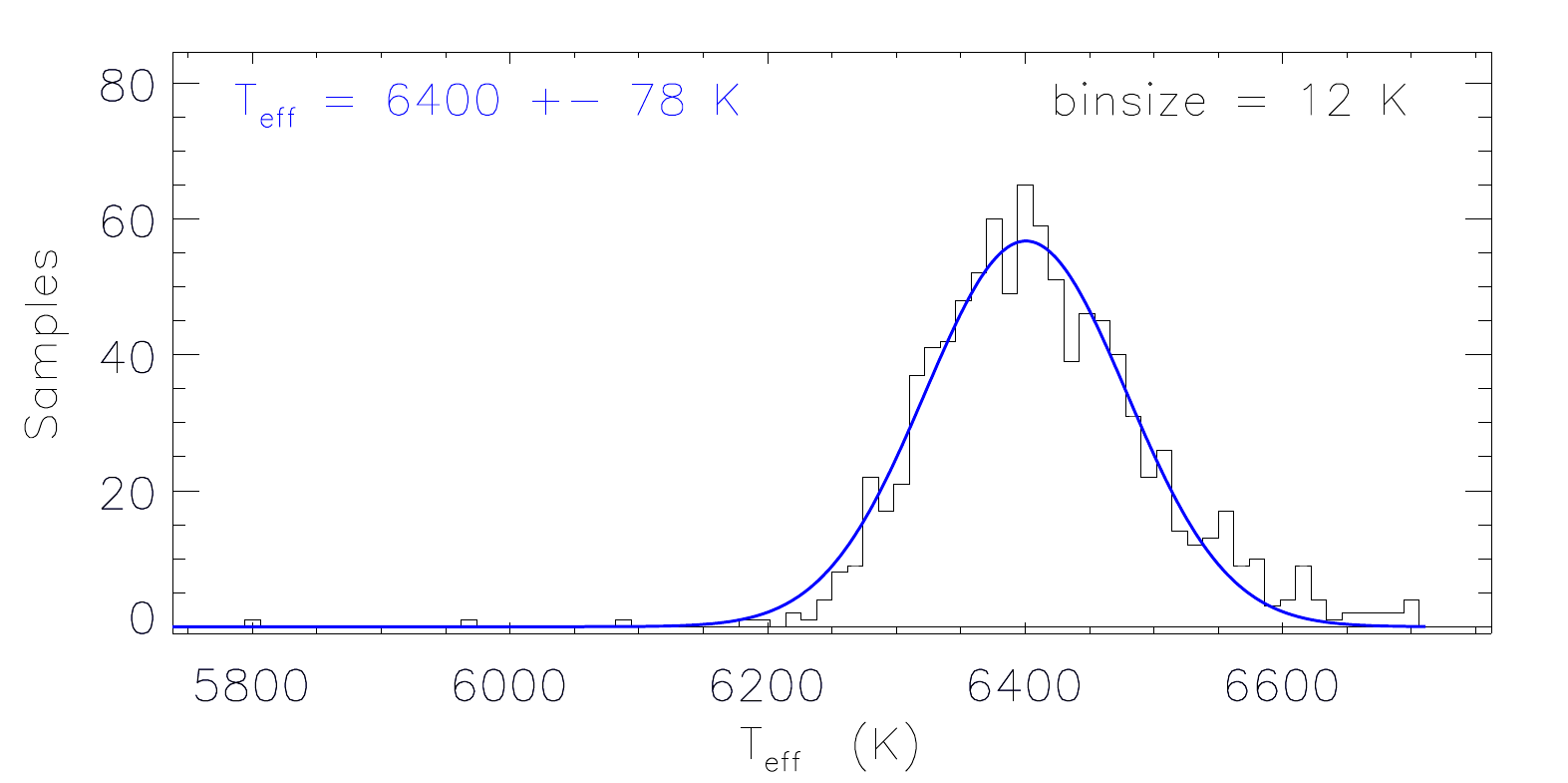}
    \caption{\centering Last normalization-fitting for HD~84937 with HARPS. The elements in the plots are same as in Fig.~\ref{fig:extension_error}.}
    \label{fig:extension_fixed}
\end{figure*}

The first and last steps of this procedure are exemplified in Figs.~\ref{fig:extension_error} and~\ref{fig:extension_fixed}, for the case of a HARPS spectrum of HD~84937, one of the GBS. 
In the first normalization, the observed profile is seen to be slightly shallower than the synthetic one in the external regions. 
Normalization errors become more evident in the outer regions of the wings because the profile sensitivity to the temperature vanishes as the wavelength approaches the continuum. 
The H$\alpha$ profile fits of the \titan, from which \teff\ was determined, are compiled in the Appendix~\ref{app:titans}.

\begin{table*}
\caption{Fundamental parameters and metallicities of the Titans.}
\label{tab:CAMK_stars}
\centering
\tiny 
\begin{threeparttable}
\begin{tabular}{lcccccccccc}
\hline\hline
Name & \teff\ (K) & \logg & $\mathcal{M}$ ($\mathcal{M}_\odot$) & $R$ ($R_\odot$) & Age (Gyr) & [\ion{Fe}{i}/H] & [\ion{Fe}{ii}/H] & $V$ & $E(B-V)$ \\
\hline
BD+24 1676 & $6440 \pm 33$ & $4.09 \pm 0.04$ & $ 0.79 \pm 0.01 $ & $ 1.31 \pm 0.05 $ & $ 13.07 \pm 0.67 $ & $ -2.26$ & $-2.24$ & $10.81$ & $0.015$  \\
BD$-10$ 388 & $6293 \pm 43$ & $3.90 \pm 0.04$ & $ 0.81 \pm 0.01 $ & $ 1.67 \pm 0.07 $ & $ 12.21 \pm 0.54 $ & $-2.21$ & $-2.22$ & $10.35$ & $0.029$  \\
BPS CS22166-030 & $6255 \pm 39$ & $3.76 \pm 0.04$ & $ 0.80 \pm 0.01 $ & $ 1.89 \pm 0.08 $ & $ 13.25 \pm 0.66 $ & $-2.98$ & $-3.11$ & $13.62$ & $0.023$ \\
CD$-33\,3337^\S$ & $6087 \pm 36$ & $3.97 \pm 0.04$ & $ 0.85 \pm 0.01 $ & $ 1.58 \pm 0.06 $ & $ 11.42 \pm 0.65 $ & $-1.33$ & $-1.34$ & $9.11$ & $0.002$ \\
CD$-35$ 14849 & $  6377 \pm 34$ & $4.37 \pm 0.04$ & $ 0.78 \pm 0.02 $ & $ 0.95 \pm 0.03 $ & $ 11.12 \pm 1.73 $ & $-2.17$& $-2.15$ & $10.56$ & $0.060$  \\
HD 16031 & $6266 \pm 31$ & $4.27 \pm 0.04$ & $ 0.80 \pm 0.02 $ & $ 1.09 \pm 0.04 $ & $ 11.78 \pm 1.23 $ & $-1.63$ & $-1.63$ & $9.77$ & $0.003$  \\
HD 166913 & $6172 \pm 35$ & $4.23 \pm 0.04$ & $ 0.78 \pm 0.02 $ & $ 1.13 \pm 0.04 $ & $ 12.97 \pm 1.19 $ & $-1.51$ & $-1.51$ & $8.22$ & $0.006$ \\
HD 218502 & $6323 \pm 32$ & $4.10 \pm 0.04$ & $ 0.80 \pm 0.01 $ & $ 1.31 \pm 0.05 $ & $ 12.43 \pm 0.68 $ & $-1.75$ & $-1.74$ & $8.25$ & $0.002$ \\
HD $34328^\S$ & $5997 \pm 36$ & $4.48 \pm 0.04$ & $ 0.74 \pm 0.02 $ & $ 0.82 \pm 0.03 $ & $ 11.72 \pm 2.27 $ & $-1.56$ & $-1.64$ & $9.46$ & $0.001$ \\
HD 59392 & $6160 \pm 39$ & $4.02 \pm 0.04$ & $ 0.82 \pm 0.01 $ & $ 1.47 \pm 0.06 $ & $ 12.21 \pm 0.70 $ & $-1.52$ & $-1.53$ & $9.76$ & $0.005$ \\
BD$+02\,4651^\S$ & $6274 \pm 49$ & $4.07 \pm 0.04$ & $ 0.81 \pm 0.01 $ & $ 1.37 \pm 0.06 $ & $ 12.13 \pm 0.80 $ & $-1.63$ & $-1.62$ & $10.20$ & $0.037$ \\
BD+03 740 & $6385 \pm 48$ & $3.99 \pm 0.04$ & $ 0.78 \pm 0.01 $ & $ 1.48 \pm 0.06 $ & $ 13.94 \pm 0.60 $ & $-2.56$ & $-2.52$ & $9.81$ & $0.019$ \\
BD$-13$ 3442 & $6446 \pm 34$ & $4.00 \pm 0.04$ & $ 0.79 \pm 0.01 $ & $ 1.48 \pm 0.06 $ & $ 13.14 \pm 0.57 $ & $-2.60$ & $-2.61$ & $10.27$ & $0.026$ \\
BD+26 2621 & $6457 \pm 31$ & $4.31 \pm 0.04$ & $ 0.78 \pm 0.02 $ & $ 1.02 \pm 0.04 $ & $ 11.97 \pm 1.30 $ & $-2.43$ & $-2.42$ & $11.02$ & $0.031$ \\
BD+26 4251 & $6107 \pm 33$ & $4.24 \pm 0.04$ & $ 0.80 \pm 0.02 $ & $ 1.12 \pm 0.04 $ & $ 12.72 \pm 1.27 $ & $-1.24$ & $-1.28$ & $10.05$ & $0.024$ \\
BD+29 2091 & $5879 \pm 32$ & $4.56 \pm 0.04$ & $ 0.70 \pm 0.02 $ & $ 0.72 \pm 0.02 $ & $ 11.82 \pm 2.42 $ & $-1.86$ & $-1.96$ & $10.24$ & $0.004$ \\
CD$-30$ 18140 & $6317 \pm 33$ & $4.18 \pm 0.04$ & $ 0.78 \pm 0.01 $ & $ 1.19 \pm 0.05 $ & $ 13.07 \pm 0.94 $ & $-1.83$ & $-1.79$ & $9.93$ & $0.007$ \\
CD$-33$ 1173 & $6602 \pm 32$ & $4.30 \pm 0.04$  & $ 0.70 \pm 0.01 $ & $ 1.05 \pm 0.04 $ & $ 11.23 \pm 1.06 $ & $-2.67$ & $-2.72$ & $10.90$ & $0.005$ \\
CD$-48\,2445^\S$ & [$6545 \pm 32$] & $3.95 \pm 0.04$ & $ 0.90 \pm 0.01 $ & $ 1.65 \pm 0.06 $ & $ 8.97 \pm 0.46 $ & $-1.71$ & $-1.88$ & $9.64$ & $0.023$ \\
CD$-71$ 1234 & $6432 \pm 31$ & $4.28 \pm 0.04$ & $ 0.78 \pm 0.02 $ & $ 1.06 \pm 0.04 $ & $ 12.30 \pm 1.22 $ & $-2.24$ & $-2.23$ & $10.44$ & $0.028$ \\
G~24$-3$ & $6076 \pm 31$ & $4.36 \pm 0.04$ & $ 0.76 \pm 0.01 $ & $ 0.95 \pm 0.03 $ & $ 13.05 \pm 1.42 $ & $-1.50$ & $-1.50$ & $10.46$ & $0.040$ \\
HD~108177 & $6274 \pm 34$ & $4.36 \pm 0.05$ & $ 0.82 \pm 0.02 $ & $ 1.00 \pm 0.04 $ & $ 9.48 \pm 1.64 $ & $-1.56$ & $-1.58$ & $9.66$ & $0.004$ \\
HD~$116064^\S$ & [$6022 \pm 37$] & $4.33 \pm 0.04$ & $0.74 \pm 0.01$  & $0.97 \pm 0.03$ & $ 14.32 \pm 0.67$ & $-2.00$ & $-1.90$ & $8.81$ & $0.005$ \\
HD~$122196^\S$ & $6097 \pm 32$ & $3.93 \pm 0.04$ & $ 0.81 \pm 0.01 $ & $ 1.61 \pm 0.06 $ & $ 12.85 \pm 0.66 $ & $-1.69$ & $-1.65$ & $8.75$ & $0.012$ \\
HD~126681 & $5605 \pm 31$ & $4.61 \pm 0.03$ & $ 0.70 \pm 0.02 $ & $ 0.69 \pm 0.02 $ & $ 11.20 \pm 3.09 $ & $-1.22$ & $-1.22$ & $9.54$ & $0.002$ \\
HD~132475 & $5786 \pm 41$ & $3.83 \pm 0.04$ & $ 0.81 \pm 0.01 $ & $ 1.80 \pm 0.07 $ & $ 13.51 \pm 0.77 $ & $-1.41$ & $-1.44$ & $8.55$ & $0.015$ \\
HD~160617 & $6102 \pm 43$ & $3.92 \pm 0.04$ & $ 0.80 \pm 0.01 $ & $ 1.63 \pm 0.07 $ & $ 12.97 \pm 0.69 $ & $-1.69$ & $-1.69$ & $8.73$ & $0.006$ \\
HD~189558 & $5701 \pm 32$ & $3.86 \pm 0.04$ & $ 0.84 \pm 0.02 $ & $ 1.80 \pm 0.07 $ & $ 12.50 \pm 0.85 $ & $-1.13$ & $-1.20$ & $7.72$ & $0.002$  \\
HD~193901 & $5853 \pm 34$ & $4.50 \pm 0.04$ & $ 0.77 \pm 0.02 $ & $ 0.81 \pm 0.03 $ & $ 10.45 \pm 2.95 $ & $-1.10$ & $-1.12$ & $8.67$ & 0.000 \\
HD~213657 & $6275 \pm 40$ & $4.07 \pm 0.04$ & $ 0.79 \pm 0.01 $ & $ 1.36 \pm 0.06 $ & $ 13.31 \pm 0.76 $ & $-1.84$ & $-1.82$ & $9.66$ & $0.004$  \\
HD~241253 & $5929 \pm 35$ & $4.53 \pm 0.04$ & $ 0.78 \pm 0.03 $ & $ 0.79 \pm 0.03 $ & $ 8.18 \pm 3.19 $ & $-1.11$ & $-1.11$ & $10.29$ & $0.002$ \\
HD~284248 & $6270 \pm 33$ & $4.33 \pm 0.05$ & $ 0.81 \pm 0.02 $ & $ 1.01 \pm 0.04 $ & $ 10.50 \pm 1.58 $ & $-1.55$ & $-1.54$ & $9.23$ & $0.006$ \\
HD~74000 & $6408 \pm 40$ & $4.35 \pm 0.05$ & $ 0.81 \pm 0.02 $ & $ 0.99 \pm 0.04 $ & $ 9.96 \pm 1.56 $ & $-1.82$ & $-1.85$ & $9.66$ & $0.002$  \\
HD~94028 & $6097 \pm 40$ & $4.38 \pm 0.04$ & $ 0.79 \pm 0.02 $ & $ 0.95 \pm 0.04 $ & $ 11.47 \pm 2.04 $ & $-1.33$ & $-1.36$ & $8.22$ & $0.001$ \\
HE~0926$-0508$ & $6458 \pm 32$ & $4.10 \pm 0.04$ & $ 0.78 \pm 0.01 $ & $ 1.30 \pm 0.06 $ & $ 13.33 \pm 0.68 $ & $-2.47$ & $-2.50$ & $12.19$ & $0.029$ \\
LP~815$-43$ & $6539 \pm 41$ & $4.11 \pm 0.04$ & $ 0.82 \pm 0.01 $ & $ 1.32 \pm 0.06 $ & $ 11.21 \pm 0.59 $ & $-2.53$ & $-2.57$ & $10.72$ & $0.029$ \\
LP~831$-70$ & $6379 \pm 35$ & $4.48 \pm 0.04$ & $ 0.78 \pm 0.03 $ & $ 0.84 \pm 0.03 $ & $ 9.14 \pm 2.48 $ & $-2.69$ & $-2.70$ & $11.80$ & $0.011$ \\
Ross~453 & $6151 \pm 32$ & $4.39 \pm 0.04$ & $ 0.74 \pm 0.02 $ & $ 0.91 \pm 0.03 $ & $ 13.18 \pm 1.44 $ & $-1.98$ & $-1.92$ & $11.09$ & $0.015$ \\
Ross~892 & $6056 \pm 33$ & $4.44 \pm 0.05$ & $ 0.81 \pm 0.03 $ & $ 0.90 \pm 0.04 $ & $ 9.27 \pm 2.38 $ & $-1.12$ & $-1.21$ & $11.03$ & $0.013$ \\
UCAC2~20056019 & $6443 \pm 33$ & $3.93 \pm 0.03$ & $ 0.81 \pm 0.01 $ & $ 1.62 \pm 0.07 $ & $ 12.16 \pm 0.52 $ & $-2.67$ & $-2.64$ & $13.48$ & $0.012$ \\
Wolf~1492 & $6592 \pm 32$ & $4.29 \pm 0.04$ & $ 0.79 \pm 0.01 $ & $ 1.05 \pm 0.04 $ & $ 11.63 \pm 1.07 $ & $-2.92$ & $-3.04$ & $11.45$ & $0.025$ \\
\hline
HD~140283 & $5810 \pm 32$ & $3.67 \pm 0.02$ & $0.79 \pm 0.01$ & $2.13 \pm 0.07$ & $13.40 \pm 0.69$ & $-2.40$ & $-2.30$ & 7.21 & 0.002\\
HD~84937 & $6358 \pm 33$ & $4.13 \pm 0.03$ & $0.77 \pm 0.01$ & $1.25 \pm 0.04$ & $13.80 \pm 0.51$ & $-2.09$ & $-2.08$ & 8.32 & 0.003\\
HD~22879 & $5894 \pm 23$ & $4.27 \pm 0.04$ & $0.77 \pm 0.01$ & $1.09 \pm 0.04$ & $13.84 \pm 0.70$  & $-0.88$  & $-0.86$ & 6.67 & 0.00 \\
HD~298986 & $6177 \pm 39$ & $4.28 \pm 0.04$ & $0.81 \pm 0.01$ & $1.08 \pm 0.04$ & $11.50 \pm 0.70$ & $-1.33$ & $-1.32$ & 10.03 & 0.008\\
HD~106038 & $6094 \pm 43$ & $4.29 \pm 0.04$ & $0.77 \pm 0.01$ & $1.05 \pm 0.04$ & $13.80 \pm 0.94$ & $-1.31$ & $-1.36$ & 10.16 & 0.008\\
HD~201891 & $6007 \pm 37$ & $4.34 \pm 0.04$ & $0.79 \pm 0.01$ & $0.99 \pm 0.03$ & $12.30 \pm 1.06$ & $-0.97$ & $-1.10$ & 7.37 & 0.001\\
HD~102200 & $6246 \pm 19$ & $4.36 \pm 0.04$ & $0.87 \pm 0.01$ & $1.01 \pm 0.03$ & $7.90 \pm 0.83$ & $-1.08$ & $-1.13$ & 8.76& 0.004\\
\hline
\end{tabular}
\begin{tablenotes}
\item{} \textbf{Notes.} {The symbol $\S$, following the star name, indicates a binary system identified in \citet{krevella2019} or \citet{hansen2017}. 
Values of \teff\ in square brackets are uncertain according to the analysis in Sec.~\ref{sec:IRFM}.} 
Uncertainties accompanying the \teff\ values are internal (i.e. the method precision). For the total uncertainties, one has to further include the estimated model accuracy error of $\pm$ 50 K (as discussed in Sec.~\ref{sec:error}).
$V$ is the apparent magnitude. 
$E(B-V)$ is the colour excess.
The bottom part of the table incorporates the metal-poor $Gaia$ benchmark stars. The new parameters determined here for these stars are likely more accurate than previously available estimates.
\end{tablenotes}
\end{threeparttable}
\end{table*}  

The fitting of the H$\alpha$ profile is done using masks that select regions without metallic or telluric lines. 
The line core (at least within $\pm1.8$~\AA\ from $\lambda6562.797$~\AA) is systematically masked out because in this region the photospheric models depart from the observational profiles, 
as the contribution of the chromosphere progressively increases as $\lambda$ approaches the line center \citep[e.g.][]{leenaarts2012}.
The choice of the masks varies because additional fitting regions become available for hotter and/or more metal-poor stars. 
For each observed profile, the fitting is done per pixel inside the regions selected by each mask. In other words, for each pixel of the observed spectrum we search for the synthetic profile that has the closest value of normalized flux. 
This results in a distribution of \teff\ values, obtained from the analysis of each pixel. Such distributions are depicted in the bottom panels of Figs.~\ref{fig:extension_error} and~\ref{fig:extension_fixed}. 
From these distributions, we obtain an estimate of the best \teff\, value, and of its error, by means of a Gaussian fit.
This approach demonstrated to be efficient to minimize the contribution of pixels contaminated by unexpected weak telluric or metallic lines in the temperature budget. Any contaminated pixel would have a lower flux beyond what would be caused by the noise pattern. These pixels are then associated to higher temperatures that make the histograms depart from a normal distribution on its right tail.
Hence, the fitting of a Gaussian profile to the temperature distribution tends to clip the contribution of contaminated pixels.
An example of this performance can be observed in Fig.~\ref{fig:extension_error} and \ref{fig:extension_fixed}, where the contaminated pixels are located at $\sim6569.5$~\AA.

When possible, for each star, several observed H$\alpha$ line profiles were used to derive separate estimates of \teff. This helps in clipping outliers related to distorted wing profiles. The final best estimate of \teff\ was then computed with a weighted mean of the values obtained from each individual spectrum, after applying a 2$\sigma$ clipping. The weights are given by the square inverse of the 1$\sigma$ errors of the \teff\ from each fit. For the stars that only have one good spectrum available (HE~0926-0508, Ross~453, Ross~892, UCAC2~20056019, and Wolf~1492), in order to improve the robustness of the estimated \teff\ and of its uncertainty, we performed a simple Bayesian analysis. We used a normal distribution prior over a Monte Carlo sampling of the histograms with $10^6$ points. The reported values were set to fall inside a confidence interval of 95\%.

\subsection{Metallicity, micro and macroturbulence}
\label{sec:metallicity}
Metallicity was determined by fitting synthetic spectra using iSpec to run  the the 1D LTE radiative transfer code Turbospectrum \citep{turbospectrum}. 
MARCS model atmospheres \citep{gustafson2008}, computed under LTE, were used. 
The procedure was performed in two steps. In the first step, we relied on the statistical power of many lines to constrain $v_{mic}$ and $v_{mac}$. In the second step, we refined the Fe lines selection, and used only the best ones to constrain the metallicity.

For the first step we adopted a selection of transitions, based on the Gaia-ESO line list \citep{2015PhyS...90e4010H,2021A&A...645A.106H}, that is provided inside iSpec\footnote{The selection "turbospectrum\_synth\_good\_for\_params\_all".} and combined it with the \ion{Fe}{ii} line list of \citet{melendez_barbuy2009}. We chose a total of 35 \ion{Fe}{i} and \ion{Fe}{ii} lines that were visually found to be well shaped and not extremely blended in two high-quality spectra of HD~140283 and HD~84937. 
Following \citet{2019MNRAS.486.2075B}, we always fixed \vsini\ to the value used in that work for the Sun, i.e. $v\mathrm{sin}\,i_{\odot}$ = 1.6~km s$^{-1}$, and let only $v_{mic}$ and $v_{mac}$ vary freely. 
As discussed in \citet{2019MNRAS.486.2075B} and elsewhere \citep[e.g.][]{2016A&A...592A.156D}, the effects of \vsini\ and $v_{mac}$ are difficult to disentangle from the spectral analysis only. By fixing a value for \vsini\ we can, by direct line fitting, find the optimal combination of parameters that is needed to reproduce the broadening of the lines. We will discuss tests regarding the degeneracy between these parameters in more detail in the second paper of the series, where we plan to present chemical abundances for the \titan\ (Giribaldi \& Smiljanic, in preparation). 

For the second step we adopted the selection of good \ion{Fe}{i} lines of metal-poor dwarfs from \citet[][]{jofre2014} and, separately, selected isolated \ion{Fe}{ii} lines from \citet{melendez_barbuy2009}. 
Thus, two sets of [Fe/H] values were derived, one from the neutral lines and another from the ionized lines. With the broadening fixed to the parameters determined in the first step, only the Fe abundance was allowed to vary to fit each line. The best estimate of [Fe/H] was determined by the average of the individual line abundances, applying a 2$\sigma$-clipping. 
For stars with more than one spectrum under analysis, the final metallicity was computed by the weighted mean of the metallicities obtained from the analysis of each spectrum. 
This procedure was validated by recovering the [Fe/H] values (under LTE) for the three metal-poor GBS HD~140283, HD~84937, and HD~22879, as given in Table~3 of \citet[]{jofre2015}, to better than $\pm$0.05 dex (when using the \ion{Fe}{i} lines). 
We did not attempt to rederive metallicities for the metal-rich GBS, as our focus is on metal-poor stars and analyzing the remaining GBS would require a different selection of Fe lines.

Metallicity values from neutral and ionized lines for the \titan\ are listed in Table.~\ref{tab:CAMK_stars}. 
Both sets of [Fe/H] are compatible, the mean difference is zero and the $\sigma$ dispersion is $\pm$0.05~dex. Further, there is no correlation of the [Fe/H] difference with \teff, [Fe/H], and \logg.
According to the analysis of \citet{amarsi2016} on HD~84937 and HD~140283, hydrodynamical 3D non-LTE models shift the 1D LTE abundances from \ion{Fe}{ii} lines by $+0.1$~dex, on average, with larger corrections for the \ion{Fe}{i} lines. Because of this, we adopted [\ion{Fe}{ii}/H] for the \teff(H$\alpha$) determination, although without any offset correction. The impact of this decision on the \teff\ error budget is discussed in Sec.~\ref{sec:error}. 
We postpone a more detailed discussion about the metallicities (and possible systematic corrections) for the second paper in the series (Giribaldi \& Smiljanic, in preparation).

\subsection{Surface gravity}
\label{sec:logg}

Surface gravity, \logg$_{\mathrm{iso}}$, was determined from theoretical isochrones with the Python code $q^2$ (qoyllur-quipu)\footnote{\url{https://github.com/astroChasqui/q2}} developed by \citet{ram_2014}. It makes use of Yonsei-Yale isochrones \citep{kim2002,yi2003} and requires parallax, $V$ magnitude, \teff, [Fe/H], and their respective uncertainties as inputs. The code computes a probability distribution of the stellar mass using a frequentist approach.
$q^2$ also provides the associated stellar $\mathcal{M}$, $R$, and Age as outputs. They were compiled along with the atmospheric parameters in Table~\ref{tab:CAMK_stars}.

Parallaxes from the Gaia EDR3 catalogue were used. $V$ magnitudes were mostly retrieved from the SIMBAD database or, when not available, from \citet{Casagrande2010} or from the Hipparcos catalogue \citep{Perryman}.
The $V$ magnitudes were corrected from extinction using reddening values, $E(B-V)$, computed with the online tool \textit{Stilism}\footnote{\url{stilism.obspm.fr}} \citep{Capitanio2017} and adopting the relation $A_{V} = 3.1 \times E(B-V)$. When the stellar distance (estimated here as the inverse of the parallax) exceeds the limits of \textit{Stilism}, the $E(B-V)$ was alternatively computed using the Infrared Science Archive online tool.\footnote{\url{https://irsa.ipac.caltech.edu/applications/DUST/}} This tool provides estimates of Galactic dust extinction based on \citet{Schlafly2011}. To these reddening values, that represent the total for a line of sight, we applied the corrections given by \citet{Beers}. The $V$ magnitudes and reddening values that were used are also listed in Table~\ref{tab:CAMK_stars}.

To calculate the \logg\ value in the first of the \teff--\logg\ loop series, we assumed 0.2~dex and 0.05~mag as the uncertainties of [Fe/H] and $V$ (as needed by the code $q^2$). We also added 20~K to the internal uncertainty of \teff, to account possible additional uncertainties through the loops. Once the error analysis described in Sec.~\ref{sec:error} was finished, we recomputed \logg$_{\mathrm{iso}}$ and its uncertainty using the correct value determined with Eq.~\ref{eq:uncertainty}.

\subsection{Uncertainty analysis}
\label{sec:error}
The total uncertainty of the \teff\ value is given by:

\begin{equation}
    \label{eq:uncertainty}
    \centering
    \delta(T_{\mathrm{eff}})^{2} = 
    \delta_{T_{\mathrm{eff}}-\mathrm{model}}^{2} +
    \delta_{T_{\mathrm{eff}}-\mathrm{fit}}^{2} +
    \delta_{T_{\mathrm{eff}}-\mathrm{inst}}^{2} +
    \delta_{T_{\mathrm{eff}}-\mathrm{log}\;g}^{2} +
    \delta_{T_{\mathrm{eff}}-\mathrm{[Fe/H]}}^{2}
\end{equation}

\noindent
Where $\delta_{T_{\mathrm{eff}}-\mathrm{model}}$ is the uncertainty of the model profile, 
$\delta_{T_{\mathrm{eff}}-\mathrm{fit}}$ is the uncertainty of the fitting procedure, 
$\delta_{T_{\mathrm{eff}}-\mathrm{inst}}$ is the uncertainty induced by any instrument residual pattern,
$\delta_{T_{\mathrm{eff}}-\mathrm{log}\;g}$ is the uncertainty related to the interdependence between \teff\ and \logg, and 
$\delta_{T_{\mathrm{eff}}-\mathrm{[Fe/H]}}$ is the uncertainty related to the interdependence between \teff\ and [Fe/H].

Based on a comparison with the accurate scale defined by the GBS (see Sec.~\ref{sec:accuracy}), we estimated a typical 1$\sigma$ value for $\delta_{T_{\mathrm{eff}}-\mathrm{model}} = 50$~K, at the metal-poor regime. 
The uncertainty in the fitting procedure, $\delta_{T_{\mathrm{eff}}-\mathrm{fit}}$, is related to the SNR of each spectrum. 
Since the normalization and fitting are part of one single procedure, this value includes also the error in locating the continuum. 
When only one spectrum is available,  $\delta_{T_{\mathrm{eff}}-\mathrm{fit}}$ is given by the 1$\sigma$ dispersion of the Gaussian fit of its temperature histogram (details are given in Sec.~\ref{sec:teff}). When more spectra are available for one given star, $\delta_{T_{\mathrm{eff}}-\mathrm{fit}}$ is given by the 1$\sigma$ dispersion of the mean temperatures associated to analysis of each spectrum.
For the \titan, $\delta_{T_{\mathrm{eff}}-\mathrm{fit}}$ values are between 3 and 38~K with a typical value of $15\pm9$~K. 
As the uncertainty caused by instrument residual patterns, $\delta_{T_{\mathrm{eff}}-\mathrm{inst}}$, we adopted the standard deviation of the comparison made in the consistency analysis reported in Sec.~\ref{sec:suitability}. This analysis compares \teff\ values derived from high-quality spectra obtained with the HARPS and UVES instruments for the same stars. 
We estimated $\delta_{T_{\mathrm{eff}}-\mathrm{inst}} = 33$~K.

To map the interdependence between \teff\ and \logg, we used the grid H$\alpha$ profiles with [Fe/H]~$= -2$~dex. We assumed that their behavior can act as a conservative proxy for the whole metallicity range. The impact that a variation in \logg\ has in \teff\ decreases progressively for higher [Fe/H] and remains nearly constant for lower values of metallicity.

We proceeded as follows. First, we introduced artificial Poisson noise equivalent to SNR = 250 to the synthetic line profiles. Then, we performed the profile fitting analysis using the very sensitive wavelength regions located between 2.5 and 7~\AA\ to the blue and red sides of the H$\alpha$ core 6562.797~\AA. The fittings were performed assuming as input \logg, values that are different by $\pm0.1$~dex than the correct value. In this way, we can map the influence on \teff\ of a given $\Delta$\logg.

\begin{figure}
    \centering
    \includegraphics[width=1\linewidth]{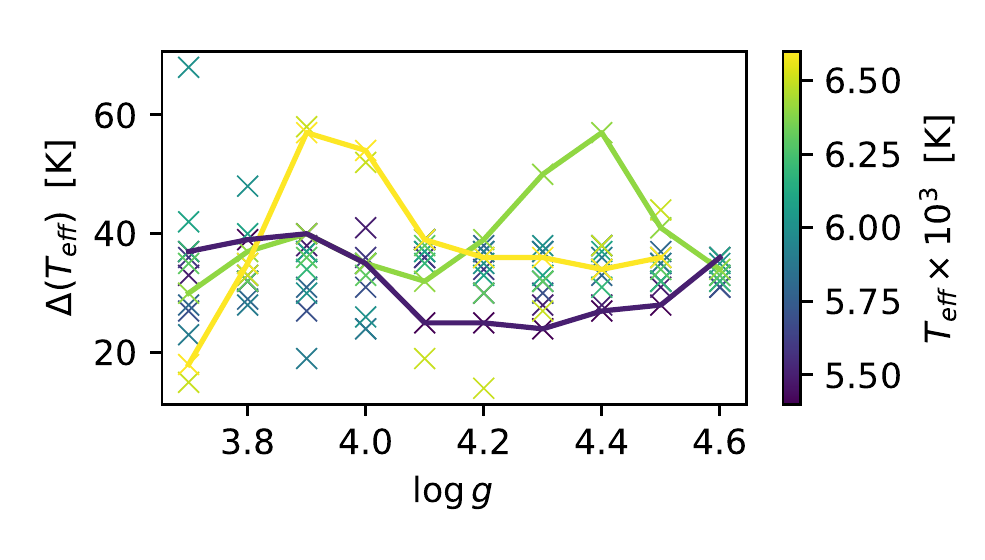}
    \includegraphics[width=1\linewidth]{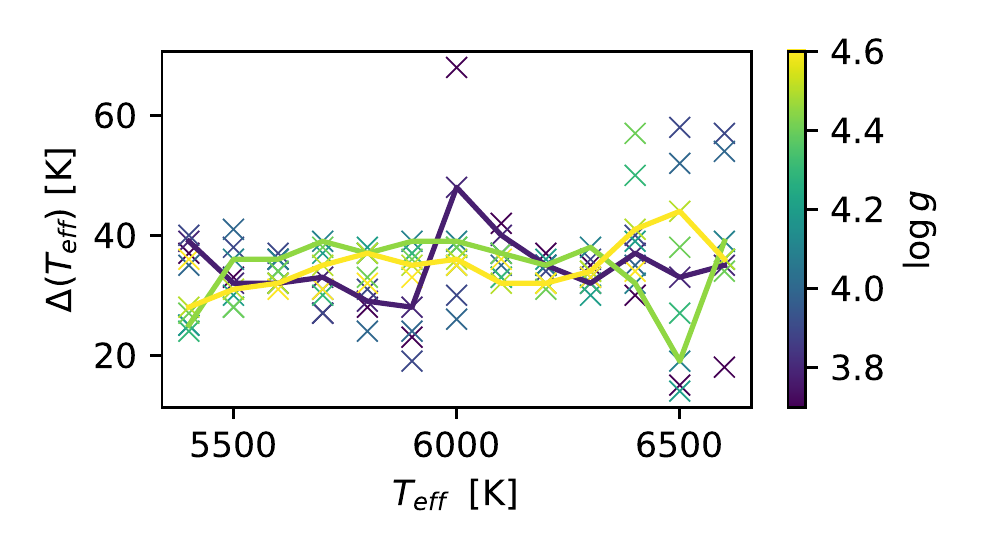}
    \caption{Temperature difference between our best value and the one that is recovered by assuming an offset of $-0.1$~dex in \logg. The difference is shown as function of \logg\ (\textit{top panel}) and of the model  temperature (\textit{bottom panel}).
    In the top panel, cases for 5400, 6400, and 6600~K are highlighted as the lines in blue, green, and yellow, respectively. In the bottom panel, cases for 3.8, 4.1, and 4.5~dex, are highlighted as the lines in blue, green, and yellow, respectively.}
    \label{fig:errortandlogg}
\end{figure}

Fig.~\ref{fig:errortandlogg} shows the distribution of $\Delta$\teff\ obtained in this exercise as function of \teff\ and \logg. These plots correspond to the negative variation of \logg, which induces positive temperature variations. However, we observed that $\Delta$\teff\ behaves almost symmetrically. We excluded from the analysis the data with \teff\ and \logg\ combinations of little physical meaning, i.e., \teff~$> 6500$~K and \logg~$>4.5$~dex.

The figure shows that $\Delta$\teff\ mostly remains below 40~K, except for a few profiles with \teff $>$ 6300~K, and for one profile with \logg~$= 3.7$~dex and \teff~$= 6000$~K. Such high variations correspond to simulated profiles that are highly asymmetric. From the plots in Fig.~\ref{fig:errortandlogg}, we conclude that a variation of $\Delta$\logg~$= \pm0.1$~dex creates an effect of $\Delta$\teff~$\approx \mp35$~K.

The average Gaia EDR3 parallax precision for the stars in this work is of $\pm0.02$~mas. The \logg\ values computed by isochrone fitting have then a precision of about $0.03$~dex, with input temperatures of $\pm$50~K precision. With a \logg\ uncertainty of this magnitude, and the $\Delta$\logg\ and $\Delta$\teff\ relation determined above, we estimated that the temperatures computed in the end of the last loop will be affected by 0.03 $\times$ $\Delta$\teff\ / $\Delta$\logg\, $\equiv \delta_{T_{\mathrm{eff}}-\mathrm{log}\;g} \approx 11$~K.

We followed the same reasoning to map the \teff--[Fe/H] interdependence. 
We found that the impact of [Fe/H] offsets are eliminated through the first loop when the H$\alpha$ profile fitting and the Fe lines fitting are performed. 
The typical uncertainty in [Fe/H] is of $\pm$0.05~dex on average. This includes the effect of a typical \teff\ error of $\pm$60 K, a typical \logg\ error of $\pm$0.05 dex, and a line-by-line fitting error of $\pm$0.01 dex. This line-by-line fitting error is the typical scatter obtained from lines measured across multiple spectra of the same star. A variation of $\pm$0.05 dex in [Fe/H] has in impact of only a few Kelvin, in extreme cases, on \teff(H$\alpha$). Thus, we assume $\delta_{T_{\mathrm{eff}}-\mathrm{[Fe/H]}} = 0$ K.
In Sec.~\ref{sec:metallicity}, we briefly mentioned that the [Fe/H] values from ionized Fe lines might need a 3D non-LTE correction of about $+0.1$~dex. Such change in [Fe/H] will induce a typical \teff\ variation of $-5$~K for a metal-poor turn-off star (like HD~84937) or even smaller for a metal-poor subgiant (like HD~140283). The largest variations (up to $-20$~K) is seen in models with parameter combinations that have little physical meaning, i.e. those with \teff~$> 6500$~K and \logg~$> 4.5$~dex. Therefore, we conclude that any offset in \teff\ caused by our use of 1D LTE [Fe/H] values is of the order of  $+5$~K.

The average of the total uncertainty of the \teff\ values is thus determined to be of $\pm$ 60 K. 
We remark that in Table~\ \ref{tab:CAMK_stars} we give \teff\ errors that do not include $\delta_{T_{\mathrm{eff}}-\mathrm{model}}$ when Eq.~\ref{eq:uncertainty} is applied. 
We preferred to include in the table what we consider to be the internal errors (i.e., the precision) of the method. The $\delta_{T_{\mathrm{eff}}-\mathrm{model}}$ component of the total error is, instead, related to the method accuracy (or trueness).

\section{Accuracy of the parameter scale}
\label{sec:accuracy}

\subsection{Consistency between HARPS and UVES spectra}
\label{sec:suitability}

It was shown in \citet{giribaldi2019_Ha} that HARPS spectra present negligible variation of the instrumental residual patterns with time. It was then concluded that such spectra are fully suitable for H$\alpha$ profile temperature diagnostics. Instead, in this work, we mostly rely on UVES spectra, as these allowed a bigger sample of metal-poor stars to be built. It is thus important for us to establish that consistent results can be obtained irrespective of whether our analysis method is applied to HARPS or UVES spectra.

Analysing a broad line, such as H$\alpha$, in echelle data is challenging because the line spans several spectral orders which need to be correctly merged. Although the flat-fielding procedure eliminates most of the instrumental patterns, the shape of the extracted spectrum may remain convoluted with subtle residuals \citep[see for instance][]{skoda2008}. These residuals can manifest as small incompatibilities between observed and synthetic H$\alpha$ profiles, causing offsets in the derived temperature values.

\begin{figure}
    \centering
    \includegraphics[width=0.95\linewidth]{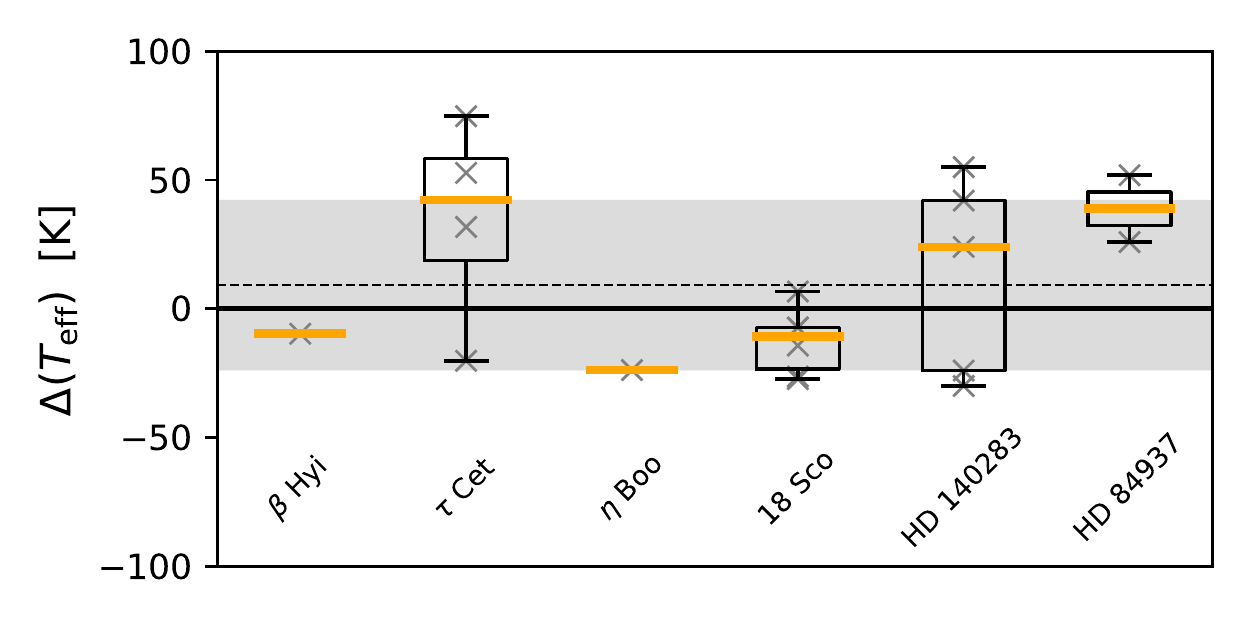}
    \caption{Temperatures derived from UVES spectra relative to the average temperatures derived from HARPS data for each of the GBS labeled in the plot. Gray symbols represent the individual values from multiple spectra.
    The box-bars extend from the lower to upper quartiles, with an orange line representing the median. The whiskers show the complete range of the values. The dashed line indicates the average difference, $+9$~K,  and the shaded area encloses the standard deviation of $\pm33$~K.}
    \label{fig:UVES_VS_HARPS}
\end{figure}

For this comparison, we employed six GBS stars for which we could retrieve good-quality HARPS and UVES spectra (see Tab.~\ref{tab:results}). The results are shown in Fig.~\ref{fig:UVES_VS_HARPS}. 
The average of the differences is close to the perfect agreement, at $+9$~K. 
This shows that, collectively, 
the temperatures obtained from both sets of spectra are compatible. 
The dispersion of $33$~K is very good, considering that spectra of low SNR were included for $\tau$~Cet.

The eight HARPS spectra of $\tau$ Cet that we used have SNR~$> 500$. This excellent spectral quality permitted the derivation of temperatures with individual fitting errors of about $\pm30$~K. On the other hand, among the four UVES spectra of this star, only one is of good quality, with SNR = 282. For it, we derived a \teff\ with an offset of $-22$~K. The remaining three spectra have SNR~$\sim50$. Looking at the fits to the profiles of $\tau$~Cet in Appendix~\ref{app:gbs}, it can be seen that the individual fitting errors of the low quality spectra are $\pm150$, $\pm134$, and $\pm133$~K. This suggests that SNR is responsible for the realatively poorer agreement for this star as seen in Fig.~\ref{fig:UVES_VS_HARPS}.

The temperature values for HD~140283 also present a relatively large dispersion, which seems connected to the SNR as well. 
Its spectra have SNR = 108, 139, 145, 273, and 386 (while, for example, all the spectra of 18~Sco exceed SNR = 200). For the UVES spectrum of HD~140283 with best SNR, we obtain a  difference of only 24~K in comparison to the temperature obtained from its HARPS spectrum. This allows us to conclude that, when using UVES spectra without obvious distortions and of good SNR, temperature offsets induced by residual instrumental patterns are negligible. It is thus possible to obtain accurate \teff\ values from the analysis UVES spectra, which are compatible with the values obtained from the analysis of HARPS spectra \citep[whose accuracy, in turn, was demonstrated in][]{giribaldi2019_Ha}.

\subsection{Comparison with the \textit{Gaia} benchmark stars}
\label{sec:GBS_test}

\begin{table*}
\centering
\normalsize
\caption{Atmospheric parameters and Galactic membership for the sample of \textit{Gaia} benchmark stars.}
\label{tab:results}
\begin{threeparttable}
\footnotesize
\begin{tabular}{l c c c c c c c c c}
\hline\hline 
Star & \teff & \logg & [Fe/H]$_{\mathrm{non-LTE}}$ & \teff(H$\alpha$) & \logg$_{\mathrm{iso}}$ & Spectra & Instrument & Galactic\\
 & (K) & (dex) & (dex) & (K) & (dex) & & & membership\\
\hline
Sun  & $5771^{a} \pm 1$ & $4.44$ & $+0.03$ & $5767 \pm 10$ & -- & 47 & HARPS & --\\
\hline
$\beta$~Hyi & $5873^{b} \pm 45$ &  $3.98$ & $-0.04$ & $5763 \pm 19$ & $3.94 \pm 0.03$ & 4 / 1 & HARPS / UVES & --\\
$\tau$~Cet & $5414^{b} \pm 21$ &  -- & $-0.49$ & $5366 \pm 24$ & $4.52 \pm 0.03$ & 8 / 4 & HARPS / UVES & --\\ 
$\delta$~Eri & $4954^{b} \pm 30$ & $3.76$ & $+0.06$ & $5007 \pm 26$ & $3.74 \pm 0.04$& 5 & HARPS & --\\  
$\beta$~Vir & $6083^{b} \pm 41$ & $4.10$ & $+0.24$ & $6068 \pm 18$ & $4.08 \pm 0.04$ &3 & HARPS & --\\
$\eta$~Boo & $6099^{b} \pm 28$ & $3.79$ & $+0.32$ & $6027 \pm 21$ & -- & 5 / 1 & HARPS / UVES & --\\
$\alpha$~Cen~A & $5792^{b} \pm 16$ &  $4.31$ & $+0.26$ & $5762 \pm 16$ & -- & 1 & HARPS & --\\
18~Sco & $5810^{b} \pm 80$ & $4.44$ & $+0.03$ & $5775 \pm 26$ & $4.41 \pm 0.04$ &7 / 6 & HARPS / UVES & --\\
$\beta$~Gem & $4858^{b} \pm 60$ & $2.90$ & $+0.13$ & $4860 \pm 18$ & -- & 2 & HARPS & --\\
$\xi$~Hya & $5044^{b} \pm 40$ & $2.87$ & $+0.16$ & $5043 \pm 26$ & $2.85 \pm 0.03$& 6 & HARPS & --\\
$\epsilon$~Vir & $4983^{b} \pm 61$ & $2.77$ & $+0.15$ & $5036 \pm 16$ & $2.77 \pm 0.04$ & 2 & HARPS & --\\
HD~49933 & $6635^{b} \pm 91$ & $4.20$ & $-0.41$ & $6631 \pm 18$ & $4.22 \pm 0.04$ & 3 & HARPS & --\\
HD~140283 & $5792^{b\,\blacklozenge} \pm 55$ & $3.65^{\blacklozenge}$ & $-2.28^{\spadesuit}$ & $5810 \pm 32$ & $3.67 \pm 0.02$ &1 / 5 & HARPS / UVES & GE$\dagger$\\
\hline
$\epsilon$~For & $5123^{c} \pm 78$ &  -- & $-0.60$ & $5042 \pm 17$ & $3.64 \pm 0.04$ & 3 & HARPS & --\\ 
HD~107328  & $4496^{c} \pm 59$ & $2.09$ & $-0.33$ & $4498 \pm 18$ & $2.07 \pm 0.07$ &1 & HARPS & --\\
HD~84937 & $6356^{c} \pm 97$ & $4.06$ & $-2.03$ & $6358 \pm 33$ & $4.13 \pm 0.03$ & 3 / 2 & HARPS/ UVES & GE$\dagger$\\
HD~22879 & $5868^{c} \pm 89$ & $4.27$ & $-0.86$ & $5894 \pm 23$ & $4.27 \pm 0.04$ & 7 & HARPS & --\\
\hline
HD~298986 & $6223^{d} \pm 100$ & $4.19$ & $-1.26$ & $6177 \pm 39$ & $4.28 \pm 0.04$ & 3 & UVES & HS$\dagger$\\
HD~106038 & $6121^{d} \pm 80$ & $4.55$ & $-1.25$ & $6094 \pm 43$ & $4.29 \pm 0.04$ & 5 & UVES & GE$\dagger$/ Sq $\ddagger$\\
HD~201891 & $5948^{d} \pm 80$ & $4.30$ & $-0.97$ & $6007 \pm 37$ & $4.34 \pm 0.04$ & 1 & UVES & --\\
HD~102200& $6155^{d} \pm 80$ & $4.22$ & $-1.12$ & $6246 \pm 19$ & $4.36 \pm 0.04$ & 2 & HARPS & --\\
\hline
\end{tabular}
\begin{tablenotes}
\item{} \textbf{Notes.} Second, third and fourth columns list the reference \teff, \logg, and [Fe/H] (this last one under non-LTE) as listed in \citet{2018RNAAS...2..152J}.
The codes indicate: $^{(a)}$ Direct measurement of the stellar radius and irradiance.
$^{(b)}$ Interferometric measurements of the stellar angular diameter.
$^{(c)}$ Photometric calibrations based on interferometric measurements.
$^{(d)}$ IRFM.
$^{(\blacklozenge)}$ Interferometric measurements from \citet{karovicova2020}.
$^{(\spadesuit)}$ 3D non-LTE abundances from \ion{Fe}{ii} lines from \citet{amarsi2016}.
Fifth column lists the temperatures determined here by H$\alpha$ fitting, with their internal uncertainties For the total uncertainties, one has to further include the estimated model accuracy error of $\pm$ 50 K (as discussed in Sec.~\ref{sec:error}). 
Seventh column lists the number of spectra used in the analysis, followed by a column that lists the instrument related to the spectra. In the last column, the Galactic membership is provided. 
GE stands for \textit{Gaia}-Enceladus, HS for Helmi Streams and Sq for Sequoia (see Section \ref{sec:substructures}). The remaining GBS belong to the thin/thick disk. ($\dagger$) Membership according to the criterion of \citet{Massari2019}. ($\ddagger$) Membership according to the criterion of \citep{Myeong2019}.
\end{tablenotes}
\end{threeparttable}
\end{table*}

Having established that the \teff\ values derived from the analysis of UVES and HARPS spectra are compatible with each other, we now present a more extensive accuracy test of the methodology applied to a larger number of GBS. 
The derived parameter values are listed in Table~\ref{tab:results} along with the type and number of spectra used for the determinations, and the reference parameters. 
The table also indicates the method by which the reference \teff\ values were derived, and it is divided ranking the most direct method on the top, to the less direct one on the bottom.
We refer the reader to the original sources in the table notes for detailed information about the methods.
Missing values in the \loggiso\ column are due to the lack of \textit{Gaia} parallax measurements for the corresponding stars. For the metal-rich GBS, we did not rederive the [Fe/H] values but instead fixed the value to the reference ones from \citet{jofre2014}. 

We did not include the star HD~122563, although it has one HARPS spectrum of high SNR and without obvious order merging distortions because we did not find a synthetic profile in the grid that reproduces the very asymmetric shape of its observational profile. 
This was the only spectrum for a star of its type (i.e. a very metal-poor giant) that we tried to analyze. The reasons for this discrepancy are unclear and uncovering the explanation is beyond the scope of this paper. Possibilities include artifacts in the spectrum, some stellar peculiarity, and problems in the modelling of this type of giants.
HD~175305 is another benchmark that was not included in the comparison because it lacks HARPS or UVES spectra. 

\begin{figure}
    \centering
    \includegraphics[width=0.95\linewidth]{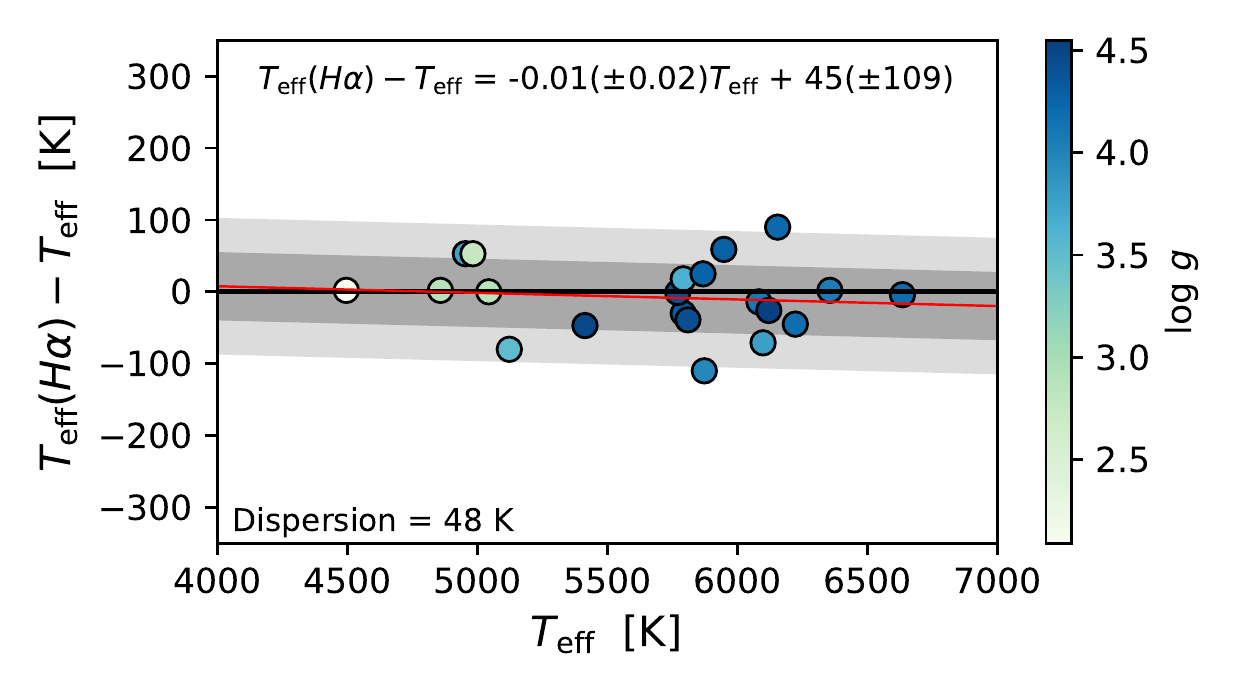}
    \includegraphics[width=0.95\linewidth]{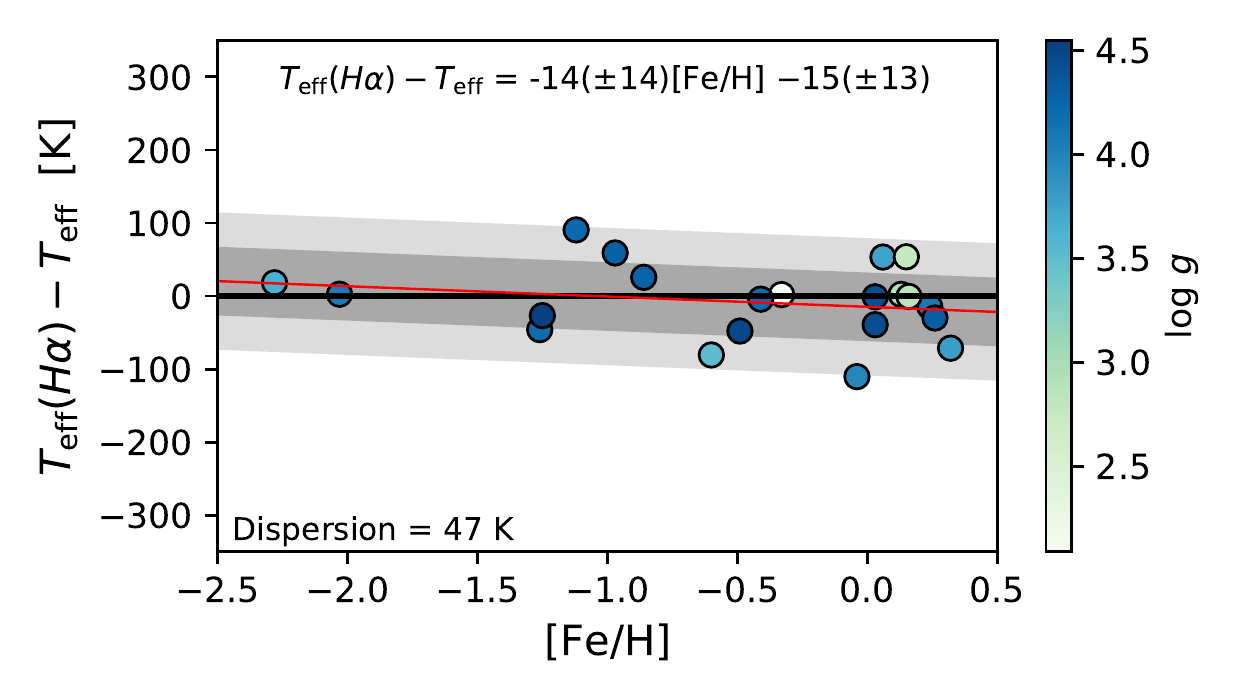}
    \includegraphics[width=0.95\linewidth]{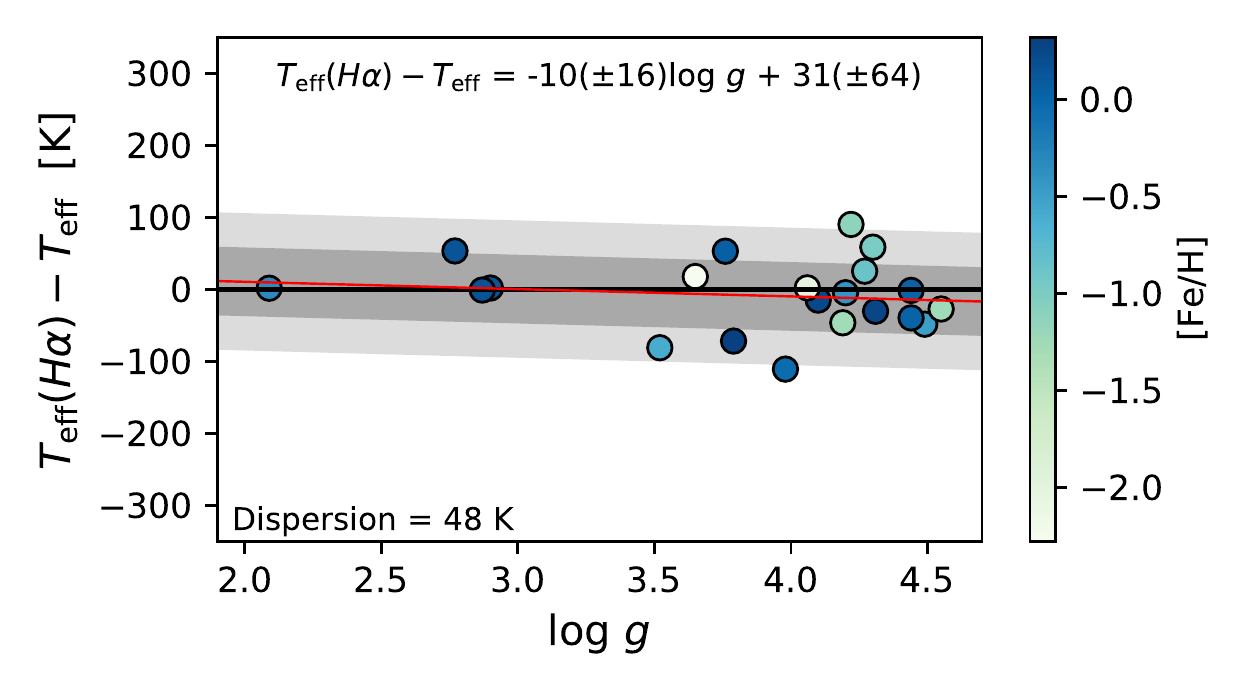}
    \caption{ For the GBS of Table \ref{tab:results}, difference between \teff(H$\alpha$) and the reference \teff\ as function of the reference \teff, [Fe/H], and \logg\ from the top to the bottom, respectively.
    The red lines are linear regressions, and the shaded areas indicate 1$\sigma$ and 2$\sigma$ dispersions.}
    \label{fig:Teff_accuracy}
\end{figure}

\begin{figure}
    \centering
    \includegraphics[width=0.95\linewidth]{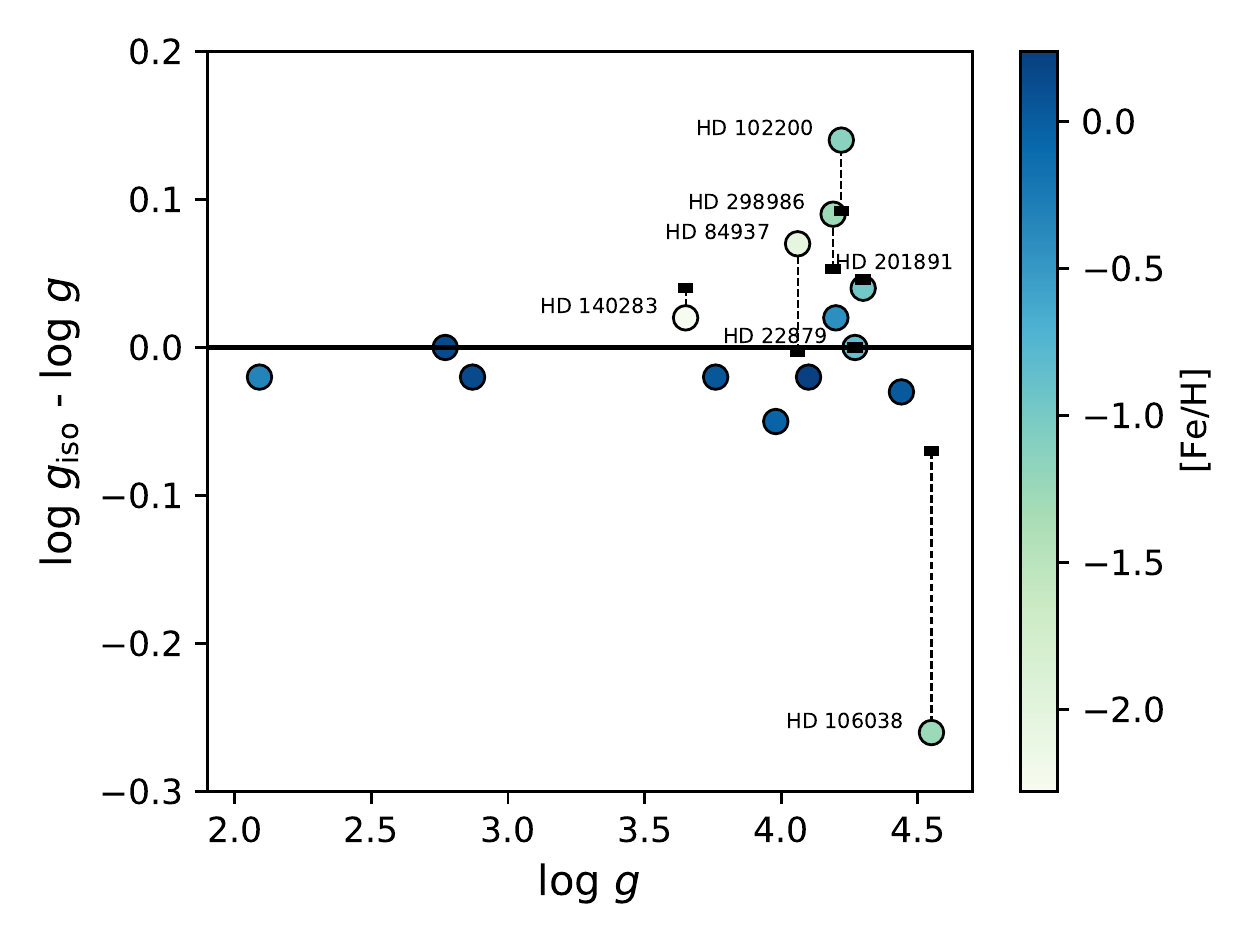}
    \caption{For the GBS of Table \ref{tab:results}, difference between \loggiso\ and reference \logg\ as function of the reference value. The labels identify the metal-poor GBS. The circles represent the values for which \loggiso\ was derived using the \textit{Gaia} parallaxes. The dashes represent the values for which \loggiso\ was derived using the same (Hipparcos) parallaxes that were used to derive the reference \logg\ from \citet{2018RNAAS...2..152J}.}
    \label{fig:logg_dispersion}
\end{figure}

The fits to the H$\alpha$ profile of each benchmark are shown in the Appendix \ref{app:gbs}\footnote{To improve visibility of the plots, we chose to display only the right wing of the line.}. 
Our \teff\ values are compared to the reference ones in Fig.~\ref{fig:Teff_accuracy}. The individual values agree with each other within their 3$\sigma$ uncertainties for all stars, and within 2$\sigma$ dispersions except for $\beta$~Hyi and HD~102200 that slightly exceed the 2$\sigma$ limits by few Kelvin.
No significant correlation is seen between the \teff\ differences and any of the atmospheric parameters. We can thus conclude that our methodology provides unbiased results of \teff\ for stars in a large range of parameters, including also giants. It is our plan to extend the \titan\ sample with giants in forthcoming papers of the series. 
The dispersion of the \teff\ differences, of about $\pm$50~K, seems to be dominated by the uncertainty of the GBS scale, which is of the same order, $\pm$61~K on average. 
We decided to adopt the $\pm$50~K dispersion as the 1$\sigma$ uncertainty associated to the accuracy of the \teff\ measurements that are possible with the current model of H$\alpha$ lines. 
This quantity is used as $\delta_{T_{\mathrm{eff}}-\mathrm{model}}$ in Equation \ref{eq:uncertainty}.
We stress that this is likely an upper-bound estimate that mostly reflects the uncertainty of the GBS \teff\ values used as reference. 
The accuracy limit can be better constrained in case a sample of accurate standards with \teff\ of superb precision (i.e. $\sigma$(\teff)~$\pm20$~K, at most) becomes available. This would permit the proper mapping of the accuracy of the H$\alpha$ models across the parameter space.

In the metal-poor regime, only HD~140283 has a reference \teff\ value based on a measurement of its angular diameter ($\theta$) \citep{karovicova2020}. 
For HD~84937, the reference value of \teff\ was derived from $\theta$--color relations \citep{kervella2004}. For the remaining GBS, HD~298986, HD~102200, HD~106038, and HD~201891, reference \teff\ values are those derived using the IRFM by \citet{2016A&A...592A..70H}. 
These \teff\ values are in the same scale of the IRFM implementation from \citet{Casagrande2010} and were shown to agree with the \teff\ derived using the $\theta$--color relations of \citet{kervella2004} to within 100~K. The fact that we derived compatible temperatures in the metal-poor regime, where the reference scale is shared between IRFM and interferometry, provides confidence on the accuracy of the H$\alpha$-based \teff\ scale.
The temperatures we derive for HD~140283 and HD~84937 are also compatible with those derived by \citet{amarsi2018}. 
For the Sun, the authors find a slightly cooler value, $5721 \pm 26$~K caused by small normalization distortion of the solar profile that they used \citep{wall2011}. This distortion was discussed in  \citet{giribaldi2019_Ha}.

A comparison between our \logg\ values and the reference ones for the GBS is presented in Fig.~\ref{fig:logg_dispersion}.
The plot suggests an increased dispersion affecting exactly the metal-poor dwarfs.
It happens, however, that the reference \logg\ values of the GBS were computed with parallaxes from the new reduction of Hipparcos data \citep{2007A&A...474..653V},
with the only exception of HD~140283 whose reference \logg\ was derived with a parallax value from $Gaia$ DR2.
We derived \loggiso\ with improved parallax values from \textit{Gaia} EDR3.
Thus, for a fair comparison, we recomputed our \loggiso\ values, adopting instead the older parallaxes. 
As expected, we obtain a better agreement with the reference \logg\ and the large scatter disappears
(see Fig.~\ref{fig:logg_dispersion}).
Given the improved precision of the \textit{Gaia} parallaxes, we consider that our new \loggiso\ determination, available in Table~\ref{tab:results}, supersedes the original \logg\ reference values of the metal-poor GBS. We are thus confident to have employed a method that achieves very good accuracy in the determination of \logg.

We note that the \logg\ reference values for $\tau$~Cet and $\epsilon$~For were not included in Table~\ref{tab:results} because they were evaluated to be uncertain in \citet{Heiter}, hence not recommended for calibration or validation purposes.
These values are $4.49 \pm 0.02$~dex and $3.52 \pm 0.08$~dex for each star, respectively, as displayed in Table~10 in that paper. Our \loggiso\ values in the Table~\ref{tab:results} are actually compatible with these determinations.

\subsection{Comparison with IRFM temperatures}
\label{sec:IRFM}

A comparison with \teff\ values derived using the IRFM is an important step in validating our temperature scale. 
The IRFM is the only technique that can offer a nearly model-independent determination of \teff\ for distant stars because interferometric measurements are not feasible for them. 
The accuracy of the IRFM has been discussed and progressively improved in several works \citep[e.g.][]{alonso1996, ram2005, gonzalez2009, Casagrande2010}. 
A comparison against \teff\ values based on interferometric data by \citet{casagrande2014}  suggested that a typical 1\% can be achieved by direct application of IRFM. 
New color--\teff\ relations, for use with Gaia photometry, have recently been implemented by \citet{Casagrande2020}, not only for dwarfs and subgiants but also for giants.

\begin{figure}
    \centering
    \includegraphics[width=0.95\linewidth]{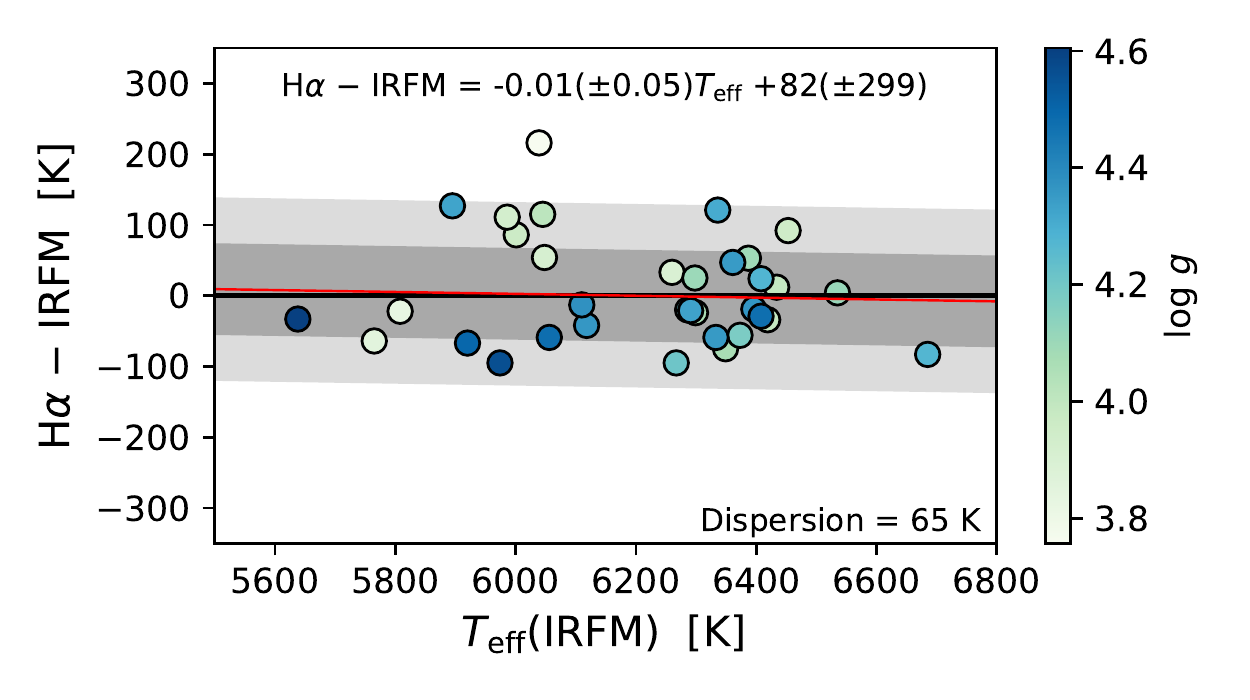}
    \includegraphics[width=0.95\linewidth]{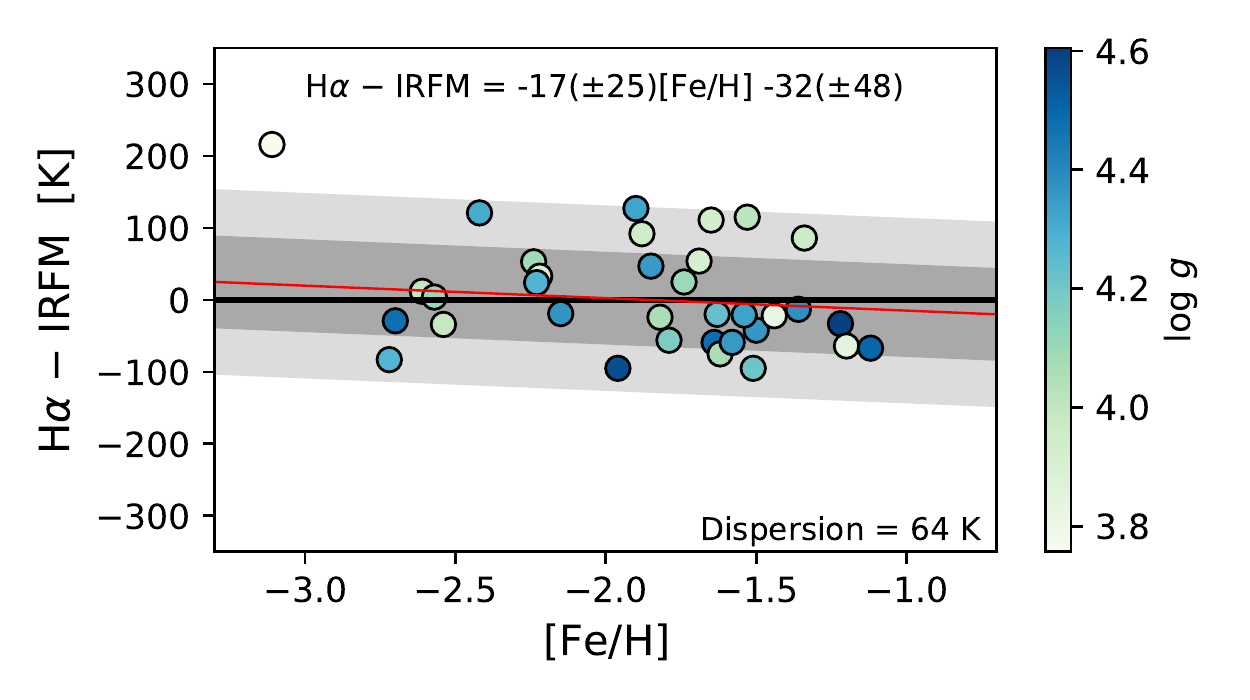}
    \includegraphics[width=0.95\linewidth]{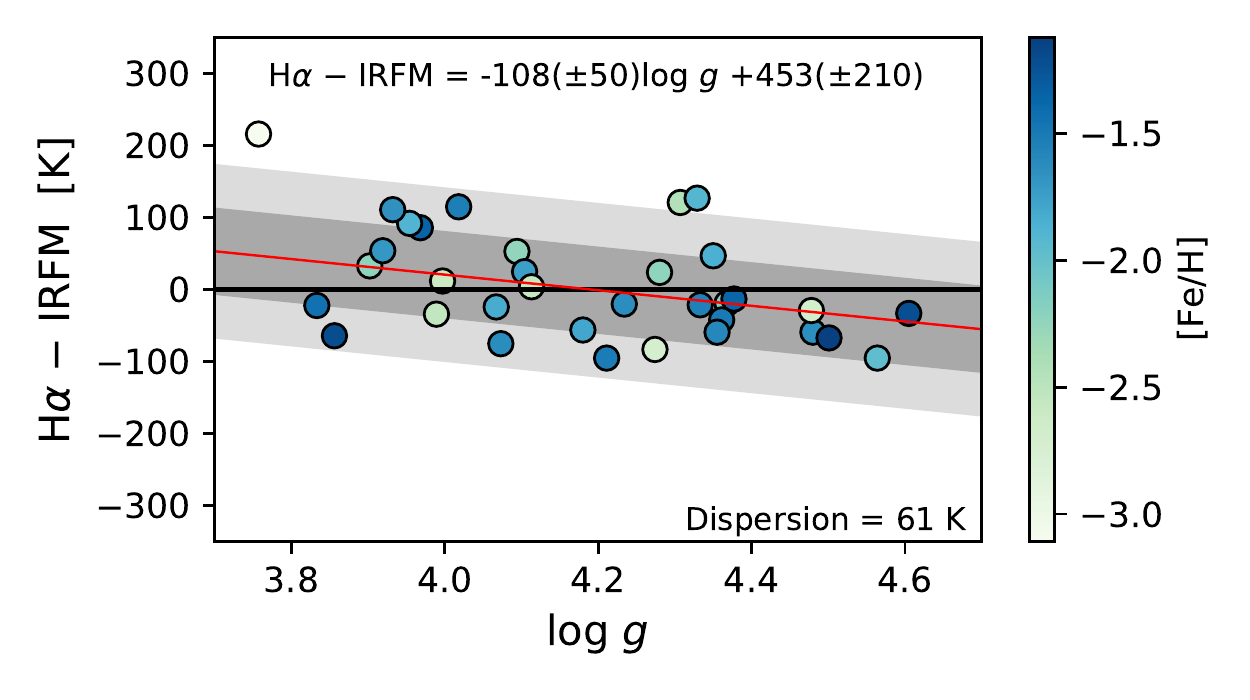}
    \caption{For the \titan, the difference between temperatures derived by H$\alpha$ fitting and those by the IRFM, as function of \teff, [Fe/H], and \logg\ from the top to the bottom, respectively. 
    The red lines are linear regressions, and the shaded areas indicate 1$\sigma$ and 2$\sigma$ dispersions.
    Same vertical scale as in Fig.~\ref{fig:Teff_accuracy} is used.}
    \label{fig:IRFM_comparison}
\end{figure}

The comparison with the IRFM \teff\ values is shown in Fig.~\ref{fig:IRFM_comparison}. 
The figure includes 35 stars (80\% of our sample) for which IRFM \teff\ were extracted from \citet[][Table~8]{Casagrande2010}; they are compiled in Table~\ref{tab:preliminar_parameters} and noted by the code Ca10.
These are the original values derived directly by the technique, and not by color--\teff\ relations,
thus they have negligible influence from the [Fe/H] determination.
A small correlation, above the 1$\sigma$ errors, is observed only with \logg\ (bottom panel of Fig.~\ref{fig:IRFM_comparison}). The trend with \logg\ is mainly driven by five stars with \logg~$> 4.4$~dex, all of them with cooler temperatures inferred from H$\alpha$, and one star with $\logg <$ 3.8, which had a hotter H$\alpha$-based \teff.

The five stars with \logg\ $>$ 4.4 are HD~34328, BD+29~2091, HD~126681, HD~193901, and \mbox{LP~831-70}. Except for the last, they are all located within 100~pc. Thus, uncertainties in the photometric magnitudes caused by interstellar extinction are likely not the source of the differences. For two of them, HD~126681 and \mbox{LP~831-70}, we used $V$ magnitudes that are larger by 0.2~mag with respect to those used to derive the IRFM~\teff. This difference indeed increases \logg\ by 0.05~dex, resulting in a decrease of 20~K in \teff(H$\alpha$). This essentially removes the discrepancies, leaving differences of $-13$ and $-9$~K for the two stars, respectively. Of the other three stars, we note that BD+29~2091 lacks $R$ and $I$ magnitudes which are needed to compute bolometric corrections for the IRFM implementation. The use of $V$ alone gives less precise results \citep[][Sec.~7]{casagrande2006}. We believe this can likely explain part of the $-95$~K difference observed between ours and the IRFM \teff\ values. We could not find similar issues for the remaining two stars, HD~34328 and HD~193901, which present temperature differences of $-59$ and $-67$~K. These differences, however, are still consistent with the total uncertainty that we estimated for our \teff\ determination.

At the low extreme of \logg, the star \hbox{BPS~CS22166-030} is a particular case, presenting the largest temperature difference (216~K). About 100~K of this difference can be explained by the more precise \textit{Gaia} parallax, which results in a \logg\ that is smaller by 0.24~dex than the value used in the IRFM determination. Further, as BD+29~2091, this star lacks $R$ and $I$ magnitudes, which increases the uncertainty of its IRFM \teff. We could not find any collective change of all the parameters required for the isochrone fitting (\teff, [Fe/H], $V$, and parallax) that, within their 3$\sigma$ uncertainties, would conciliate our \logg\ and \teff(H$\alpha$) values with the literature values. We consider our values more robust, as they are supported by the \textit{Gaia} parallax measurements.

Three of the \titan, CD$-71\,1234$, $\mathrm{G}\,24-3$, and HE 0926$-0508$, were analysed by \citet{Casagrande2020} where the values $6605\pm77$~K, $6668\pm74$~K, and $6458\pm75$~K were derived for each one, respectively.
Our \teff(H$\alpha$) values almost exactly match the new IRFM \teff\ for the the last two stars. For the first one, our \teff(H$\alpha$) is $-173$~K cooler, however it agrees with the original IRFM \teff\ of $6408\pm69$~K.

Doppler shifts in the spectra of spectroscopic binaries are a possible source of error in the H$\alpha$ fitting, as they can cause an additional enhancement of the line. We searched for known binary systems in the catalog of \citet{krevella2019}. Their method is based on detecting anomalies in the long-term proper motion of stars in common between the $Gaia$ DR2 \citep{Gai(c)} and Hipparcos catalogs. 
This is the case of HD~116064, identified as a double-lined spectroscopic binary \citep{smith1998}, for which we find a moderately large temperature difference of 127~K (\teff(H$\alpha$) being hotter). 
In fact, this star has the most imprecise parallax of the sample, $14.40 \pm 0.14$~mas, which affects the accuracy of \loggiso, and therefore the accuracy of \teff(H$\alpha$).
Another binary in this sample is CD$-48\,2445$ \citep{hansen2017}, for which the temperature difference is 92~K, again with the \teff(H$\alpha$) hotter than the one inferred with the IRFM. 
These stars are flagged in Table~\ref{tab:CAMK_stars}. We conclude that unrecognized binarity can be a source of additional scatter in the \teff\ comparison. Its general influence in this sample, however, must be small, as the comparison have zero offset and a dispersion of $\sim$65 K, very close to the total \teff\ error that we estimated. 

\section{Galactic substructure membership}\label{sec:substructures}

With the advent of \textit{Gaia}, it became possible to study the kinematic and dynamical properties of large stellar samples, and thus to unfold the merger history of the Milky Way. This development has lead to the recent discovery of major halo substructures like \textit{Gaia} Enceladus or \textit{Gaia} Sausage \citep{helmi2018,Belokurov2018} and Sequoia \citep{Myeong2019,Barba2019}, among other suggestions \citep[e.g.][and references therein]{2019A&A...631L...9K,2021ApJ...907...10L}. At least the major structures among these new discoveries are thought to be the result of early merger events, and thus, to come from smaller galaxies or stellar systems with different chemical evolution histories, when compared to the \textit{in-situ} Galactic halo.

\begin{table}
\caption{Membership classification of the Titans.}
\label{tab:membership}
\centering
\scriptsize
\begin{threeparttable}
\begin{tabular}{lccc}
\hline\hline
Star & parallax (mas) & $v_{helio}$ ($kms^{-1}$)& Membership \\
\hline
BD+24 1676 &  $3.964 \pm 0.017$ & $-237.48\pm0.95$ & GE$\dagger\ddagger$ \\
BD$-10$ 388  & $4.229\pm0.024$ & $36.79\pm0.86$ & -- \\
BPS CS22166-030  & $0.842\pm0.010$ &$92.0\pm2.4$& GE$\dagger$ \\
CD$-33\,3337^\S$  & $8.173\pm0.016$ &$73.79\pm0.25$& HS$\dagger$ \\
CD$-35$ 14849 & $6.296\pm0.034$ &$108.18\pm0.92$& GE$\dagger$/Sq$\ddagger$ \\
HD 16031 & $8.230\pm0.016$ &$24.72\pm0.39$& -- \\
HD 166913  & $16.707\pm0.029$ &$-48.61\pm0.30$& -- \\
HD 218502  & $13.374\pm0.038$ &$-29.48\pm0.48$& -- \\
HD $34328^\S$  & $13.759\pm0.010$ &$235.25\pm0.28$& Sq$\dagger\ddagger$ \\
HD 59392  & $6.382\pm0.013$ &$268.42\pm0.33$& GE$\dagger$/Sq$\ddagger$ \\
BD$+02\,4651^\S$  & $5.595\pm0.020$ &$-251.11\pm0.38$& GE$\dagger$/Sq$\ddagger$ \\
BD+03 740  & $5.893\pm0.026$ &$174.5\pm1.6$& GE$\dagger$ \\
BD$-13$ 3442  & $4.703\pm0.031$ &$116.1\pm1.5$& GE$\dagger$ \\
BD+26 2621  & $4.804\pm0.018$ &$-62.2\pm1.3$& GE$\dagger$/Sq$\ddagger$ \\
BD+26 4251  & $7.595\pm0.016$ &$-93.89\pm0.23$& -- \\
BD+29 2091  & $11.371\pm0.019$ &$83.23\pm0.36$& GE$\dagger$ \\
CD$-30$ 18140  & $6.897\pm0.023$ &$17.91\pm0.49$& GE$\dagger$ \\
CD$-33$ 1173  & $4.558\pm0.016$ &$47.7\pm1.9$& GE$\dagger$ \\
CD$-48\,2445^\S$  & $5.369\pm0.014$ &$319.56\pm0.60$& GE$\dagger$ \\
CD$-71$ 1234  & $6.110\pm0.029$ &$235.4\pm1.0$& GE$\dagger$ \\
G~24$-3$  & $7.648\pm0.020$ &$-208.18\pm0.29$& GE$\dagger$ \\
HD~108177  & $9.405\pm0.018$ &$155.31\pm0.37$& GE$\dagger$ \\
HD~$116064^\S$  & $14.40\pm0.14$ &$143.70\pm0.52$& GE$\dagger$ \\
HD~$122196^\S$  & $9.635\pm0.034$ &$-27.78\pm0.38$& -- \\
HD~126681  & $17.816\pm0.023$ &$45.40\pm0.13$& -- \\
HD~132475  & $10.671\pm0.025$ &$176.67\pm0.22$& Sq$\ddagger$ \\
HD~160617  & $9.525\pm0.021$ &$100.65\pm0.35$& GE$\dagger$ \\
HD~189558 & $15.765\pm0.024$ &$-12.77\pm0.14$& -- \\
HD~193901  & $20.854\pm0.023$ &$-171.43\pm0.14$& GE$\dagger$ \\
HD~213657 & $6.967\pm0.016$ &$51.82\pm0.49$& -- \\
HD~241253  & $10.068\pm0.016$ &$-15.91\pm0.16$& -- \\
HD~284248  & $11.378\pm0.019$  &$339.13\pm0.36$& -- \\
HD~74000 & $9.081\pm0.017$ &$206.17\pm0.55$& GE$\dagger$ \\
HD~94028  & $20.208\pm0.026$ &$66.00\pm0.25$& -- \\
HE~0926$-0508$  & $2.190\pm0.061$ &$189.6\pm1.6$& GE$\dagger\ddagger$ \\
LP~815$-43$  & $4.156\pm0.018$ &$-4.1\pm1.6$& GE$\dagger\ddagger$ \\
LP~831$-70$  & $4.162\pm0.022$ &$-47.6\pm1.7$& -- \\
Ross~453  & $5.619\pm0.020$ &$99.14\pm0.54$& Sq$\ddagger$ \\
Ross~892  & $6.056\pm0.018$ &$88.53\pm0.18$& GE$\dagger$ \\
UCAC2~20056019  & $1.017\pm0.017$ &$-7.69\pm1.7$& GE$\dagger$ \\
Wolf~1492  & $3.649\pm0.028$ &$449.8\pm3.4$ & Sq$\dagger$ \\
\hline
HD~140283 & $16.267 \pm 0.026$ & $-170.37\pm 0.15$ & GE$\dagger$ \\
HD~84937  & $13.500 \pm 0.044$ & $-15.69 \pm 0.43$ & GE$\dagger$ \\
HD~22879  & $38.325 \pm 0.031$ & $120.54 \pm 0.21$ & --\\
HD~298986 & $7.489 \pm 0.012$  & $198.21 \pm 0.27$ & HS$\dagger$\\
HD~106038 & $7.449 \pm 0.030$  & $99.52 \pm  0.26$  & GE$\dagger$/Sq$\ddagger$\\
HD~201891 & $29.877 \pm 0.020$ & $-44.15 \pm 0.19$ & --\\
HD~102200 & $13.861 \pm 0.014$ & $161.44 \pm 0.27$ & --\\
\hline
\end{tabular}
\begin{tablenotes}
\item{} \textbf{Notes.} {The symbol $\S$, following the star name, indicates a binary system identified in \citet{krevella2019} or \citet{hansen2017}.  
Parallax values are from the $Gaia$ EDR3 catalog. 
In the membership column, GE means \textit{Gaia}-Enceladus, HS Helmi-Streams, and Sq Sequoia (see Section \ref{sec:substructures}). The symbol $\dagger$ means likely membership according to the criterion of \citet{Massari2019} and $\ddagger$ means according to the criterion of \citet{Myeong2019}. The bottom part of the table incorporates the metal-poor $Gaia$ benchmark stars.}
\end{tablenotes}
\end{threeparttable}
\end{table}

\begin{figure}
    \centering
    \includegraphics[width=0.95\linewidth]{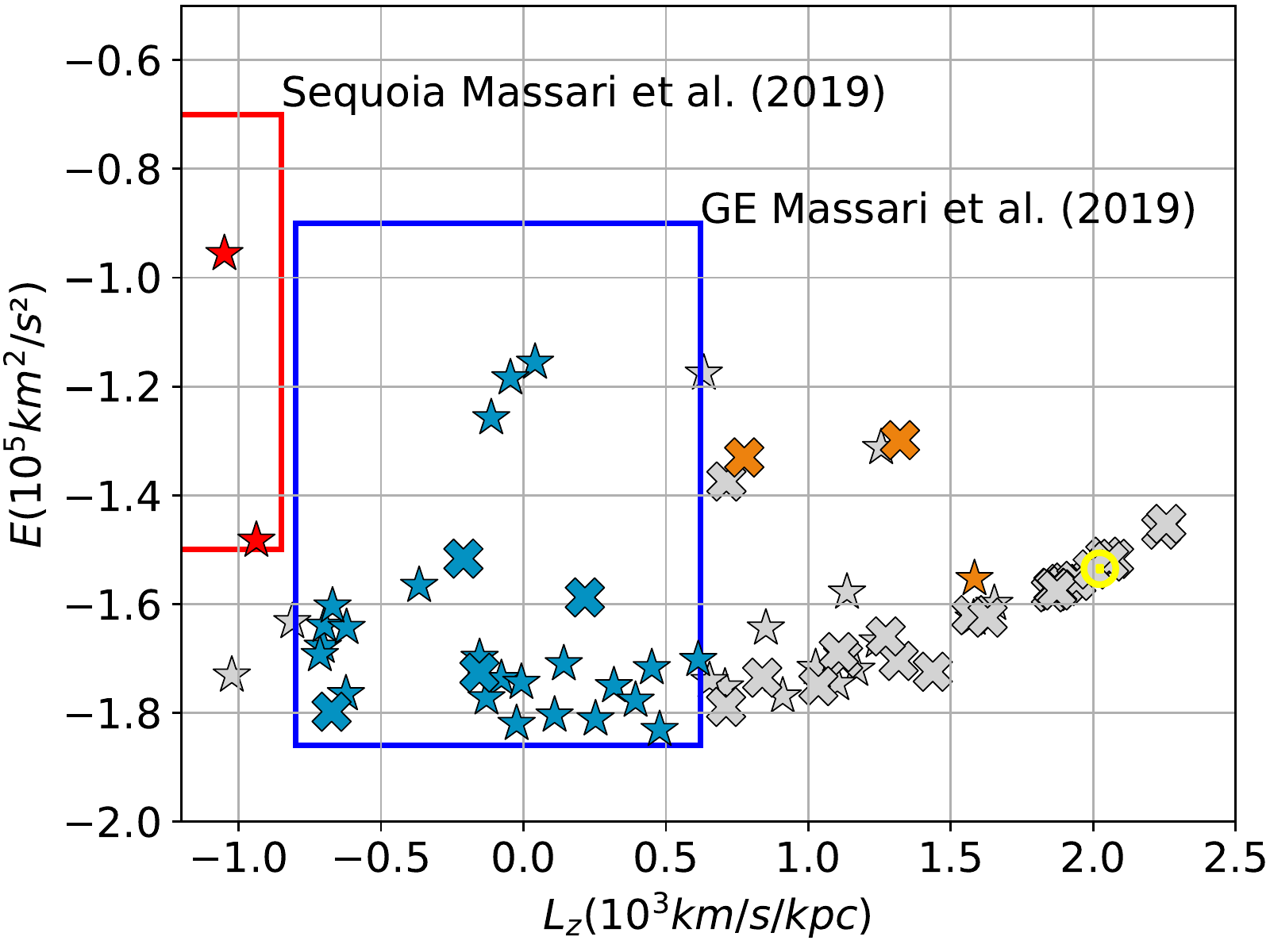}
    \includegraphics[width=0.8\linewidth]{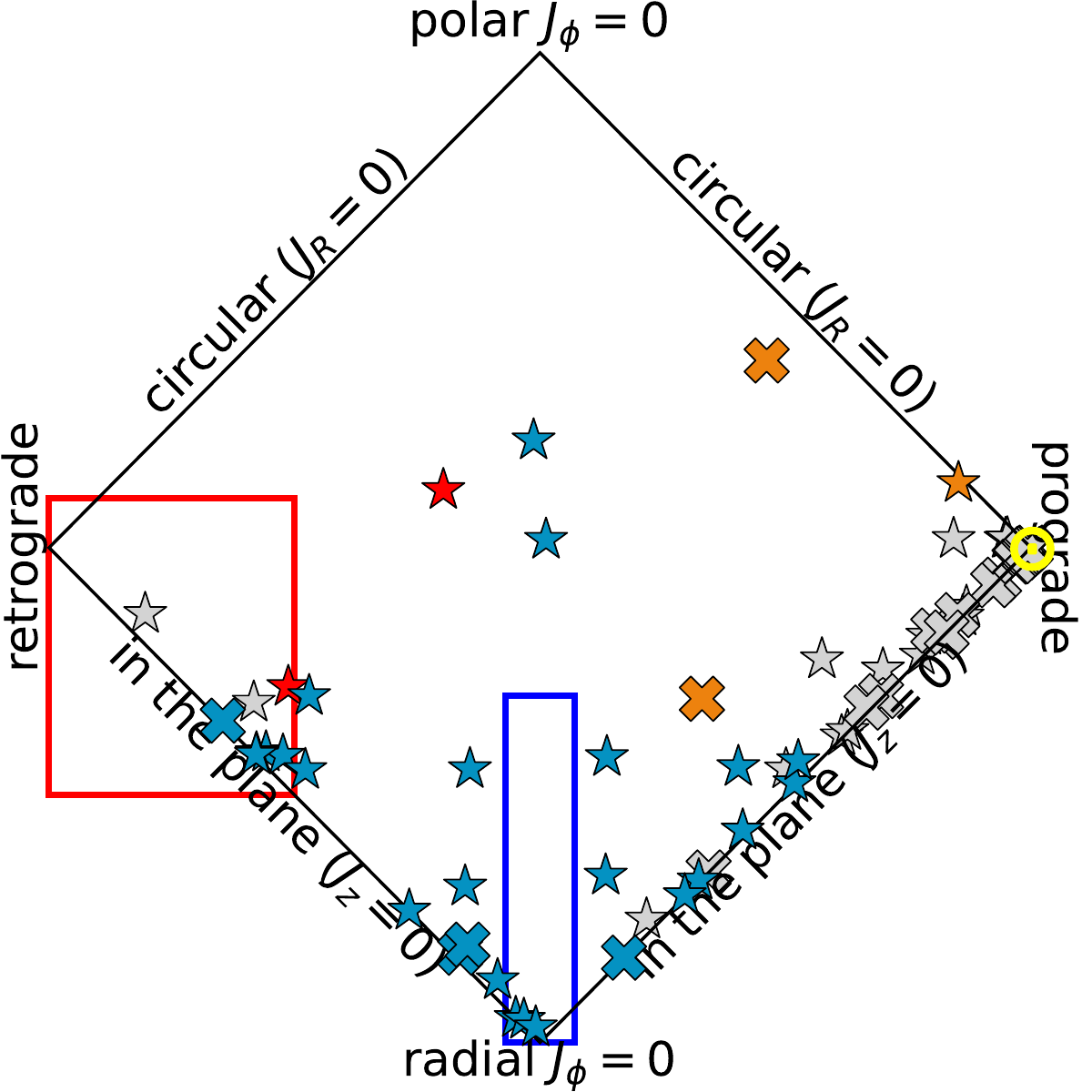}
    \caption{\textit{Top}: Lindblad diagram -- total orbital binding energy as a function of the angular momentum component in the direction of the Galactic north pole. The colours show the \citet{Massari2019} criteria to identify Sequoia (the red box and red symbols), \textit{Gaia} Enceladus (the blue box and blue symbols), and Helmi streams (orange symbols). Unclassified stars are shown in gray. The Sun is shown as the yellow $\odot$ symbol. 
    The \titan\ are shown as stars and the GBS as crosses. \textit{Bottom}: Action space map -- difference of action in the z and r directions by the action in the $\phi$ direction, all divided by total action \citep[see Fig.9 in][]{Myeong2019}. The blue box shows the limits for the \textit{Gaia}-Enceladus merger and the red box the limits for Sequoia, according to the \cite{Myeong2019} criteria. The symbols and colors for the stars are the same as in the top panel.}
    \label{fig:myeongplot}
\end{figure}

We decided to investigate whether our sample of metal-poor reference stars includes any candidate member of these stellar structures. Such stars, having accurate stellar atmospheric parameters, can be used for the determination of high-quality chemical abundances. This development can pave the way for the accurate chemical characterization of the stellar systems that merged with the Milky Way. We used the orbital parameters of the \titan\ and the GBS, computed as described in Section \ref{sec:orbits}, to classify the stars adopting the criteria defined by \citet[][criterion I henceforth]{Massari2019} and \citet[][criterion II henceforth]{Myeong2019}. We considered both criteria as, currently, there is no consensus on the boundaries that enclose the Sequoia and the \textit{Gaia}-Enceladus remnant stars.
These criteria, involving orbital energy, angular momentum, and orbital actions, are shown as boxes in the panels of Fig.~\ref{fig:myeongplot}. 

As can be noted, the two criteria do not always agree on which stars can be associated to each structure. It is possible that chemical abundances could help to disentangle the structures. We will report on chemo-dynamic analysis of stellar properties performed with that aim elsewhere (da Silva et al., in preparation). Using criterion I, several \titan\ become candidate members of \textit{Gaia}-Enceladus. Only 
the stars BD$+24$~1676, HE~0926-0508, and LP~815$-43$ are classified as within \textit{Gaia}-Enceladus by both criteria. Stars  BD+26~2621, HD~59392, and BD+02~4651 are classified as within \textit{Gaia}-Enceladus by the criterion of \citep{Massari2019} but as within Sequoia by the criterion of \citep{Myeong2019}. Both criteria agree on classifying star HD~34328 as a Sequoia member. Star Wolf~1492 is part of Sequoia by criterion I, but not by criterion II. The stars Ross~453 and HD~132475 are classified within Sequoia by criterion II, but remain unclassified by criterion I. Finally, CD-33~3337 is a possible Helmi Stream candidate member, as is one of the GBS, HD~298986 (\citet{Myeong2019} does not provide criteria for the Helmi Streams). 
From the metal-poor GBS, HD~140283 and HD~84937 belong to the \textit{Gaia}-Enceladus, and HD~298986 to the Helmi Streams by criterion I. 
HD~106038, is classified as a member of \textit{Gaia}-Enceladus by criterion I but as a member of Sequoia by criterion II. This information is summarized in Table~\ref{tab:membership}.\par

\section{Conclusions}
\label{sec:conclusions}

In response to the current demand of reference stars for all-sky-surveys, we introduced the \titan: a selection of metal-poor stars with accurate parameters intended as calibration objects for spectroscopic analysis. In this first paper we provide \teff\ and \logg\ values for dwarf and subgiants. 
We estimated the typical 1$\sigma$  accuracy of the \teff\ values to be of the order of 50~K while the average precision is of $\sim$40~K. The accuracy and precision of \logg\ remains within 0.04~dex.

The atmospheric parameters presented here are model-dependent, inasmuch as the \teff\ values depend on computations of synthetic H$\alpha$ profiles and the \logg\ values depend on stellar evolution models. Therefore, we scrutinized the methodology against the \textit{Gaia} benchmark stars. The reference \teff\ of the benchmarks were inferred either quasi-directly, by interferometric measurements, or by photometric calibrations based on interferometric measurements or, in last instance, by the IRFM, which is almost model independent. 

The reliability of the \titan' parameters, presented in Table~\ref{tab:CAMK_stars}, thus stands on the ability of our methodology to recover the reference parameters of the \textit{Gaia} benchmarks. The prime result, in this regard, is showcased in the comparison of Fig.~\ref{fig:Teff_accuracy}. We did not find significant correlations or offsets with respect to any atmospheric parameter.
This allowed us to conclude that the method permits the derivation of accurate \teff\ values within the whole parameter space analysed here. 
From this comparison, we determined the accuracy of the 3D non-LTE H$\alpha$ model used to derive \teff\ to be 50~K.
We note that this value is unavoidably tied to (and perhaps dominated by) the uncertainties of the reference \teff\ scale, the average value of which is  60~K.
Thus, this is likely the main source of the dispersion in the plots. Similarly, we found that our method recovers values of surface gravity that are very close to the standard ones, when the same parallaxes are used (Fig.~\ref{fig:logg_dispersion}).

A direct comparison with \teff\ values obtained using the IRFM for the \titan\ confirms the results above (Fig.~\ref{fig:IRFM_comparison}). A weak correlation of the temperature differences was found only with respect to \logg. Nevertheless, we showed this to be partially explained by photometric or parallax uncertainties, affecting the IRFM analysis.
Still, the differences are small and of the same magnitude of the uncertainties of both sets of temperatures. The important conclusion then is that \teff\ scales based on interferometry, on the IRFM, and on 3D non-LTE H$\alpha$ models are consistent to within 1\%. This conclusion is valid not only for metal-rich stars \citep[as shown in][]{giribaldi2019_Ha}, but also for metal-poor stars (as demonstrated here).

From the perspective of precision spectroscopy, the series of consistent analysis loops, as implemented here, is an efficient methodology to eliminate biases due to the inter-dependence between the atmospheric parameters. There were two keys for the success of our methodology. The first one was the state-of-the-art of the H$\alpha$ models based on 3D non-LTE atmospheric models \citep{amarsi2018}. The second one was the availability of parallax measurements of superb quality provided by the \textit{Gaia} mission \citep{2020arXiv201201533G}. Both were important since, for metal-poor stars, the fine tuning of \teff\ from H$\alpha$ fitting requires accurate \logg\ values, which could only be obtained via the accurate and precise parallaxes.

Finally, we remark that the \titan\ are a reliable new sample to be used as templates for calibrating spectroscopic techniques, in the analysis of large star samples, and/or the application of strictly differential spectroscopic methods as done, for example, by \citet{reggiani2016,reggiani2017} for metal-poor stars, and \citet{yong2013} for stellar clusters.

\begin{acknowledgements}
    The authors acknowledge the anonymous referee for the careful reading of the manuscript and the constructive criticism that has helped us to improve the presentation of the results. 
    REG, ARS, and RS acknowledge support by the National Science Centre, Poland, through project 2018/31/B/ST9/01469. 
    REG acknowledges Ana Giribaldi and Margarita Santamar\'{i}a for the the initial support of this work.
    DC acknowledges partial support by the Coordena\c{c}\~{a}o de Aperfei\c{c}oamento de Pessoal de N\'ivel Superior - Brasil (CAPES) - Finace code 001. This work is based on observations collected at the European Southern Observatory under ESO programmes: 67.D-0439(A), 095.D-0504(A), 093.D-0095(A), 68.B-0475(A), 074.B-0639(A), 65.L-0507(A), 076.B-0133(A), 077.D-0299(A), 68.D-0094(A), 71.B-0529(A), 67.D-0086(A), 077.B-0507(A), 70.D-0474(A), 072.B-0585(A), 076.A-0463(A), 086.D-0871(A), 090.B-0605(A), 67.D-0554(A), 188.B-3002(H), 188.B-3002(B), 266.D-5655(A), 072.C-0513(B), 075.D-0760(A), 096.C-0053(A), 081.C-0802(C), 082.C-0427(A), 086.C-0145(A), 072.C-0488(E), 073.D-0590(A), 081.D-0531(A), 183.C-0972(A), 183.D-0729(A), 080.D-0347(A), 078.C-0233(B), 082.B-0610(A), 085.C-0063(A), 190.C-0027(A), 192.C-0852(A), 196.C-0042(E), 295.C-5031(A), 60.A-9036(A), 60.A-9700(G), 082.B-0610(A), 086.C-0284(A), 086.C-0448(A), 289.D-5015(A), 60.A-9036(A).
    Use was made of the Simbad database, operated at the CDS, Strasbourg, France, and of NASA’s Astrophysics Data System Bibliographic Services. 
     This research has made use of the services of the ESO Science Archive Facility.
    This publication makes use of data products from the Two Micron
    All Sky Survey, which is a joint project of the University of
    Massachusetts and the Infrared Processing and Analysis
    Center/California Institute of Technology, funded by the National Aeronautics and Space Administration and the National Science Foundation.
    This work made use of TOPCAT \citep{TOPCAT}.
    This research made use of Astropy\footnote{http://www.astropy.org}, a community-developed core Python package for Astronomy \citep{astropy:2013, astropy:2018}. 
    This work presents results from the European Space Agency (ESA)
    space mission Gaia. Gaia data are being processed by the Gaia Data Processing and Analysis Consortium (DPAC). Funding for the DPAC is provided by national institutions, in particular the institutions participating in the Gaia MultiLateral Agreement (MLA). The Gaia mission website is \url{https://www.cosmos.esa.int/gaia}. The Gaia archive website is \url{https://archives.esac.esa.int/gaia}.
\end{acknowledgements}
    
\bibliographystyle{aa} 
\bibliography{riano}

\begin{thebibliography}{137}
\expandafter\ifx\csname natexlab\endcsname\relax\def\natexlab#1{#1}\fi

\bibitem[{{Alonso} {et~al.}(1996){Alonso}, {Arribas}, \&
  {Martinez-Roger}}]{alonso1996}
{Alonso}, A., {Arribas}, S., \& {Martinez-Roger}, C. 1996, \aaps, 117, 227

\bibitem[{{Amarsi} {et~al.}(2016){Amarsi}, {Lind}, {Asplund}, {Barklem}, \&
  {Collet}}]{amarsi2016}
{Amarsi}, A.~M., {Lind}, K., {Asplund}, M., {Barklem}, P.~S., \& {Collet}, R.
  2016, \mnras, 463, 1518

\bibitem[{{Amarsi} {et~al.}(2019){Amarsi}, {Nissen}, {Asplund}, {Lind}, \&
  {Barklem}}]{amarsi2019}
{Amarsi}, A.~M., {Nissen}, P.~E., {Asplund}, M., {Lind}, K., \& {Barklem},
  P.~S. 2019, \aap, 622, L4

\bibitem[{{Amarsi} {et~al.}(2018){Amarsi}, {Nordlander}, {Barklem}, {Asplund},
  {Collet}, \& {Lind}}]{amarsi2018}
{Amarsi}, A.~M., {Nordlander}, T., {Barklem}, P.~S., {et~al.} 2018, \aap, 615,
  A139

\bibitem[{{Astropy Collaboration} {et~al.}(2018){Astropy Collaboration},
  {Price-Whelan}, {Sip{\H{o}}cz}, {G{\"u}nther}, {Lim}, {Crawford}, {Conseil},
  {Shupe}, {Craig}, {Dencheva}, {Ginsburg}, {VanderPlas}, {Bradley},
  {P{\'e}rez-Su{\'a}rez}, {de Val-Borro}, {Aldcroft}, {Cruz}, {Robitaille},
  {Tollerud}, {Ardelean}, {Babej}, {Bach}, {Bachetti}, {Bakanov}, {Bamford},
  {Barentsen}, {Barmby}, {Baumbach}, {Berry}, {Biscani}, {Boquien}, {Bostroem},
  {Bouma}, {Brammer}, {Bray}, {Breytenbach}, {Buddelmeijer}, {Burke},
  {Calderone}, {Cano Rodr{\'\i}guez}, {Cara}, {Cardoso}, {Cheedella}, {Copin},
  {Corrales}, {Crichton}, {D'Avella}, {Deil}, {Depagne}, {Dietrich}, {Donath},
  {Droettboom}, {Earl}, {Erben}, {Fabbro}, {Ferreira}, {Finethy}, {Fox},
  {Garrison}, {Gibbons}, {Goldstein}, {Gommers}, {Greco}, {Greenfield},
  {Groener}, {Grollier}, {Hagen}, {Hirst}, {Homeier}, {Horton}, {Hosseinzadeh},
  {Hu}, {Hunkeler}, {Ivezi{\'c}}, {Jain}, {Jenness}, {Kanarek}, {Kendrew},
  {Kern}, {Kerzendorf}, {Khvalko}, {King}, {Kirkby}, {Kulkarni}, {Kumar},
  {Lee}, {Lenz}, {Littlefair}, {Ma}, {Macleod}, {Mastropietro}, {McCully},
  {Montagnac}, {Morris}, {Mueller}, {Mumford}, {Muna}, {Murphy}, {Nelson},
  {Nguyen}, {Ninan}, {N{\"o}the}, {Ogaz}, {Oh}, {Parejko}, {Parley}, {Pascual},
  {Patil}, {Patil}, {Plunkett}, {Prochaska}, {Rastogi}, {Reddy Janga},
  {Sabater}, {Sakurikar}, {Seifert}, {Sherbert}, {Sherwood-Taylor}, {Shih},
  {Sick}, {Silbiger}, {Singanamalla}, {Singer}, {Sladen}, {Sooley},
  {Sornarajah}, {Streicher}, {Teuben}, {Thomas}, {Tremblay}, {Turner},
  {Terr{\'o}n}, {van Kerkwijk}, {de la Vega}, {Watkins}, {Weaver}, {Whitmore},
  {Woillez}, {Zabalza}, \& {Astropy Contributors}}]{astropy:2018}
{Astropy Collaboration}, {Price-Whelan}, A.~M., {Sip{\H{o}}cz}, B.~M., {et~al.}
  2018, \aj, 156, 123

\bibitem[{{Astropy Collaboration} {et~al.}(2013){Astropy Collaboration},
  {Robitaille}, {Tollerud}, {Greenfield}, {Droettboom}, {Bray}, {Aldcroft},
  {Davis}, {Ginsburg}, {Price-Whelan}, {Kerzendorf}, {Conley}, {Crighton},
  {Barbary}, {Muna}, {Ferguson}, {Grollier}, {Parikh}, {Nair}, {Unther},
  {Deil}, {Woillez}, {Conseil}, {Kramer}, {Turner}, {Singer}, {Fox}, {Weaver},
  {Zabalza}, {Edwards}, {Azalee Bostroem}, {Burke}, {Casey}, {Crawford},
  {Dencheva}, {Ely}, {Jenness}, {Labrie}, {Lim}, {Pierfederici}, {Pontzen},
  {Ptak}, {Refsdal}, {Servillat}, \& {Streicher}}]{astropy:2013}
{Astropy Collaboration}, {Robitaille}, T.~P., {Tollerud}, E.~J., {et~al.} 2013,
  \aap, 558, A33

\bibitem[{{Bailer-Jones}(2015)}]{BailerJones2015}
{Bailer-Jones}, C. A.~L. 2015, \pasp, 127, 994

\bibitem[{{Ballester} {et~al.}(2000){Ballester}, {Modigliani}, {Boitquin},
  {Cristiani}, {Hanuschik}, {Kaufer}, \& {Wolf}}]{2000Msngr.101...31B}
{Ballester}, P., {Modigliani}, A., {Boitquin}, O., {et~al.} 2000, The
  Messenger, 101, 31

\bibitem[{{Barb{\'a}} {et~al.}(2019){Barb{\'a}}, {Minniti}, {Geisler},
  {Alonso-Garc{\'\i}a}, {Hempel}, {Monachesi}, {Arias}, \&
  {G{\'o}mez}}]{Barba2019}
{Barb{\'a}}, R.~H., {Minniti}, D., {Geisler}, D., {et~al.} 2019, \apjl, 870,
  L24

\bibitem[{{Barklem}(2016)}]{2016A&ARv..24....9B}
{Barklem}, P.~S. 2016, \aapr, 24, 9

\bibitem[{{Bazot} {et~al.}(2011){Bazot}, {Ireland}, {Huber}, {Bedding},
  {Broomhall}, {Campante}, {Carfantan}, {Chaplin}, {Elsworth}, {Mel{\'e}ndez},
  {Petit}, {Th{\'e}ado}, {Van Grootel}, {Arentoft}, {Asplund}, {Castro},
  {Christensen-Dalsgaard}, {Do Nascimento}, {Dintrans}, {Dumusque}, {Kjeldsen},
  {McAlister}, {Metcalfe}, {Monteiro}, {Santos}, {Sousa}, {Sturmann},
  {Sturmann}, {ten Brummelaar}, {Turner}, \& {Vauclair}}]{bazot2011}
{Bazot}, M., {Ireland}, M.~J., {Huber}, D., {et~al.} 2011, \aap, 526, L4

\bibitem[{{Beers} {et~al.}(2002){Beers}, {Drilling}, {Rossi}, {Chiba}, {Rhee},
  {F{\"u}hrmeister}, {Norris}, \& {von Hippel}}]{Beers}
{Beers}, T.~C., {Drilling}, J.~S., {Rossi}, S., {et~al.} 2002, \aj, 124, 931

\bibitem[{{Beers} {et~al.}(2017){Beers}, {Placco}, {Carollo}, {Rossi}, {Lee},
  {Frebel}, {Norris}, {Dietz}, \& {Masseron}}]{2017ApJ...835...81B}
{Beers}, T.~C., {Placco}, V.~M., {Carollo}, D., {et~al.} 2017, \apj, 835, 81

\bibitem[{{Belokurov} {et~al.}(2018){Belokurov}, {Erkal}, {Evans}, {Koposov},
  \& {Deason}}]{Belokurov2018}
{Belokurov}, V., {Erkal}, D., {Evans}, N.~W., {Koposov}, S.~E., \& {Deason},
  A.~J. 2018, \mnras, 478, 611

\bibitem[{{Blackwell} {et~al.}(1980){Blackwell}, {Petford}, \&
  {Shallis}}]{Blackwell1980}
{Blackwell}, D.~E., {Petford}, A.~D., \& {Shallis}, M.~J. 1980, \aap, 82, 249

\bibitem[{{Blackwell} \& {Shallis}(1977)}]{Blackwell1977}
{Blackwell}, D.~E. \& {Shallis}, M.~J. 1977, \mnras, 180, 177

\bibitem[{{Blackwell} {et~al.}(1979){Blackwell}, {Shallis}, \&
  {Selby}}]{Blackwell1979}
{Blackwell}, D.~E., {Shallis}, M.~J., \& {Selby}, M.~J. 1979, \mnras, 188, 847

\bibitem[{{Blanco-Cuaresma}(2019)}]{2019MNRAS.486.2075B}
{Blanco-Cuaresma}, S. 2019, \mnras, 486, 2075

\bibitem[{{Blanco-Cuaresma} {et~al.}(2014){Blanco-Cuaresma}, {Soubiran},
  {Heiter}, \& {Jofr{\'e}}}]{blanco-cuaresma2014}
{Blanco-Cuaresma}, S., {Soubiran}, C., {Heiter}, U., \& {Jofr{\'e}}, P. 2014,
  \aap, 569, A111

\bibitem[{{Bohlin} {et~al.}(2014){Bohlin}, {Gordon}, \&
  {Tremblay}}]{2014PASP..126..711B}
{Bohlin}, R.~C., {Gordon}, K.~D., \& {Tremblay}, P.~E. 2014, \pasp, 126, 711

\bibitem[{{Bonifacio} {et~al.}(2018){Bonifacio}, {Caffau}, {Ludwig}, {Steffen},
  {Castelli}, {Gallagher}, {Ku{\v{c}}inskas}, {Prakapavi{\v{c}}ius}, {Cayrel},
  {Freytag}, {Plez}, \& {Homeier}}]{2018A&A...611A..68B}
{Bonifacio}, P., {Caffau}, E., {Ludwig}, H.~G., {et~al.} 2018, \aap, 611, A68

\bibitem[{{Bovy}(2015)}]{Bovy2015}
{Bovy}, J. 2015, \apjs, 216, 29

\bibitem[{{Brown}(2021)}]{brown2021microarcsecond}
{Brown}, A.~G.~A. 2021, arXiv e-prints, arXiv:2102.11712

\bibitem[{{Brucalassi} {et~al.}(2021){Brucalassi}, {Tsantaki}, {Magrini},
  {Sousa}, {Danielski}, {Biazzo}, {Casali}, {Van der Swaelmen}, {Rainer},
  {Adibekyan}, {Delgado-Mena}, \& {Sanna}}]{2021arXiv210102242B}
{Brucalassi}, A., {Tsantaki}, M., {Magrini}, L., {et~al.} 2021, arXiv e-prints,
  arXiv:2101.02242

\bibitem[{{Capitanio} {et~al.}(2017){Capitanio}, {Lallement}, {Vergely},
  {Elyajouri}, \& {Monreal-Ibero}}]{Capitanio2017}
{Capitanio}, L., {Lallement}, R., {Vergely}, J.~L., {Elyajouri}, M., \&
  {Monreal-Ibero}, A. 2017, \aap, 606, A65

\bibitem[{{Casagrande} {et~al.}(2020){Casagrande}, {Lin}, {Rains}, {Liu},
  {Buder}, {Horner}, {Asplund}, {Lewis}, {Martell}, {Nordlander}, {Stello},
  {Ting}, {Wittenmyer}, {Bland-Hawthorn}, {Casey}, {De Silva}, {D'Orazi},
  {Freeman}, {Hayden}, {Kos}, {Lind}, {Schlesinger}, {Sharma}, {Simpson},
  {Zucker}, \& {Zwitter}}]{Casagrande2020}
{Casagrande}, L., {Lin}, J., {Rains}, A.~D., {et~al.} 2020, arXiv e-prints,
  arXiv:2011.02517

\bibitem[{{Casagrande} {et~al.}(2006){Casagrande}, {Portinari}, \&
  {Flynn}}]{casagrande2006}
{Casagrande}, L., {Portinari}, L., \& {Flynn}, C. 2006, \mnras, 373, 13

\bibitem[{{Casagrande} {et~al.}(2014){Casagrande}, {Portinari}, {Glass},
  {Laney}, {Silva Aguirre}, {Datson}, {Andersen}, {Nordstr{\"o}m}, {Holmberg},
  {Flynn}, \& {Asplund}}]{casagrande2014}
{Casagrande}, L., {Portinari}, L., {Glass}, I.~S., {et~al.} 2014, \mnras, 439,
  2060

\bibitem[{{Casagrande} {et~al.}(2010){Casagrande}, {Ram{\'{\i}}rez},
  {Mel{\'e}ndez}, {Bessell}, \& {Asplund}}]{Casagrande2010}
{Casagrande}, L., {Ram{\'{\i}}rez}, I., {Mel{\'e}ndez}, J., {Bessell}, M., \&
  {Asplund}, M. 2010, \aap, 512, A54

\bibitem[{{Cayrel} {et~al.}(1985){Cayrel}, {Cayrel de Strobel}, \&
  {Campbell}}]{cayrel1985}
{Cayrel}, R., {Cayrel de Strobel}, G., \& {Campbell}, B. 1985, \aap, 146, 249

\bibitem[{{Cayrel de Strobel}(1985)}]{1985IAUS..111..137C}
{Cayrel de Strobel}, G. 1985, in Calibration of Fundamental Stellar Quantities,
  ed. D.~S. {Hayes}, L.~E. {Pasinetti}, \& A.~G.~D. {Philip}, Vol. 111,
  137--162

\bibitem[{{Cui} {et~al.}(2012){Cui}, {Zhao}, {Chu}, {Li}, {Li}, {Zhang}, {Su},
  {Yao}, {Wang}, {Xing}, {Li}, {Zhu}, {Wang}, {Gu}, {Luo}, {Xu}, {Zhang},
  {Liu}, {Zhang}, {Yang}, {Cao}, {Chen}, {Chen}, {Chen}, {Chen}, {Chu}, {Feng},
  {Gong}, {Hou}, {Hu}, {Hu}, {Hu}, {Jia}, {Jiang}, {Jiang}, {Jiang}, {Jin},
  {Li}, {Li}, {Li}, {Liu}, {Liu}, {Lu}, {Mao}, {Men}, {Qi}, {Qi}, {Shi},
  {Tang}, {Tao}, {Wang}, {Wang}, {Wang}, {Wang}, {Wang}, {Wang}, {Wang},
  {Wang}, {Wang}, {Wang}, {Wang}, {Wang}, {Xu}, {Xu}, {Yang}, {Yu}, {Yuan},
  {Yuan}, {Zhai}, {Zhang}, {Zhang}, {Zhang}, {Zhao}, {Zhou}, {Zhou}, {Zhu}, \&
  {Zou}}]{2012RAA....12.1197C}
{Cui}, X.-Q., {Zhao}, Y.-H., {Chu}, Y.-Q., {et~al.} 2012, Research in Astronomy
  and Astrophysics, 12, 1197

\bibitem[{{Davis} {et~al.}(2000){Davis}, {Tango}, \&
  {Booth}}]{2000MNRAS.318..387D}
{Davis}, J., {Tango}, W.~J., \& {Booth}, A.~J. 2000, \mnras, 318, 387

\bibitem[{{de Jong} {et~al.}(2019){de Jong}, {Agertz}, {Berbel}, {Aird},
  {Alexander}, {Amarsi}, {Anders}, {Andrae}, {Ansarinejad}, {Ansorge},
  {Antilogus}, {Anwand -Heerwart}, {Arentsen}, {Arnadottir}, {Asplund},
  {Auger}, {Azais}, {Baade}, {Baker}, {Baker}, {Balbinot}, {Baldry}, {Banerji},
  {Barden}, {Barklem}, {Barth{\'e}l{\'e}my-Mazot}, {Battistini}, {Bauer},
  {Bell}, {Bellido-Tirado}, {Bellstedt}, {Belokurov}, {Bensby}, {Bergemann},
  {Bestenlehner}, {Bielby}, {Bilicki}, {Blake}, {Bland-Hawthorn}, {Boeche},
  {Boland}, {Boller}, {Bongard}, {Bongiorno}, {Bonifacio}, {Boudon}, {Brooks},
  {Brown}, {Brown}, {Br{\"u}ggen}, {Brynnel}, {Brzeski}, {Buchert},
  {Buschkamp}, {Caffau}, {Caillier}, {Carrick}, {Casagrande}, {Case}, {Casey},
  {Cesarini}, {Cescutti}, {Chapuis}, {Chiappini}, {Childress}, {Christlieb},
  {Church}, {Cioni}, {Cluver}, {Colless}, {Collett}, {Comparat}, {Cooper},
  {Couch}, {Courbin}, {Croom}, {Croton}, {Daguis{\'e}}, {Dalton}, {Davies},
  {Davis}, {de Laverny}, {Deason}, {Dionies}, {Disseau}, {Doel}, {D{\"o}scher},
  {Driver}, {Dwelly}, {Eckert}, {Edge}, {Edvardsson}, {Youssoufi}, {Elhaddad},
  {Enke}, {Erfanianfar}, {Farrell}, {Fechner}, {Feiz}, {Feltzing}, {Ferreras},
  {Feuerstein}, {Feuillet}, {Finoguenov}, {Ford}, {Fotopoulou}, {Fouesneau},
  {Frenk}, {Frey}, {Gaessler}, {Geier}, {Fusillo}, {Gerhard}, {Giannantonio},
  {Giannone}, {Gibson}, {Gillingham}, {Gonz{\'a}lez-Fern{\'a}ndez},
  {Gonzalez-Solares}, {Gottloeber}, {Gould}, {Grebel}, {Gueguen}, {Guiglion},
  {Haehnelt}, {Hahn}, {Hansen}, {Hartman}, {Hauptner}, {Hawkins}, {Haynes},
  {Haynes}, {Heiter}, {Helmi}, {Aguayo}, {Hewett}, {Hinton}, {Hobbs}, {Hoenig},
  {Hofman}, {Hook}, {Hopgood}, {Hopkins}, {Hourihane}, {Howes}, {Howlett},
  {Huet}, {Irwin}, {Iwert}, {Jablonka}, {Jahn}, {Jahnke}, {Jarno}, {Jin},
  {Jofre}, {Johl}, {Jones}, {J{\"o}nsson}, {Jordan}, {Karovicova}, {Khalatyan},
  {Kelz}, {Kennicutt}, {King}, {Kitaura}, {Klar}, {Klauser}, {Kneib}, {Koch},
  {Koposov}, {Kordopatis}, {Korn}, {Kosmalski}, {Kotak}, {Kovalev}, {Kreckel},
  {Kripak}, {Krumpe}, {Kuijken}, {Kunder}, {Kushniruk}, {Lam}, {Lamer},
  {Laurent}, {Lawrence}, {Lehmitz}, {Lemasle}, {Lewis}, {Li}, {Lidman}, {Lind},
  {Liske}, {Lizon}, {Loveday}, {Ludwig}, {McDermid}, {Maguire}, {Mainieri},
  {Mali}, {Mandel}, {Mandel}, {Mannering}, {Martell}, {Martinez Delgado},
  {Matijevic}, {McGregor}, {McMahon}, {McMillan}, {Mena}, {Merloni}, {Meyer},
  {Michel}, {Micheva}, {Migniau}, {Minchev}, {Monari}, {Muller}, {Murphy},
  {Muthukrishna}, {Nandra}, {Navarro}, {Ness}, {Nichani}, {Nichol}, {Nicklas},
  {Niederhofer}, {Norberg}, {Obreschkow}, {Oliver}, {Owers}, {Pai},
  {Pankratow}, {Parkinson}, {Paschke}, {Paterson}, {Pecontal}, {Parry},
  {Phillips}, {Pillepich}, {Pinard}, {Pirard}, {Piskunov}, {Plank},
  {Pl{\"u}schke}, {Pons}, {Popesso}, {Power}, {Pragt}, {Pramskiy}, {Pryer},
  {Quattri}, {Queiroz}, {Quirrenbach}, {Rahurkar}, {Raichoor}, {Ramstedt},
  {Rau}, {Recio-Blanco}, {Reiss}, {Renaud}, {Revaz}, {Rhode}, {Richard},
  {Richter}, {Rix}, {Robotham}, {Roelfsema}, {Romaniello}, {Rosario},
  {Rothmaier}, {Roukema}, {Ruchti}, {Rupprecht}, {Rybizki}, {Ryde}, {Saar},
  {Sadler}, {Sahl{\'e}n}, {Salvato}, {Sassolas}, {Saunders}, {Saviauk},
  {Sbordone}, {Schmidt}, {Schnurr}, {Scholz}, {Schwope}, {Seifert}, {Shanks},
  {Sheinis}, {Sivov}, {Sk{\'u}lad{\'o}ttir}, {Smartt}, {Smedley}, {Smith},
  {Smith}, {Sorce}, {Spitler}, {Starkenburg}, {Steinmetz}, {Stilz}, {Storm},
  {Sullivan}, {Sutherland}, {Swann}, {Tamone}, {Taylor}, {Teillon}, {Tempel},
  {ter Horst}, {Thi}, {Tolstoy}, {Trager}, {Traven}, {Tremblay}, {Tresse},
  {Valentini}, {van de Weygaert}, {van den Ancker}, {Veljanoski}, {Venkatesan},
  {Wagner}, {Wagner}, {Walcher}, {Waller}, {Walton}, {Wang}, {Winkler},
  {Wisotzki}, {Worley}, {Worseck}, {Xiang}, {Xu}, {Yong}, {Zhao}, {Zheng},
  {Zscheyge}, \& {Zucker}}]{2019Msngr.175....3D}
{de Jong}, R.~S., {Agertz}, O., {Berbel}, A.~A., {et~al.} 2019, The Messenger,
  175, 3

\bibitem[{{De Silva} {et~al.}(2015){De Silva}, {Freeman}, {Bland-Hawthorn},
  {Martell}, {de Boer}, {Asplund}, {Keller}, {Sharma}, {Zucker}, {Zwitter},
  {Anguiano}, {Bacigalupo}, {Bayliss}, {Beavis}, {Bergemann}, {Campbell},
  {Cannon}, {Carollo}, {Casagrande}, {Casey}, {Da Costa}, {D'Orazi}, {Dotter},
  {Duong}, {Heger}, {Ireland}, {Kafle}, {Kos}, {Lattanzio}, {Lewis}, {Lin},
  {Lind}, {Munari}, {Nataf}, {O'Toole}, {Parker}, {Reid}, {Schlesinger},
  {Sheinis}, {Simpson}, {Stello}, {Ting}, {Traven}, {Watson}, {Wittenmyer},
  {Yong}, \& {{\v{Z}}erjal}}]{2015MNRAS.449.2604D}
{De Silva}, G.~M., {Freeman}, K.~C., {Bland-Hawthorn}, J., {et~al.} 2015,
  \mnras, 449, 2604

\bibitem[{{Dekker} {et~al.}(2000){Dekker}, {D'Odorico}, {Kaufer}, {Delabre}, \&
  {Kotzlowski}}]{dekker2000}
{Dekker}, H., {D'Odorico}, S., {Kaufer}, A., {Delabre}, B., \& {Kotzlowski}, H.
  2000, in Society of Photo-Optical Instrumentation Engineers (SPIE) Conference
  Series, Vol. 4008, Optical and IR Telescope Instrumentation and Detectors,
  ed. M.~{Iye} \& A.~F. {Moorwood}, 534--545

\bibitem[{{dos Santos} {et~al.}(2016){dos Santos}, {Mel{\'e}ndez}, {do
  Nascimento}, {Bedell}, {Ram{\'\i}rez}, {Bean}, {Asplund}, {Spina},
  {Dreizler}, {Alves-Brito}, \& {Casagrande}}]{2016A&A...592A.156D}
{dos Santos}, L.~A., {Mel{\'e}ndez}, J., {do Nascimento}, J.-D., {et~al.} 2016,
  \aap, 592, A156

\bibitem[{{Fabbro} {et~al.}(2018){Fabbro}, {Venn}, {O'Briain}, {Bialek},
  {Kielty}, {Jahandar}, \& {Monty}}]{2018MNRAS.475.2978F}
{Fabbro}, S., {Venn}, K.~A., {O'Briain}, T., {et~al.} 2018, \mnras, 475, 2978

\bibitem[{{Freeman} \& {Bland-Hawthorn}(2002)}]{2002ARA&A..40..487F}
{Freeman}, K. \& {Bland-Hawthorn}, J. 2002, \araa, 40, 487

\bibitem[{{Fuhrmann}(1998)}]{fuhrmann1998}
{Fuhrmann}, K. 1998, \aap, 338, 161

\bibitem[{{Fuhrmann} {et~al.}(1994){Fuhrmann}, {Axer}, \&
  {Gehren}}]{fuhrmann1994}
{Fuhrmann}, K., {Axer}, M., \& {Gehren}, T. 1994, \aap, 285, 585

\bibitem[{{Gaia Collaboration} {et~al.}(2018){Gaia Collaboration}, {Brown},
  {Vallenari}, {Prusti}, {de Bruijne}, {Babusiaux}, {Bailer-Jones}, {Biermann},
  {Evans}, {Eyer}, \& et~al.}]{Gai(c)}
{Gaia Collaboration}, {Brown}, A.~G.~A., {Vallenari}, A., {et~al.} 2018, \aap,
  616, A1

\bibitem[{{Gaia Collaboration} {et~al.}(2020){Gaia Collaboration}, {Brown},
  {Vallenari}, {Prusti}, {de Bruijne}, {Babusiaux}, \&
  {Biermann}}]{2020arXiv201201533G}
{Gaia Collaboration}, {Brown}, A.~G.~A., {Vallenari}, A., {et~al.} 2020, arXiv
  e-prints, arXiv:2012.01533

\bibitem[{{Gaia Collaboration} {et~al.}(2016){Gaia Collaboration}, {Prusti},
  {de Bruijne}, {Brown}, {Vallenari}, {Babusiaux}, {Bailer-Jones}, {Bastian},
  {Biermann}, {Evans}, \& et~al.}]{Gai(a)}
{Gaia Collaboration}, {Prusti}, T., {de Bruijne}, J.~H.~J., {et~al.} 2016,
  \aap, 595, A1

\bibitem[{{Gavel} {et~al.}(2019){Gavel}, {Gruyters}, {Heiter}, {Korn}, {Lind},
  \& {Nordlander}}]{2019A&A...629A..74G}
{Gavel}, A., {Gruyters}, P., {Heiter}, U., {et~al.} 2019, \aap, 629, A74

\bibitem[{{Gehren} {et~al.}(2006){Gehren}, {Shi}, {Zhang}, {Zhao}, \&
  {Korn}}]{2006A&A...451.1065G}
{Gehren}, T., {Shi}, J.~R., {Zhang}, H.~W., {Zhao}, G., \& {Korn}, A.~J. 2006,
  \aap, 451, 1065

\bibitem[{{Gilmore} {et~al.}(2012){Gilmore}, {Randich}, {Asplund}, {Binney},
  {Bonifacio}, {Drew}, {Feltzing}, {Ferguson}, {Jeffries}, {Micela},
  {Negueruela}, {Prusti}, {Rix}, {Vallenari}, {Alfaro}, {Allende-Prieto},
  {Babusiaux}, {Bensby}, {Blomme}, {Bragaglia}, {Flaccomio}, {Fran{\c{c}}ois},
  {Irwin}, {Koposov}, {Korn}, {Lanzafame}, {Pancino}, {Paunzen},
  {Recio-Blanco}, {Sacco}, {Smiljanic}, {Van Eck}, {Walton}, {Aden}, {Aerts},
  {Affer}, {Alcala}, {Altavilla}, {Alves}, {Antoja}, {Arenou}, {Argiroffi},
  {Asensio Ramos}, {Bailer-Jones}, {Balaguer-Nunez}, {Bayo}, {Barbuy},
  {Barisevicius}, {Barrado y Navascues}, {Battistini}, {Bellas Velidis},
  {Bellazzini}, {Belokurov}, {Bergemann}, {Bertelli}, {Biazzo}, {Bienayme},
  {Bland-Hawthorn}, {Boeche}, {Bonito}, {Boudreault}, {Bouvier}, {Brandao},
  {Brown}, {de Bruijne}, {Burleigh}, {Caballero}, {Caffau}, {Calura},
  {Capuzzo-Dolcetta}, {Caramazza}, {Carraro}, {Casagrande}, {Casewell},
  {Chapman}, {Chiappini}, {Chorniy}, {Christlieb}, {Cignoni}, {Cocozza},
  {Colless}, {Collet}, {Collins}, {Correnti}, {Covino}, {Crnojevic}, {Cropper},
  {Cunha}, {Damiani}, {David}, {Delgado}, {Duffau}, {Edvardsson}, {Eldridge},
  {Enke}, {Eriksson}, {Evans}, {Eyer}, {Famaey}, {Fellhauer}, {Ferreras},
  {Figueras}, {Fiorentino}, {Flynn}, {Folha}, {Franciosini}, {Frasca},
  {Freeman}, {Fremat}, {Friel}, {Gaensicke}, {Gameiro}, {Garzon}, {Geier},
  {Geisler}, {Gerhard}, {Gibson}, {Gomboc}, {Gomez}, {Gonzalez-Fernandez},
  {Gonzalez Hernandez}, {Gosset}, {Grebel}, {Greimel}, {Groenewegen},
  {Grundahl}, {Guarcello}, {Gustafsson}, {Hadrava}, {Hatzidimitriou}, {Hambly},
  {Hammersley}, {Hansen}, {Haywood}, {Heber}, {Heiter}, {Held}, {Helmi},
  {Hensler}, {Herrero}, {Hill}, {Hodgkin}, {Huelamo}, {Huxor}, {Ibata},
  {Jackson}, {de Jong}, {Jonker}, {Jordan}, {Jordi}, {Jorissen}, {Katz},
  {Kawata}, {Keller}, {Kharchenko}, {Klement}, {Klutsch}, {Knude}, {Koch},
  {Kochukhov}, {Kontizas}, {Koubsky}, {Lallement}, {de Laverny}, {van Leeuwen},
  {Lemasle}, {Lewis}, {Lind}, {Lindstrom}, {Lobel}, {Lopez Santiago}, {Lucas},
  {Ludwig}, {Lueftinger}, {Magrini}, {Maiz Apellaniz}, {Maldonado}, {Marconi},
  {Marino}, {Martayan}, {Martinez-Valpuesta}, {Matijevic}, {McMahon},
  {Messina}, {Meyer}, {Miglio}, {Mikolaitis}, {Minchev}, {Minniti}, {Moitinho},
  {Momany}, {Monaco}, {Montalto}, {Monteiro}, {Monier}, {Montes}, {Mora},
  {Moraux}, {Morel}, {Mowlavi}, {Mucciarelli}, {Munari}, {Napiwotzki},
  {Nardetto}, {Naylor}, {Naze}, {Nelemans}, {Okamoto}, {Ortolani}, {Pace},
  {Palla}, {Palous}, {Parker}, {Penarrubia}, {Pillitteri}, {Piotto}, {Posbic},
  {Prisinzano}, {Puzeras}, {Quirrenbach}, {Ragaini}, {Read}, {Read}, {Reyle},
  {De Ridder}, {Robichon}, {Robin}, {Roeser}, {Romano}, {Royer}, {Ruchti},
  {Ruzicka}, {Ryan}, {Ryde}, {Santos}, {Sanz Forcada}, {Sarro Baro},
  {Sbordone}, {Schilbach}, {Schmeja}, {Schnurr}, {Schoenrich}, {Scholz},
  {Seabroke}, {Sharma}, {De Silva}, {Smith}, {Solano}, {Sordo}, {Soubiran},
  {Sousa}, {Spagna}, {Steffen}, {Steinmetz}, {Stelzer}, {Stempels},
  {Tabernero}, {Tautvaisiene}, {Thevenin}, {Torra}, {Tosi}, {Tolstoy}, {Turon},
  {Walker}, {Wambsganss}, {Worley}, {Venn}, {Vink}, {Wyse}, {Zaggia},
  {Zeilinger}, {Zoccali}, {Zorec}, {Zucker}, {Zwitter}, \& {Gaia-ESO Survey
  Team}}]{2012Msngr.147...25G}
{Gilmore}, G., {Randich}, S., {Asplund}, M., {et~al.} 2012, {The Gaia-ESO
  Public Spectroscopic Survey}

\bibitem[{{Giribaldi} {et~al.}(2019){Giribaldi}, {Ubaldo-Melo}, {Porto de
  Mello}, {Pasquini}, {Ludwig}, {Ulmer-Moll}, \&
  {Lorenzo-Oliveira}}]{giribaldi2019_Ha}
{Giribaldi}, R.~E., {Ubaldo-Melo}, M.~L., {Porto de Mello}, G.~F., {et~al.}
  2019, \aap, 624, A10

\bibitem[{{Gonz{\'a}lez Hern{\'a}ndez} \& {Bonifacio}(2009)}]{gonzalez2009}
{Gonz{\'a}lez Hern{\'a}ndez}, J.~I. \& {Bonifacio}, P. 2009, \aap, 497, 497

\bibitem[{{Gray} \& {Corbally}(1994)}]{1994AJ....107..742G}
{Gray}, R.~O. \& {Corbally}, C.~J. 1994, \aj, 107, 742

\bibitem[{{Guiglion} {et~al.}(2020){Guiglion}, {Matijevi{\v{c}}}, {Queiroz},
  {Valentini}, {Steinmetz}, {Chiappini}, {Grebel}, {McMillan}, {Kordopatis},
  {Kunder}, {Zwitter}, {Khalatyan}, {Anders}, {Enke}, {Minchev}, {Monari},
  {Wyse}, {Bienaym{\'e}}, {Bland-Hawthorn}, {Gibson}, {Navarro}, {Parker},
  {Reid}, {Seabroke}, \& {Siebert}}]{2020A&A...644A.168G}
{Guiglion}, G., {Matijevi{\v{c}}}, G., {Queiroz}, A.~B.~A., {et~al.} 2020,
  \aap, 644, A168

\bibitem[{{Gustafsson} {et~al.}(2008){Gustafsson}, {Edvardsson}, {Eriksson},
  {J{\o}rgensen}, {Nordlund}, \& {Plez}}]{gustafson2008}
{Gustafsson}, B., {Edvardsson}, B., {Eriksson}, K., {et~al.} 2008, \aap, 486,
  951

\bibitem[{{Hanke} {et~al.}(2020){Hanke}, {Hansen}, {Ludwig}, {Cristallo},
  {McWilliam}, {Grebel}, \& {Piersanti}}]{hanke2020}
{Hanke}, M., {Hansen}, C.~J., {Ludwig}, H.-G., {et~al.} 2020, \aap, 635, A104

\bibitem[{{Hansen} {et~al.}(2017){Hansen}, {Jofr{\'e}}, {Koch}, {McWilliam}, \&
  {Sneden}}]{hansen2017}
{Hansen}, C.~J., {Jofr{\'e}}, P., {Koch}, A., {McWilliam}, A., \& {Sneden},
  C.~S. 2017, \aap, 598, A54

\bibitem[{{Hawkins} {et~al.}(2016){Hawkins}, {Jofr{\'e}}, {Heiter}, {Soubiran},
  {Blanco-Cuaresma}, {Casagrande}, {Gilmore}, {Lind}, {Magrini}, {Masseron},
  {Pancino}, {Randich}, \& {Worley}}]{2016A&A...592A..70H}
{Hawkins}, K., {Jofr{\'e}}, P., {Heiter}, U., {et~al.} 2016, \aap, 592, A70

\bibitem[{Heiter {et~al.}(2015)Heiter, Jofr\'{e}, Gustafsson, Korn, Soubiran,
  \& Th\'{e}venin}]{Heiter}
Heiter, U., Jofr\'{e}, P., Gustafsson, B., {et~al.} 2015, A\&A, 582, 49

\bibitem[{{Heiter} {et~al.}(2015){Heiter}, {Lind}, {Asplund}, {Barklem},
  {Bergemann}, {Magrini}, {Masseron}, {Mikolaitis}, {Pickering}, \&
  {Ruffoni}}]{2015PhyS...90e4010H}
{Heiter}, U., {Lind}, K., {Asplund}, M., {et~al.} 2015, \physscr, 90, 054010

\bibitem[{{Heiter} {et~al.}(2021){Heiter}, {Lind}, {Bergemann}, {Asplund},
  {Mikolaitis}, {Barklem}, {Masseron}, {de Laverny}, {Magrini}, {Edvardsson},
  {J{\"o}nsson}, {Pickering}, {Ryde}, {Bayo Ar{\'a}n}, {Bensby}, {Casey},
  {Feltzing}, {Jofr{\'e}}, {Korn}, {Pancino}, {Damiani}, {Lanzafame}, {Lardo},
  {Monaco}, {Morbidelli}, {Smiljanic}, {Worley}, {Zaggia}, {Randich}, \&
  {Gilmore}}]{2021A&A...645A.106H}
{Heiter}, U., {Lind}, K., {Bergemann}, M., {et~al.} 2021, \aap, 645, A106

\bibitem[{{Hekker}(2020)}]{2020FrASS...7....3H}
{Hekker}, S. 2020, Frontiers in Astronomy and Space Sciences, 7, 3

\bibitem[{{Helmi} {et~al.}(2018){Helmi}, {Babusiaux}, {Koppelman}, {Massari},
  {Veljanoski}, \& {Brown}}]{helmi2018}
{Helmi}, A., {Babusiaux}, C., {Koppelman}, H.~H., {et~al.} 2018, \nat, 563, 85

\bibitem[{{He{\l}miniak} {et~al.}(2019){He{\l}miniak}, {Konacki}, {Maehara},
  {Kambe}, {Ukita}, {Ratajczak}, {Pigulski}, \&
  {Koz{\l}owski}}]{2019MNRAS.484..451H}
{He{\l}miniak}, K.~G., {Konacki}, M., {Maehara}, H., {et~al.} 2019, \mnras,
  484, 451

\bibitem[{{Hinkel} {et~al.}(2014){Hinkel}, {Timmes}, {Young}, {Pagano}, \&
  {Turnbull}}]{2014AJ....148...54H}
{Hinkel}, N.~R., {Timmes}, F.~X., {Young}, P.~A., {Pagano}, M.~D., \&
  {Turnbull}, M.~C. 2014, \aj, 148, 54

\bibitem[{{Hinkel} {et~al.}(2016){Hinkel}, {Young}, {Pagano}, {Desch}, {Anbar},
  {Adibekyan}, {Blanco-Cuaresma}, {Carlberg}, {Delgado Mena}, {Liu},
  {Nordlander}, {Sousa}, {Korn}, {Gruyters}, {Heiter}, {Jofr{\'e}}, {Santos},
  \& {Soubiran}}]{2016ApJS..226....4H}
{Hinkel}, N.~R., {Young}, P.~A., {Pagano}, M.~D., {et~al.} 2016, \apjs, 226, 4

\bibitem[{{Ireland} {et~al.}(2008){Ireland}, {M{\'e}rand}, {ten Brummelaar},
  {Tuthill}, {Schaefer}, {Turner}, {Sturmann}, {Sturmann}, \&
  {McAlister}}]{ireland2008}
{Ireland}, M.~J., {M{\'e}rand}, A., {ten Brummelaar}, T.~A., {et~al.} 2008, in
  Society of Photo-Optical Instrumentation Engineers (SPIE) Conference Series,
  Vol. 7013, Optical and Infrared Interferometry, ed. M.~{Sch{\"o}ller}, W.~C.
  {Danchi}, \& F.~{Delplancke}, 701324

\bibitem[{{Jofr{\'e}} {et~al.}(2015){Jofr{\'e}}, {Petrucci}, {Saffe}, {Saker},
  {de la Villarmois}, {Chavero}, {G{\'o}mez}, \& {Mauas}}]{jofre2015}
{Jofr{\'e}}, E., {Petrucci}, R., {Saffe}, C., {et~al.} 2015, \aap, 574, A50

\bibitem[{{Jofr{\'e}} {et~al.}(2019){Jofr{\'e}}, {Heiter}, \&
  {Soubiran}}]{2019ARA&A..57..571J}
{Jofr{\'e}}, P., {Heiter}, U., \& {Soubiran}, C. 2019, \araa, 57, 571

\bibitem[{{Jofr{\'e}} {et~al.}(2014){Jofr{\'e}}, {Heiter}, {Soubiran},
  {Blanco-Cuaresma}, {Worley}, {Pancino}, {Cantat-Gaudin}, {Magrini},
  {Bergemann}, {Gonz{\'a}lez Hern{\'a}ndez}, {Hill}, {Lardo}, {de Laverny},
  {Lind}, {Masseron}, {Montes}, {Mucciarelli}, {Nordlander}, {Recio Blanco},
  {Sobeck}, {Sordo}, {Sousa}, {Tabernero}, {Vallenari}, \& {Van
  Eck}}]{jofre2014}
{Jofr{\'e}}, P., {Heiter}, U., {Soubiran}, C., {et~al.} 2014, \aap, 564, A133

\bibitem[{{Jofr{\'e}} {et~al.}(2018){Jofr{\'e}}, {Heiter}, {Tucci Maia},
  {Soubiran}, {Worley}, {Hawkins}, {Blanco-Cuaresma}, \&
  {Rodrigo}}]{2018RNAAS...2..152J}
{Jofr{\'e}}, P., {Heiter}, U., {Tucci Maia}, M., {et~al.} 2018, Research Notes
  of the American Astronomical Society, 2, 152

\bibitem[{{Jofr{\'e}} {et~al.}(2017){Jofr{\'e}}, {Heiter}, {Worley},
  {Blanco-Cuaresma}, {Soubiran}, {Masseron}, {Hawkins}, {Adibekyan}, {Buder},
  {Casamiquela}, {Gilmore}, {Hourihane}, \& {Tabernero}}]{2017A&A...601A..38J}
{Jofr{\'e}}, P., {Heiter}, U., {Worley}, C.~C., {et~al.} 2017, \aap, 601, A38

\bibitem[{{Karovicova} {et~al.}(2020){Karovicova}, {White}, {Nordlander},
  {Casagrande}, {Ireland}, {Huber}, \& {Jofr{\'e}}}]{karovicova2020}
{Karovicova}, I., {White}, T.~R., {Nordlander}, T., {et~al.} 2020, \aap, 640,
  A25

\bibitem[{{Karovicova} {et~al.}(2018){Karovicova}, {White}, {Nordlander},
  {Lind}, {Casagrande}, {Ireland}, {Huber}, {Creevey}, {Mourard}, {Schaefer},
  {Gilmore}, {Chiavassa}, {Wittkowski}, {Jofr{\'e}}, {Heiter}, {Th{\'e}venin},
  \& {Asplund}}]{karovicova2018}
{Karovicova}, I., {White}, T.~R., {Nordlander}, T., {et~al.} 2018, \mnras, 475,
  L81

\bibitem[{{Kassounian} {et~al.}(2019){Kassounian}, {Gebran}, {Paletou}, \&
  {Watson}}]{2019OAst...28...68K}
{Kassounian}, S., {Gebran}, M., {Paletou}, F., \& {Watson}, V. 2019, Open
  Astronomy, 28, 68

\bibitem[{{Kervella} {et~al.}(2019){Kervella}, {Arenou}, {Mignard}, \&
  {Th{\'e}venin}}]{krevella2019}
{Kervella}, P., {Arenou}, F., {Mignard}, F., \& {Th{\'e}venin}, F. 2019, \aap,
  623, A72

\bibitem[{{Kervella} {et~al.}(2004){Kervella}, {Th{\'e}venin}, {Di Folco}, \&
  {S{\'e}gransan}}]{kervella2004}
{Kervella}, P., {Th{\'e}venin}, F., {Di Folco}, E., \& {S{\'e}gransan}, D.
  2004, \aap, 426, 297

\bibitem[{{Kim} {et~al.}(2002){Kim}, {Demarque}, {Yi}, \&
  {Alexander}}]{kim2002}
{Kim}, Y.-C., {Demarque}, P., {Yi}, S.~K., \& {Alexander}, D.~R. 2002, \apjs,
  143, 499

\bibitem[{{Kollmeier} {et~al.}(2017){Kollmeier}, {Zasowski}, {Rix}, {Johns},
  {Anderson}, {Drory}, {Johnson}, {Pogge}, {Bird}, {Blanc}, {Brownstein},
  {Crane}, {De Lee}, {Klaene}, {Kreckel}, {MacDonald}, {Merloni}, {Ness},
  {O'Brien}, {Sanchez-Gallego}, {Sayres}, {Shen}, {Thakar}, {Tkachenko},
  {Aerts}, {Blanton}, {Eisenstein}, {Holtzman}, {Maoz}, {Nandra}, {Rockosi},
  {Weinberg}, {Bovy}, {Casey}, {Chaname}, {Clerc}, {Conroy}, {Eracleous},
  {G{\"a}nsicke}, {Hekker}, {Horne}, {Kauffmann}, {McQuinn}, {Pellegrini},
  {Schinnerer}, {Schlafly}, {Schwope}, {Seibert}, {Teske}, \& {van
  Saders}}]{2017arXiv171103234K}
{Kollmeier}, J.~A., {Zasowski}, G., {Rix}, H.-W., {et~al.} 2017, arXiv
  e-prints, arXiv:1711.03234

\bibitem[{{Koppelman} {et~al.}(2019){Koppelman}, {Helmi}, {Massari},
  {Price-Whelan}, \& {Starkenburg}}]{2019A&A...631L...9K}
{Koppelman}, H.~H., {Helmi}, A., {Massari}, D., {Price-Whelan}, A.~M., \&
  {Starkenburg}, T.~K. 2019, \aap, 631, L9

\bibitem[{{Kupka} {et~al.}(2011){Kupka}, {Dubernet}, \& {VAMDC
  Collaboration}}]{2011BaltA..20..503K}
{Kupka}, F., {Dubernet}, M.~L., \& {VAMDC Collaboration}. 2011, Baltic
  Astronomy, 20, 503

\bibitem[{{Lebzelter} {et~al.}(2012){Lebzelter}, {Heiter}, {Abia}, {Eriksson},
  {Ireland}, {Neilson}, {Nowotny}, {Maldonado}, {Merle}, {Peterson}, {Plez},
  {Short}, {Wahlgren}, {Worley}, {Aringer}, {Bladh}, {de Laverny}, {Goswami},
  {Mora}, {Norris}, {Recio-Blanco}, {Scholz}, {Th{\'e}venin}, {Tsuji},
  {Kordopatis}, {Montesinos}, \& {Wing}}]{2012A&A...547A.108L}
{Lebzelter}, T., {Heiter}, U., {Abia}, C., {et~al.} 2012, \aap, 547, A108

\bibitem[{{Leenaarts} {et~al.}(2012){Leenaarts}, {Carlsson}, \& {Rouppe van der
  Voort}}]{leenaarts2012}
{Leenaarts}, J., {Carlsson}, M., \& {Rouppe van der Voort}, L. 2012, \apj, 749,
  136

\bibitem[{{Leung} \& {Bovy}(2019)}]{2019MNRAS.483.3255L}
{Leung}, H.~W. \& {Bovy}, J. 2019, \mnras, 483, 3255

\bibitem[{{Limberg} {et~al.}(2021){Limberg}, {Rossi}, {Beers}, {Perottoni},
  {P{\'e}rez-Villegas}, {Santucci}, {Abuchaim}, {Placco}, {Lee}, {Christlieb},
  {Norris}, {Bessell}, {Ryan}, {Wilhelm}, {Rhee}, \&
  {Frebel}}]{2021ApJ...907...10L}
{Limberg}, G., {Rossi}, S., {Beers}, T.~C., {et~al.} 2021, \apj, 907, 10

\bibitem[{{Lindegren} {et~al.}(2020{\natexlab{a}}){Lindegren}, {Bastian},
  {Biermann}, {Bombrun}, {de Torres}, {Gerlach}, {Geyer}, {Hern{\'a}ndez},
  {Hilger}, {Hobbs}, {Klioner}, {Lammers}, {McMillan}, {Ramos-Lerate},
  {Steidelm{\"u}ller}, {Stephenson}, \& {van Leeuwen}}]{2020arXiv201201742L}
{Lindegren}, L., {Bastian}, U., {Biermann}, M., {et~al.} 2020{\natexlab{a}},
  arXiv e-prints, arXiv:2012.01742

\bibitem[{{Lindegren} {et~al.}(2020{\natexlab{b}}){Lindegren}, {Klioner},
  {Hern{\'a}ndez}, {Bombrun}, {Ramos-Lerate}, {Steidelm{\"u}ller}, {Bastian},
  {Biermann}, {de Torres}, {Gerlach}, {Geyer}, {Hilger}, {Hobbs}, {Lammers},
  {McMillan}, {Stephenson}, {Casta{\~n}eda}, {Davidson}, {Fabricius},
  {Gracia-Abril}, {Portell}, {Rowell}, {Teyssier}, {Torra}, {Bartolom{\'e}},
  {Clotet}, {Garralda}, {Gonz{\'a}lez-Vidal}, {Torra}, {Abbas}, {Altmann},
  {Anglada Varela}, {Balaguer-N{\'u}{\~n}ez}, {Balog}, {Barache}, {Becciani},
  {Bernet}, {Bertone}, {Bianchi}, {Bouquillon}, {Brown}, {Bucciarelli},
  {Busonero}, {Butkevich}, {Buzzi}, {Cancelliere}, {Carlucci}, {Charlot},
  {Cioni}, {Crosta}, {Crowley}, {del Peloso}, {del Pozo}, {Drimmel}, {Esquej},
  {Fienga}, {Fraile}, {Gai}, {Garcia-Reinaldos}, {Guerra}, {Hambly}, {Hauser},
  {Jan{\ss}en}, {Jordan}, {Kostrzewa-Rutkowska}, {Lattanzi}, {Liao}, {Licata},
  {Lister}, {L{\"o}ffler}, {Marchant}, {Masip}, {Mignard}, {Mints}, {Molina},
  {Mora}, {Morbidelli}, {Murphy}, {Pagani}, {Panuzzo}, {Pe{\~n}alosa Esteller},
  {Poggio}, {Re Fiorentin}, {Riva}, {Sagrist{\`a} Sell{\'e}s}, {Sanchez
  Gimenez}, {Sarasso}, {Sciacca}, {Siddiqui}, {Smart}, {Souami}, {Spagna},
  {Steele}, {Taris}, {Utrilla}, {van Reeven}, \&
  {Vecchiato}}]{2020arXiv201203380L}
{Lindegren}, L., {Klioner}, S.~A., {Hern{\'a}ndez}, J., {et~al.}
  2020{\natexlab{b}}, arXiv e-prints, arXiv:2012.03380

\bibitem[{{Luri} {et~al.}(2018){Luri}, {Brown}, {Sarro}, {Arenou},
  {Bailer-Jones}, {Castro-Ginard}, {de Bruijne}, {Prusti}, {Babusiaux}, \&
  {Delgado}}]{Luri2018}
{Luri}, X., {Brown}, A.~G.~A., {Sarro}, L.~M., {et~al.} 2018, \aap, 616, A9

\bibitem[{{Magrini} {et~al.}(2013){Magrini}, {Randich}, {Friel}, {Spina},
  {Jacobson}, {Cantat-Gaudin}, {Donati}, {Baglioni}, {Maiorca}, {Bragaglia},
  {Sordo}, \& {Vallenari}}]{2013A&A...558A..38M}
{Magrini}, L., {Randich}, S., {Friel}, E., {et~al.} 2013, \aap, 558, A38

\bibitem[{{Massari} {et~al.}(2019){Massari}, {Koppelman}, \&
  {Helmi}}]{Massari2019}
{Massari}, D., {Koppelman}, H.~H., \& {Helmi}, A. 2019, \aap, 630, L4

\bibitem[{{Mayor} {et~al.}(2003){Mayor}, {Pepe}, {Queloz}, {Bouchy},
  {Rupprecht}, {Lo Curto}, {Avila}, {Benz}, {Bertaux}, {Bonfils}, {Dall},
  {Dekker}, {Delabre}, {Eckert}, {Fleury}, {Gilliotte}, {Gojak}, {Guzman},
  {Kohler}, {Lizon}, {Longinotti}, {Lovis}, {Megevand}, {Pasquini}, {Reyes},
  {Sivan}, {Sosnowska}, {Soto}, {Udry}, {van Kesteren}, {Weber}, \&
  {Weilenmann}}]{Mayor2003}
{Mayor}, M., {Pepe}, F., {Queloz}, D., {et~al.} 2003, The Messenger, 114, 20

\bibitem[{{McMillan}(2017)}]{McMillan2017}
{McMillan}, P.~J. 2017, \mnras, 465, 76

\bibitem[{{Mel{\'e}ndez} \& {Barbuy}(2009)}]{melendez_barbuy2009}
{Mel{\'e}ndez}, J. \& {Barbuy}, B. 2009, \aap, 497, 611

\bibitem[{{Mel{\'e}ndez} \& {Ram{\'\i}rez}(2004)}]{melendez2004}
{Mel{\'e}ndez}, J. \& {Ram{\'\i}rez}, I. 2004, \apjl, 615, L33

\bibitem[{{Mel{\'e}ndez} {et~al.}(2006){Mel{\'e}ndez}, {Shchukina},
  {Vasiljeva}, \& {Ram{\'\i}rez}}]{melendez2006}
{Mel{\'e}ndez}, J., {Shchukina}, N.~G., {Vasiljeva}, I.~E., \& {Ram{\'\i}rez},
  I. 2006, \apj, 642, 1082

\bibitem[{{Miller} {et~al.}(2020){Miller}, {Maxted}, \&
  {Smalley}}]{2020MNRAS.497.2899M}
{Miller}, N.~J., {Maxted}, P.~F.~L., \& {Smalley}, B. 2020, \mnras, 497, 2899

\bibitem[{{Mishenina} \& {Kovtyukh}(2001)}]{mishenina2001}
{Mishenina}, T.~V. \& {Kovtyukh}, V.~V. 2001, \aap, 370, 951

\bibitem[{{Myeong} {et~al.}(2019){Myeong}, {Vasiliev}, {Iorio}, {Evans}, \&
  {Belokurov}}]{Myeong2019}
{Myeong}, G.~C., {Vasiliev}, E., {Iorio}, G., {Evans}, N.~W., \& {Belokurov},
  V. 2019, \mnras, 488, 1235

\bibitem[{{Ness} {et~al.}(2015){Ness}, {Hogg}, {Rix}, {Ho}, \&
  {Zasowski}}]{2015ApJ...808...16N}
{Ness}, M., {Hogg}, D.~W., {Rix}, H.~W., {Ho}, A. Y.~Q., \& {Zasowski}, G.
  2015, \apj, 808, 16

\bibitem[{{Nissen} \& {Schuster}(2010)}]{2010A&A...511L..10N}
{Nissen}, P.~E. \& {Schuster}, W.~J. 2010, \aap, 511, L10

\bibitem[{{Nordlander} {et~al.}(2019){Nordlander}, {Bessell}, {Da Costa},
  {Mackey}, {Asplund}, {Casey}, {Chiti}, {Ezzeddine}, {Frebel}, {Lind},
  {Marino}, {Murphy}, {Norris}, {Schmidt}, \& {Yong}}]{nordlander2019}
{Nordlander}, T., {Bessell}, M.~S., {Da Costa}, G.~S., {et~al.} 2019, \mnras,
  488, L109

\bibitem[{{Perryman} {et~al.}(1997){Perryman}, {Lindegren}, {Kovalevsky},
  {Hoeg}, {Bastian}, {Bernacca}, {Cr{\'e}z{\'e}}, {Donati}, {Grenon},
  {Grewing}, {van Leeuwen}, {van der Marel}, {Mignard}, {Murray}, {Le Poole},
  {Schrijver}, {Turon}, {Arenou}, {Froeschl{\'e}}, \& {Petersen}}]{Perryman}
{Perryman}, M.~A.~C., {Lindegren}, L., {Kovalevsky}, J., {et~al.} 1997, \aap,
  323, L49

\bibitem[{{Pinsonneault} {et~al.}(2018){Pinsonneault}, {Elsworth}, {Tayar},
  {Serenelli}, {Stello}, {Zinn}, {Mathur}, {Garc{\'\i}a}, {Johnson}, {Hekker},
  {Huber}, {Kallinger}, {M{\'e}sz{\'a}ros}, {Mosser}, {Stassun}, {Girardi},
  {Rodrigues}, {Silva Aguirre}, {An}, {Basu}, {Chaplin}, {Corsaro}, {Cunha},
  {Garc{\'\i}a-Hern{\'a}ndez}, {Holtzman}, {J{\"o}nsson}, {Shetrone}, {Smith},
  {Sobeck}, {Stringfellow}, {Zamora}, {Beers}, {Fern{\'a}ndez-Trincado},
  {Frinchaboy}, {Hearty}, \& {Nitschelm}}]{2018ApJS..239...32P}
{Pinsonneault}, M.~H., {Elsworth}, Y.~P., {Tayar}, J., {et~al.} 2018, \apjs,
  239, 32

\bibitem[{{Piskunov} \& {Valenti}(2017)}]{2017A&A...597A..16P}
{Piskunov}, N. \& {Valenti}, J.~A. 2017, \aap, 597, A16

\bibitem[{{Plez}(2012)}]{turbospectrum}
{Plez}, B. 2012, {Turbospectrum: Code for spectral synthesis}

\bibitem[{{Ram{\'\i}rez} \& {Mel{\'e}ndez}(2005)}]{ram2005}
{Ram{\'\i}rez}, I. \& {Mel{\'e}ndez}, J. 2005, \apj, 626, 446

\bibitem[{{Ram{\'{\i}}rez} {et~al.}(2014){Ram{\'{\i}}rez}, {Mel{\'e}ndez}, \&
  {Asplund}}]{ram_2014}
{Ram{\'{\i}}rez}, I., {Mel{\'e}ndez}, J., \& {Asplund}, M. 2014, \aap, 561, A7

\bibitem[{{Randich} {et~al.}(2013){Randich}, {Gilmore}, \& {Gaia-ESO
  Consortium}}]{2013Msngr.154...47R}
{Randich}, S., {Gilmore}, G., \& {Gaia-ESO Consortium}. 2013, Msngr, 154, 47

\bibitem[{{Ratajczak} {et~al.}(2021){Ratajczak}, {Paw{\l}aszek},
  {He{\l}miniak}, {Konacki}, {Sybilski}, {Koz{\l}owski}, {Litwicki}, {Smith},
  {Miko{\l}ajczyk}, {Anderson}, \& {Hellier}}]{2021MNRAS.500.4972R}
{Ratajczak}, M., {Paw{\l}aszek}, R.~K., {He{\l}miniak}, K.~G., {et~al.} 2021,
  \mnras, 500, 4972

\bibitem[{{Reddy} \& {Lambert}(2008)}]{2008MNRAS.391...95R}
{Reddy}, B.~E. \& {Lambert}, D.~L. 2008, \mnras, 391, 95

\bibitem[{{Reggiani} {et~al.}(2017){Reggiani}, {Mel{\'e}ndez}, {Kobayashi},
  {Karakas}, \& {Placco}}]{reggiani2017}
{Reggiani}, H., {Mel{\'e}ndez}, J., {Kobayashi}, C., {Karakas}, A., \&
  {Placco}, V. 2017, \aap, 608, A46

\bibitem[{{Reggiani} {et~al.}(2016){Reggiani}, {Mel{\'e}ndez}, {Yong},
  {Ram{\'\i}rez}, \& {Asplund}}]{reggiani2016}
{Reggiani}, H., {Mel{\'e}ndez}, J., {Yong}, D., {Ram{\'\i}rez}, I., \&
  {Asplund}, M. 2016, \aap, 586, A67

\bibitem[{{Ryan} {et~al.}(1996){Ryan}, {Norris}, \&
  {Beers}}]{1996ApJ...471..254R}
{Ryan}, S.~G., {Norris}, J.~E., \& {Beers}, T.~C. 1996, \apj, 471, 254

\bibitem[{{Sbordone} {et~al.}(2014){Sbordone}, {Caffau}, {Bonifacio}, \&
  {Duffau}}]{2014A&A...564A.109S}
{Sbordone}, L., {Caffau}, E., {Bonifacio}, P., \& {Duffau}, S. 2014, \aap, 564,
  A109

\bibitem[{{Schlafly} \& {Finkbeiner}(2011)}]{Schlafly2011}
{Schlafly}, E.~F. \& {Finkbeiner}, D.~P. 2011, \apj, 737, 103

\bibitem[{{Serenelli} {et~al.}(2017){Serenelli}, {Johnson}, {Huber},
  {Pinsonneault}, {Ball}, {Tayar}, {Silva Aguirre}, {Basu}, {Troup}, {Hekker},
  {Kallinger}, {Stello}, {Davies}, {Lund}, {Mathur}, {Mosser}, {Stassun},
  {Chaplin}, {Elsworth}, {Garc{\'\i}a}, {Handberg}, {Holtzman}, {Hearty},
  {Garc{\'\i}a-Hern{\'a}ndez}, {Gaulme}, \& {Zamora}}]{2017ApJS..233...23S}
{Serenelli}, A., {Johnson}, J., {Huber}, D., {et~al.} 2017, \apjs, 233, 23

\bibitem[{{Serenelli} {et~al.}(2020){Serenelli}, {Weiss}, {Aerts}, {Angelou},
  {Baroch}, {Bastian}, {Bergemann}, {Bestenlehner}, {Czekala}, {Elias-Rosa},
  {Escorza}, {Van Eylen}, {Feuillet}, {Gandolfi}, {Gieles}, {Girardi},
  {Lodieu}, {Martig}, {Miller Bertolami}, {Mombarg}, {Morales}, {Moya},
  {Nsamba}, {Pavlovski}, {Pedersen}, {Ribas}, {Schneider}, {Silva Aguirre},
  {Stassun}, {Tolstoy}, {Tremblay}, \& {Zwintz}}]{2020arXiv200610868S}
{Serenelli}, A., {Weiss}, A., {Aerts}, C., {et~al.} 2020, arXiv e-prints,
  arXiv:2006.10868

\bibitem[{{Smalley}(2005)}]{2005MSAIS...8..130S}
{Smalley}, B. 2005, Memorie della Societa Astronomica Italiana Supplementi, 8,
  130

\bibitem[{{Smiljanic} {et~al.}(2014){Smiljanic}, {Korn}, {Bergemann}, {Frasca},
  {Magrini}, {Masseron}, {Pancino}, {Ruchti}, {San Roman}, {Sbordone}, {Sousa},
  {Tabernero}, {Tautvai{\v{s}}ien{\.{e}}}, {Valentini}, {Weber}, {Worley},
  {Adibekyan}, {Allende Prieto}, {Barisevi{\v{c}}ius}, {Biazzo},
  {Blanco-Cuaresma}, {Bonifacio}, {Bragaglia}, {Caffau}, {Cantat-Gaudin},
  {Chorniy}, {de Laverny}, {Delgado-Mena}, {Donati}, {Duffau}, {Franciosini},
  {Friel}, {Geisler}, {Gonz{\'a}lez Hern{\'a}ndez}, {Gruyters}, {Guiglion},
  {Hansen}, {Heiter}, {Hill}, {Jacobson}, {Jofre}, {J{\"o}nsson}, {Lanzafame},
  {Lardo}, {Ludwig}, {Maiorca}, {Mikolaitis}, {Montes}, {Morel}, {Mucciarelli},
  {Mu{\~n}oz}, {Nordlander}, {Pasquini}, {Puzeras}, {Recio-Blanco}, {Ryde},
  {Sacco}, {Santos}, {Serenelli}, {Sordo}, {Soubiran}, {Spina}, {Steffen},
  {Vallenari}, {Van Eck}, {Villanova}, {Gilmore}, {Randich}, {Asplund},
  {Binney}, {Drew}, {Feltzing}, {Ferguson}, {Jeffries}, {Micela}, {Negueruela},
  {Prusti}, {Rix}, {Alfaro}, {Babusiaux}, {Bensby}, {Blomme}, {Flaccomio},
  {Fran{\c{c}}ois}, {Irwin}, {Koposov}, {Walton}, {Bayo}, {Carraro}, {Costado},
  {Damiani}, {Edvardsson}, {Hourihane}, {Jackson}, {Lewis}, {Lind}, {Marconi},
  {Martayan}, {Monaco}, {Morbidelli}, {Prisinzano}, \&
  {Zaggia}}]{2014A&A...570A.122S}
{Smiljanic}, R., {Korn}, A.~J., {Bergemann}, M., {et~al.} 2014, \aap, 570, A122

\bibitem[{{Smith} {et~al.}(1998){Smith}, {Lambert}, \& {Nissen}}]{smith1998}
{Smith}, V.~V., {Lambert}, D.~L., \& {Nissen}, P.~E. 1998, \apj, 506, 405

\bibitem[{{Sousa}(2014)}]{2014dapb.book..297S}
{Sousa}, S.~G. 2014, {ARES + MOOG: A Practical Overview of an Equivalent Width
  (EW) Method to Derive Stellar Parameters} (Springer International
  Publishing), 297--310

\bibitem[{{Tabernero} {et~al.}(2019){Tabernero}, {Marfil}, {Montes}, \&
  {Gonz{\'a}lez Hern{\'a}ndez}}]{2019A&A...628A.131T}
{Tabernero}, H.~M., {Marfil}, E., {Montes}, D., \& {Gonz{\'a}lez
  Hern{\'a}ndez}, J.~I. 2019, \aap, 628, A131

\bibitem[{{Taylor}(2005)}]{TOPCAT}
{Taylor}, M.~B. 2005, in Astronomical Society of the Pacific Conference Series,
  Vol. 347, Astronomical Data Analysis Software and Systems XIV, ed.
  P.~{Shopbell}, M.~{Britton}, \& R.~{Ebert}, 29

\bibitem[{{ten Brummelaar} {et~al.}(2005){ten Brummelaar}, {McAlister},
  {Ridgway}, {Bagnuolo}, {Turner}, {Sturmann}, {Sturmann}, {Berger}, {Ogden},
  {Cadman}, {Hartkopf}, {Hopper}, \& {Shure}}]{tenBrummelaar2005}
{ten Brummelaar}, T.~A., {McAlister}, H.~A., {Ridgway}, S.~T., {et~al.} 2005,
  \apj, 628, 453

\bibitem[{{The MSE Science Team} {et~al.}(2019){The MSE Science Team},
  {Babusiaux}, {Bergemann}, {Burgasser}, {Ellison}, {Haggard}, {Huber},
  {Kaplinghat}, {Li}, {Marshall}, {Martell}, {McConnachie}, {Percival},
  {Robotham}, {Shen}, {Thirupathi}, {Tran}, {Yeche}, {Yong}, {Adibekyan},
  {Silva Aguirre}, {Angelou}, {Asplund}, {Balogh}, {Banerjee}, {Bannister},
  {Barr{\'\i}a}, {Battaglia}, {Bayo}, {Bechtol}, {Beck}, {Beers}, {Bellinger},
  {Berg}, {Bestenlehner}, {Bilicki}, {Bitsch}, {Bland-Hawthorn}, {Bolton},
  {Boselli}, {Bovy}, {Bragaglia}, {Buzasi}, {Caffau}, {Cami}, {Carleton},
  {Casagrande}, {Cassisi}, {Catelan}, {Chang}, {Cortese}, {Damjanov}, {Davies},
  {de Grijs}, {de Rosa}, {Deason}, {di Matteo}, {Drlica-Wagner}, {Erkal},
  {Escorza}, {Ferrarese}, {Fleming}, {Font-Ribera}, {Freeman}, {G{\"a}nsicke},
  {Gabdeev}, {Gallagher}, {Gandolfi}, {Garc{\'\i}a}, {Gaulme}, {Geha},
  {Gennaro}, {Gieles}, {Gilbert}, {Gordon}, {Goswami}, {Greco}, {Grillmair},
  {Guiglion}, {H{\'e}nault-Brunet}, {Hall}, {Handler}, {Hansen}, {Hathi},
  {Hatzidimitriou}, {Haywood}, {Hern{\'a}ndez Santisteban}, {Hillenbrand},
  {Hopkins}, {Howlett}, {Hudson}, {Ibata}, {Ili{\'c}}, {Jablonka}, {Ji},
  {Jiang}, {Juneau}, {Karakas}, {Karinkuzhi}, {Kim}, {Kong}, {Konstantopoulos},
  {Krogager}, {Lagos}, {Lallement}, {Laporte}, {Lebreton}, {Lee}, {Lewis},
  {Lianou}, {Liu}, {Lodieu}, {Loveday}, {M{\'e}sz{\'a}ros}, {Makler}, {Mao},
  {Marchesini}, {Martin}, {Mateo}, {Melis}, {Merle}, {Miglio}, {Gohar
  Mohammad}, {Molaverdikhani}, {Monier}, {Morel}, {Mosser}, {Nataf}, {Necib},
  {Neilson}, {Newman}, {Nierenberg}, {Nord}, {Noterdaeme}, {O'Dea}, {Oshagh},
  {Pace}, {Palanque-Delabrouille}, {Pandey}, {Parker}, {Pawlowski}, {Peter},
  {Petitjean}, {Petric}, {Placco}, {Popovi{\'c}}, {Price-Whelan}, {Prsa},
  {Ravindranath}, {Rich}, {Ruan}, {Rybizki}, {Sakari}, {Sanderson}, {Schiavon},
  {Schimd}, {Serenelli}, {Siebert}, {Siudek}, {Smiljanic}, {Smith}, {Sobeck},
  {Starkenburg}, {Stello}, {Szab{\'o}}, {Szabo}, {Taylor}, {Thanjavur},
  {Thomas}, {Tollerud}, {Toonen}, {Tremblay}, {Tresse}, {Tsantaki},
  {Valentini}, {Van Eck}, {Variu}, {Venn}, {Villaver}, {Walker}, {Wang},
  {Wang}, {Wilson}, {Wright}, {Xu}, {Yildiz}, {Zhang}, {Zwintz}, {Anguiano},
  {Bedell}, {Chaplin}, {Collet}, {Cuillandre}, {Duc}, {Flagey}, {Hermes},
  {Hill}, {Kamath}, {Laychak}, {Ma{\l}ek}, {Marley}, {Sheinis}, {Simons},
  {Sousa}, {Szeto}, {Ting}, {Vegetti}, {Wells}, {Babas}, {Bauman}, {Bosselli},
  {C{\^o}t{\'e}}, {Colless}, {Comparat}, {Courtois}, {Crampton}, {Croom},
  {Davies}, {de Grijs}, {Denny}, {Devost}, {di Matteo}, {Driver},
  {Fernandez-Lorenzo}, {Guhathakurta}, {Han}, {Higgs}, {Hill}, {Ho}, {Hopkins},
  {Hudson}, {Ibata}, {Isani}, {Jarvis}, {Johnson}, {Jullo}, {Kaiser}, {Kneib},
  {Koda}, {Koshy}, {Mignot}, {Murowinski}, {Newman}, {Nusser}, {Pancoast},
  {Peng}, {Peroux}, {Pichon}, {Poggianti}, {Richard}, {Salmon}, {Seibert},
  {Shastri}, {Smith}, {Sutaria}, {Tao}, {Taylor}, {Tully}, {van Waerbeke},
  {Vermeulen}, {Walker}, {Willis}, {Willot}, \&
  {Withington}}]{2019arXiv190404907T}
{The MSE Science Team}, {Babusiaux}, C., {Bergemann}, M., {et~al.} 2019, arXiv
  e-prints, arXiv:1904.04907

\bibitem[{{Ting} {et~al.}(2019){Ting}, {Conroy}, {Rix}, \&
  {Cargile}}]{2019ApJ...879...69T}
{Ting}, Y.-S., {Conroy}, C., {Rix}, H.-W., \& {Cargile}, P. 2019, \apj, 879, 69

\bibitem[{{Torres} {et~al.}(2010){Torres}, {Andersen}, \&
  {Gim{\'e}nez}}]{2010A&ARv..18...67T}
{Torres}, G., {Andersen}, J., \& {Gim{\'e}nez}, A. 2010, \aapr, 18, 67

\bibitem[{{{\v S}koda} {et~al.}(2008){{\v S}koda}, {{\v S}urlan}, \&
  {Tomi{\'c}}}]{skoda2008}
{{\v S}koda}, P., {{\v S}urlan}, B., \& {Tomi{\'c}}, S. 2008, in \procspie,
  Vol. 7014, Ground-based and Airborne Instrumentation for Astronomy II, 70145X

\bibitem[{{Valentini} {et~al.}(2019){Valentini}, {Chiappini}, {Bossini},
  {Miglio}, {Davies}, {Mosser}, {Elsworth}, {Mathur}, {Garc{\'\i}a}, {Girardi},
  {Rodrigues}, {Steinmetz}, \& {Vallenari}}]{2019A&A...627A.173V}
{Valentini}, M., {Chiappini}, C., {Bossini}, D., {et~al.} 2019, \aap, 627, A173

\bibitem[{{Valentini} {et~al.}(2017){Valentini}, {Chiappini}, {Davies},
  {Elsworth}, {Mosser}, {Lund}, {Miglio}, {Chaplin}, {Rodrigues}, {Boeche},
  {Steinmetz}, {Matijevi{\v{c}}}, {Kordopatis}, {Bland-Hawthorn}, {Munari},
  {Bienaym{\'e}}, {Freeman}, {Gibson}, {Gilmore}, {Grebel}, {Helmi}, {Kunder},
  {McMillan}, {Navarro}, {Parker}, {Reid}, {Seabroke}, {Sharma}, {Siviero},
  {Watson}, {Wyse}, {Zwitter}, \& {Mott}}]{2017A&A...600A..66V}
{Valentini}, M., {Chiappini}, C., {Davies}, G.~R., {et~al.} 2017, \aap, 600,
  A66

\bibitem[{{Valentini} {et~al.}(2016){Valentini}, {Chiappini}, {Miglio},
  {Montalb{\'a}n}, {Rodrigues}, {Mosser}, {Anders}, {the CoRoT RG Group}, \&
  {GES Consortium}}]{2016AN....337..970V}
{Valentini}, M., {Chiappini}, C., {Miglio}, A., {et~al.} 2016, Astronomische
  Nachrichten, 337, 970

\bibitem[{{van Leeuwen}(2007)}]{2007A&A...474..653V}
{van Leeuwen}, F. 2007, \aap, 474, 653

\bibitem[{{Wallace} {et~al.}(2011){Wallace}, {Hinkle}, {Livingston}, \&
  {Davis}}]{wall2011}
{Wallace}, L., {Hinkle}, K.~H., {Livingston}, W.~C., \& {Davis}, S.~P. 2011,
  \apjs, 195, 6

\bibitem[{{Wang} {et~al.}(2020){Wang}, {Luo}, {Chen}, {Hou}, {Zhang}, {Zhao},
  {Li}, {Hou}, \& {LAMOST MRS Collaboration}}]{2020ApJ...891...23W}
{Wang}, R., {Luo}, A.~L., {Chen}, J.-J., {et~al.} 2020, \apj, 891, 23

\bibitem[{{Wheeler} {et~al.}(2020){Wheeler}, {Ness}, {Buder}, {Bland-Hawthorn},
  {Silva}, {Hayden}, {Kos}, {Lewis}, {Martell}, {Sharma}, {Simpson}, {Zucker},
  \& {Zwitter}}]{2020ApJ...898...58W}
{Wheeler}, A., {Ness}, M., {Buder}, S., {et~al.} 2020, \apj, 898, 58

\bibitem[{{White} {et~al.}(2013){White}, {Huber}, {Maestro}, {Bedding},
  {Ireland}, {Baron}, {Boyajian}, {Che}, {Monnier}, {Pope}, {Roettenbacher},
  {Stello}, {Tuthill}, {Farrington}, {Goldfinger}, {McAlister}, {Schaefer},
  {Sturmann}, {Sturmann}, {ten Brummelaar}, \& {Turner}}]{white2013}
{White}, T.~R., {Huber}, D., {Maestro}, V., {et~al.} 2013, \mnras, 433, 1262

\bibitem[{{Worley} {et~al.}(2016){Worley}, {de Laverny}, {Recio-Blanco},
  {Hill}, \& {Bijaoui}}]{2016A&A...591A..81W}
{Worley}, C.~C., {de Laverny}, P., {Recio-Blanco}, A., {Hill}, V., \&
  {Bijaoui}, A. 2016, \aap, 591, A81

\bibitem[{{Worley} {et~al.}(2020){Worley}, {Jofr{\'e}}, {Rendle}, {Miglio},
  {Magrini}, {Feuillet}, {Gavel}, {Smiljanic}, {Lind}, {Korn}, {Gilmore},
  {Randich}, {Hourihane}, {Gonneau}, {Francois}, {Lewis}, {Sacco}, {Bragaglia},
  {Heiter}, {Feltzing}, {Bensby}, {Irwin}, {Gonzalez Solares}, {Murphy},
  {Bayo}, {Sbordone}, {Zwitter}, {Lanzafame}, {Walton}, {Zaggia}, {Alfaro},
  {Morbidelli}, {Sousa}, {Monaco}, {Carraro}, \& {Lardo}}]{2020A&A...643A..83W}
{Worley}, C.~C., {Jofr{\'e}}, P., {Rendle}, B., {et~al.} 2020, \aap, 643, A83

\bibitem[{{Yi} {et~al.}(2003){Yi}, {Kim}, \& {Demarque}}]{yi2003}
{Yi}, S.~K., {Kim}, Y.-C., \& {Demarque}, P. 2003, \apjs, 144, 259

\bibitem[{{Yong} {et~al.}(2013){Yong}, {Mel{\'e}ndez}, {Grundahl}, {Roederer},
  {Norris}, {Milone}, {Marino}, {Coelho}, {McArthur}, {Lind}, {Collet}, \&
  {Asplund}}]{yong2013}
{Yong}, D., {Mel{\'e}ndez}, J., {Grundahl}, F., {et~al.} 2013, \mnras, 434,
  3542

\end{thebibliography}

\begin{appendix} 
\onecolumn
\section{Profile fits of the Titans}
\label{app:titans}
\begin{figure}[!ht]
    \centering
    \includegraphics[width=0.97\linewidth]{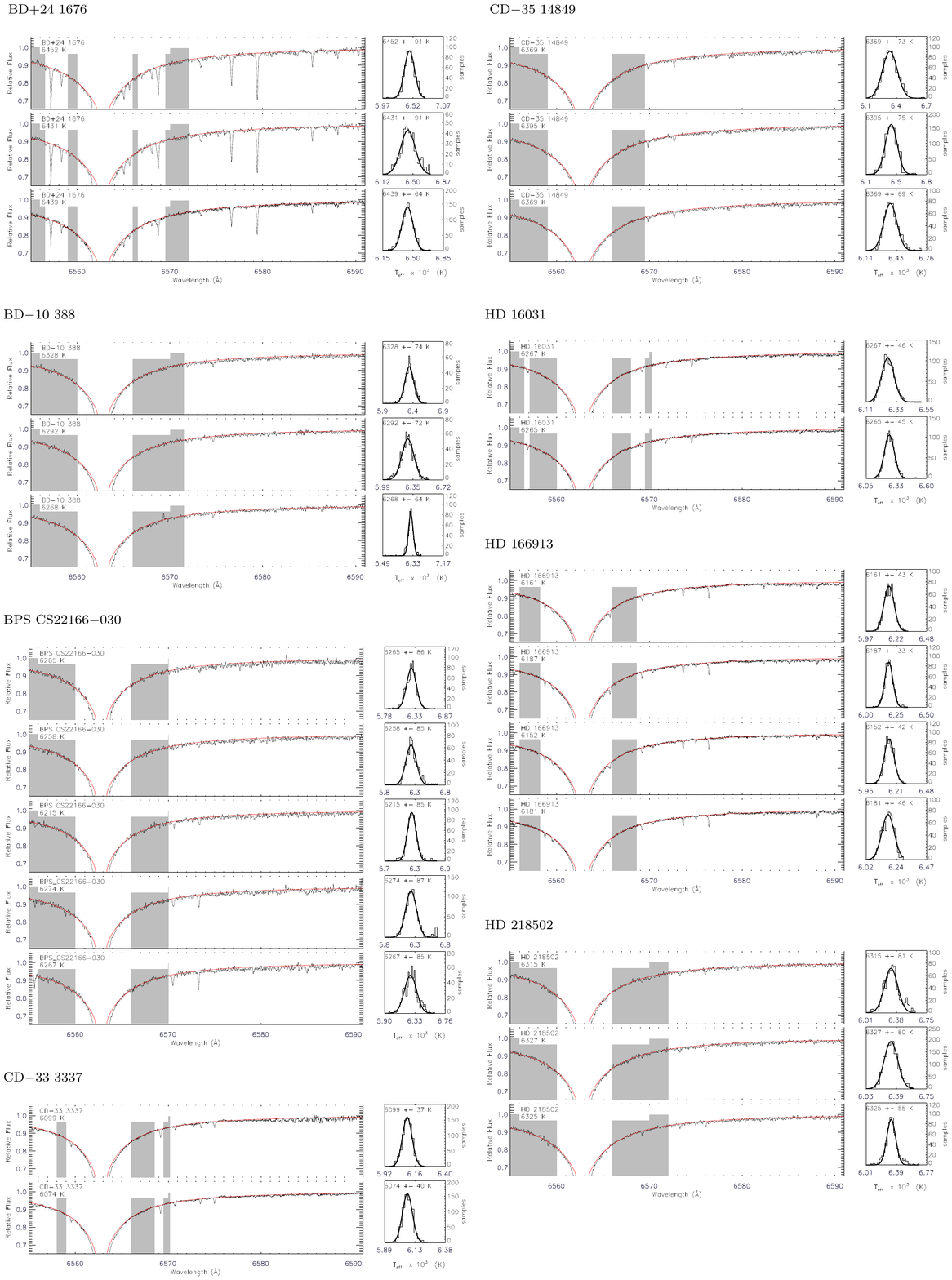}
\end{figure}
\begin{figure}[!ht]
    \centering
    \includegraphics[width=0.94\linewidth]{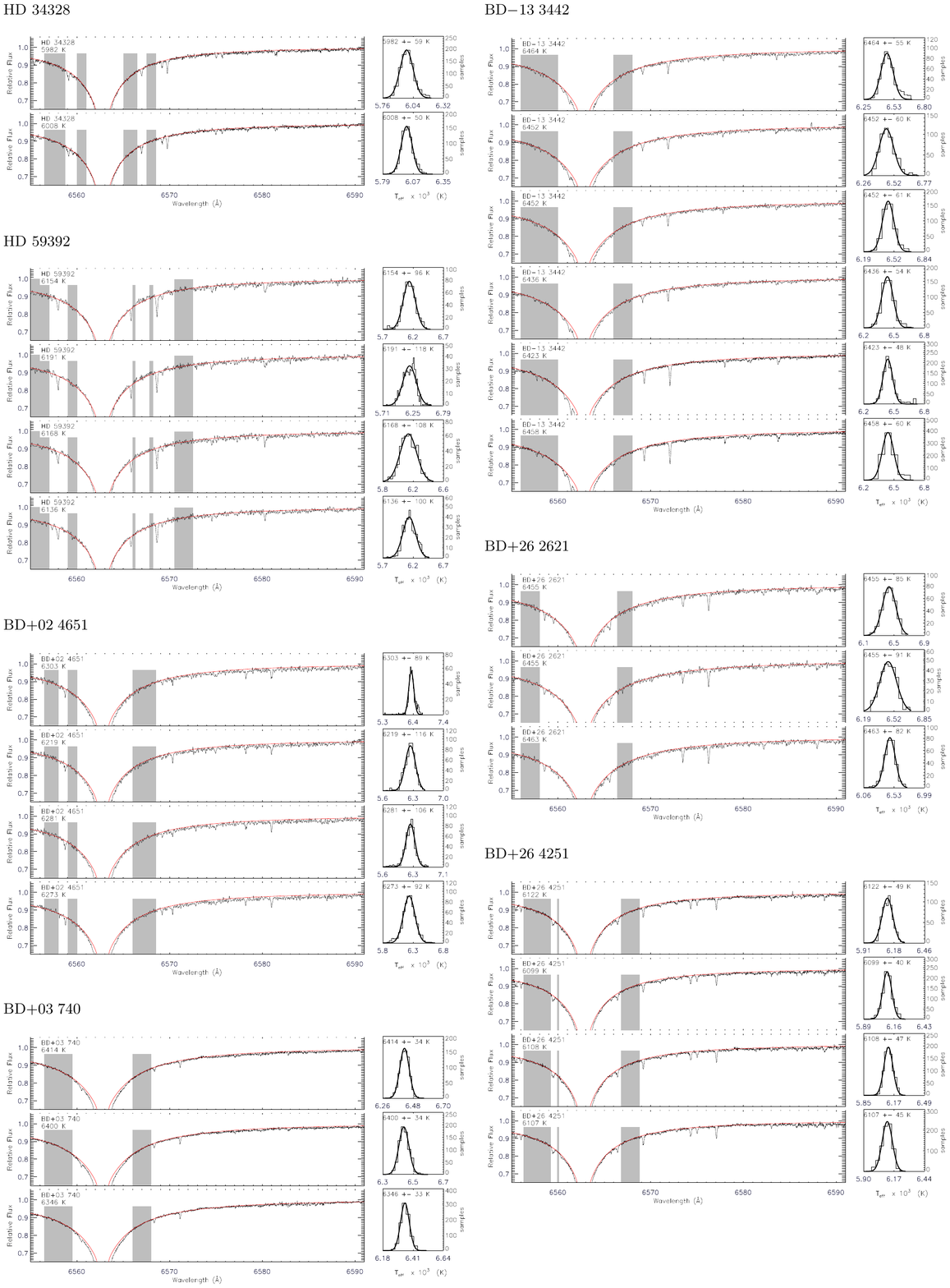}
\end{figure}
\begin{figure}[!ht]
    \centering
    \includegraphics[width=0.94\linewidth]{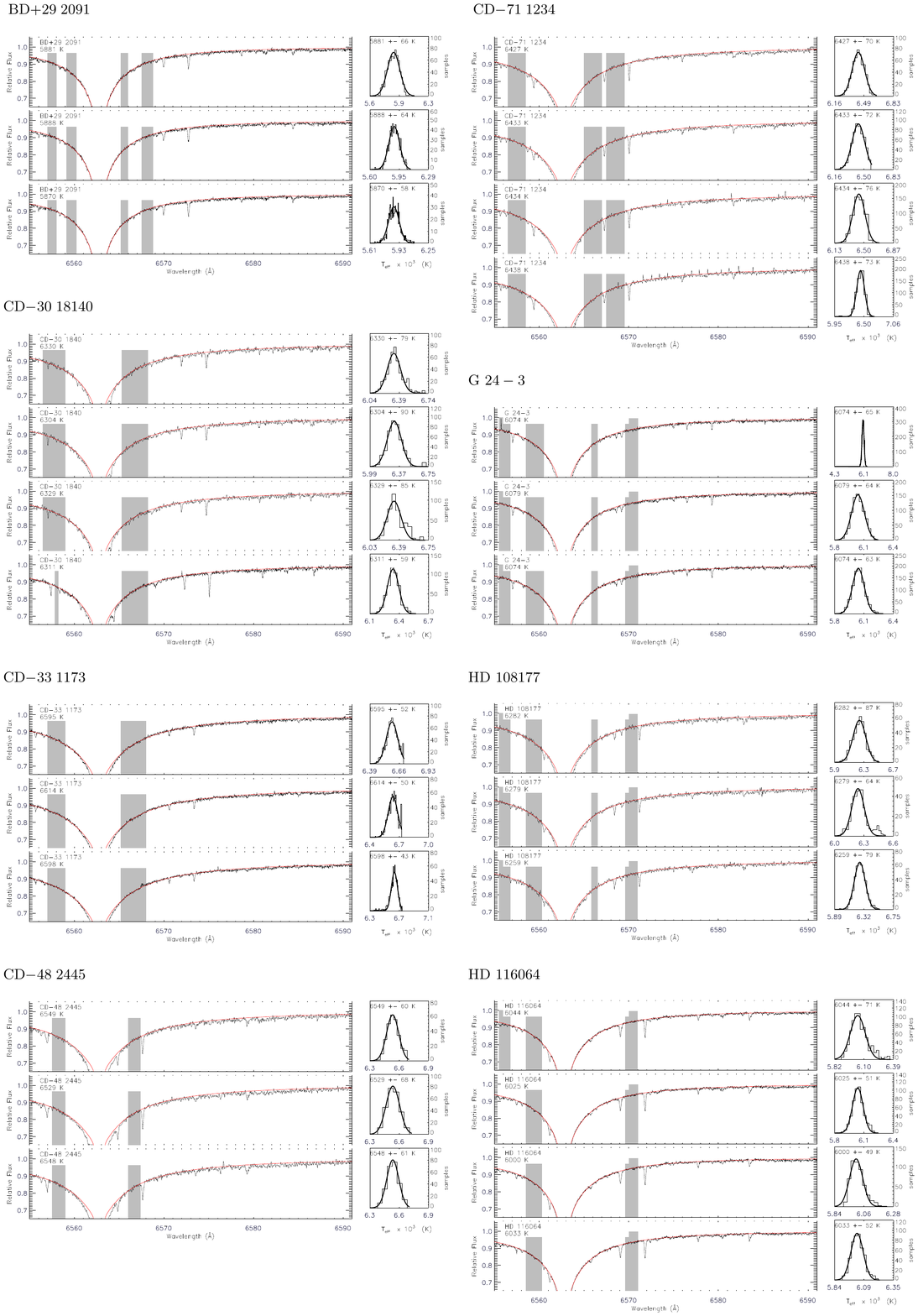}
\end{figure}
\begin{figure}[!ht]
    \centering
    \includegraphics[width=0.94\linewidth]{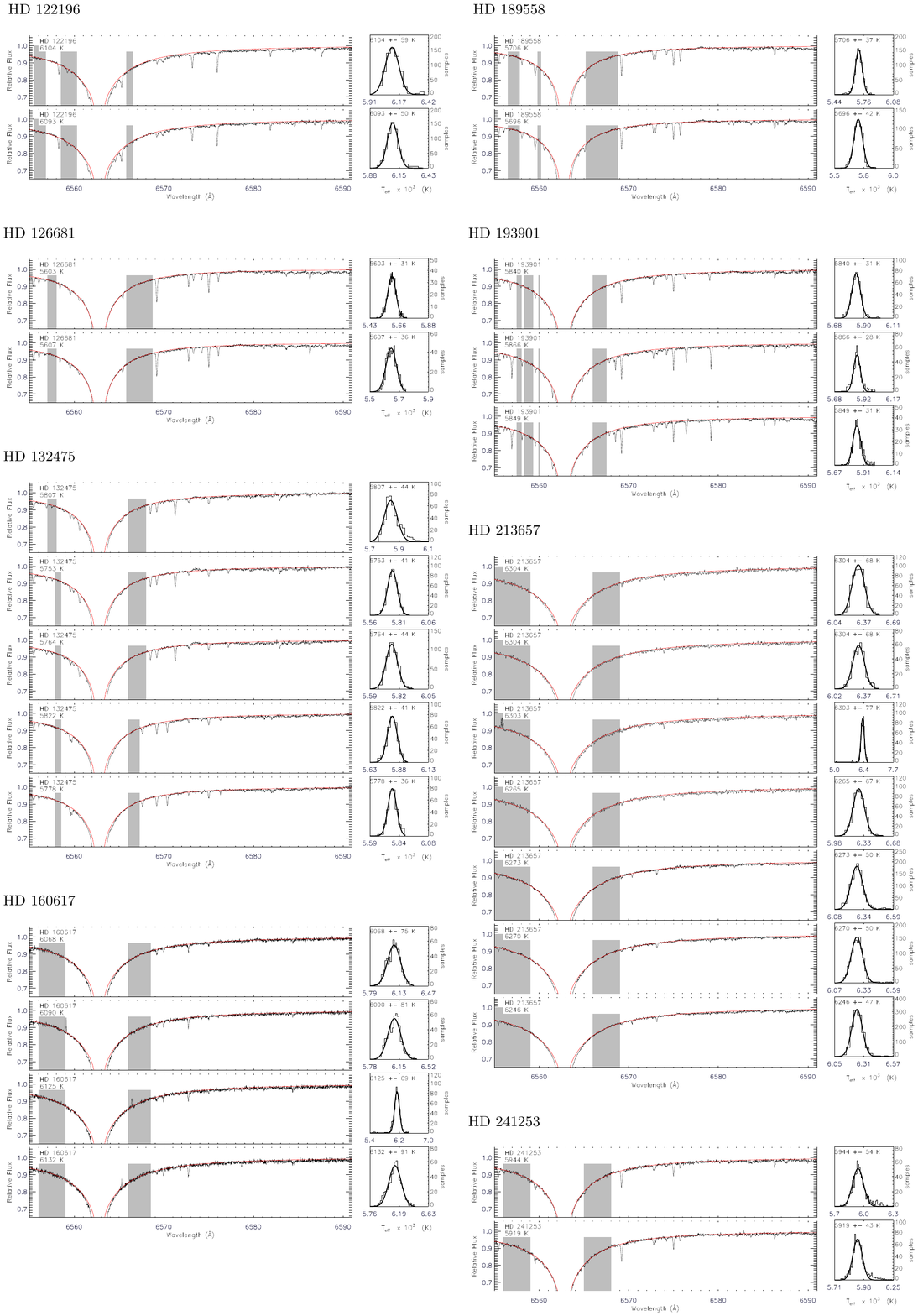}
\end{figure}
\begin{figure}[!ht]
    \centering
    \includegraphics[width=0.94\linewidth]{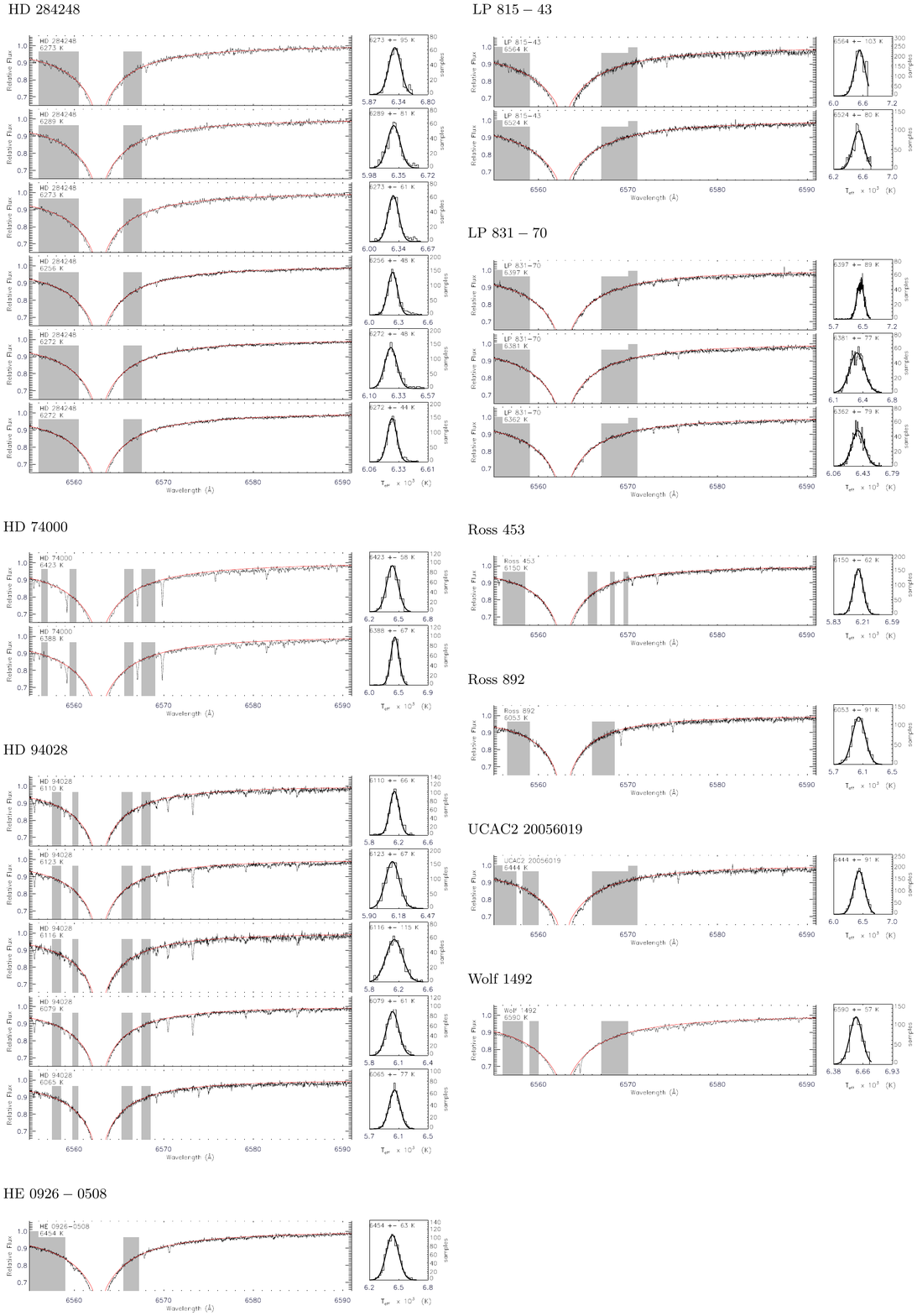}
\end{figure}

\newpage
\onecolumn
\section{Profile fits of the Gaia benchmark stars}
\label{app:gbs}
\begin{figure}[ht]
    \centering
    \includegraphics[width=0.97\linewidth]{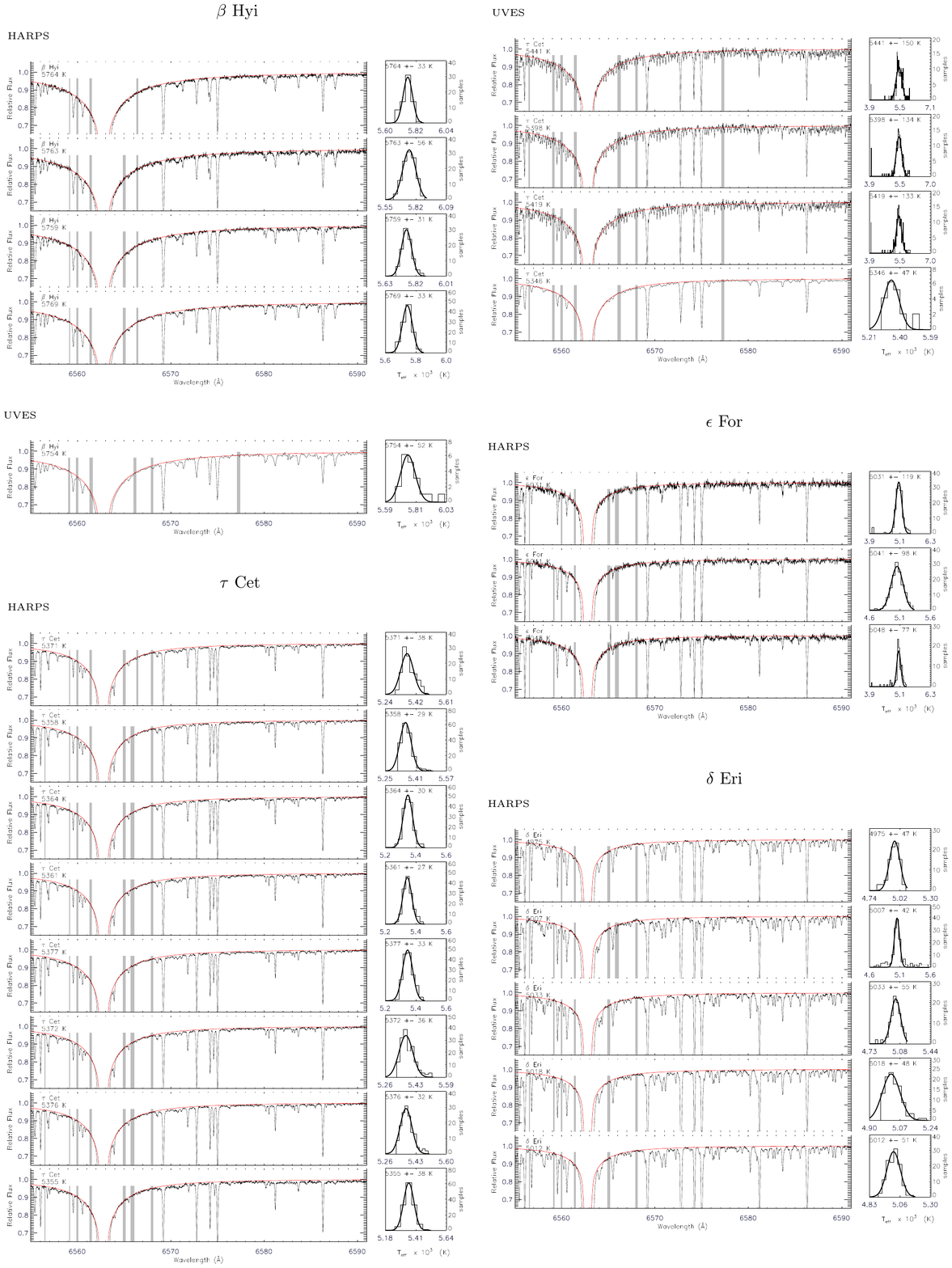}
\end{figure}
\begin{figure}[!ht]
    \centering
    \includegraphics[width=0.94\linewidth]{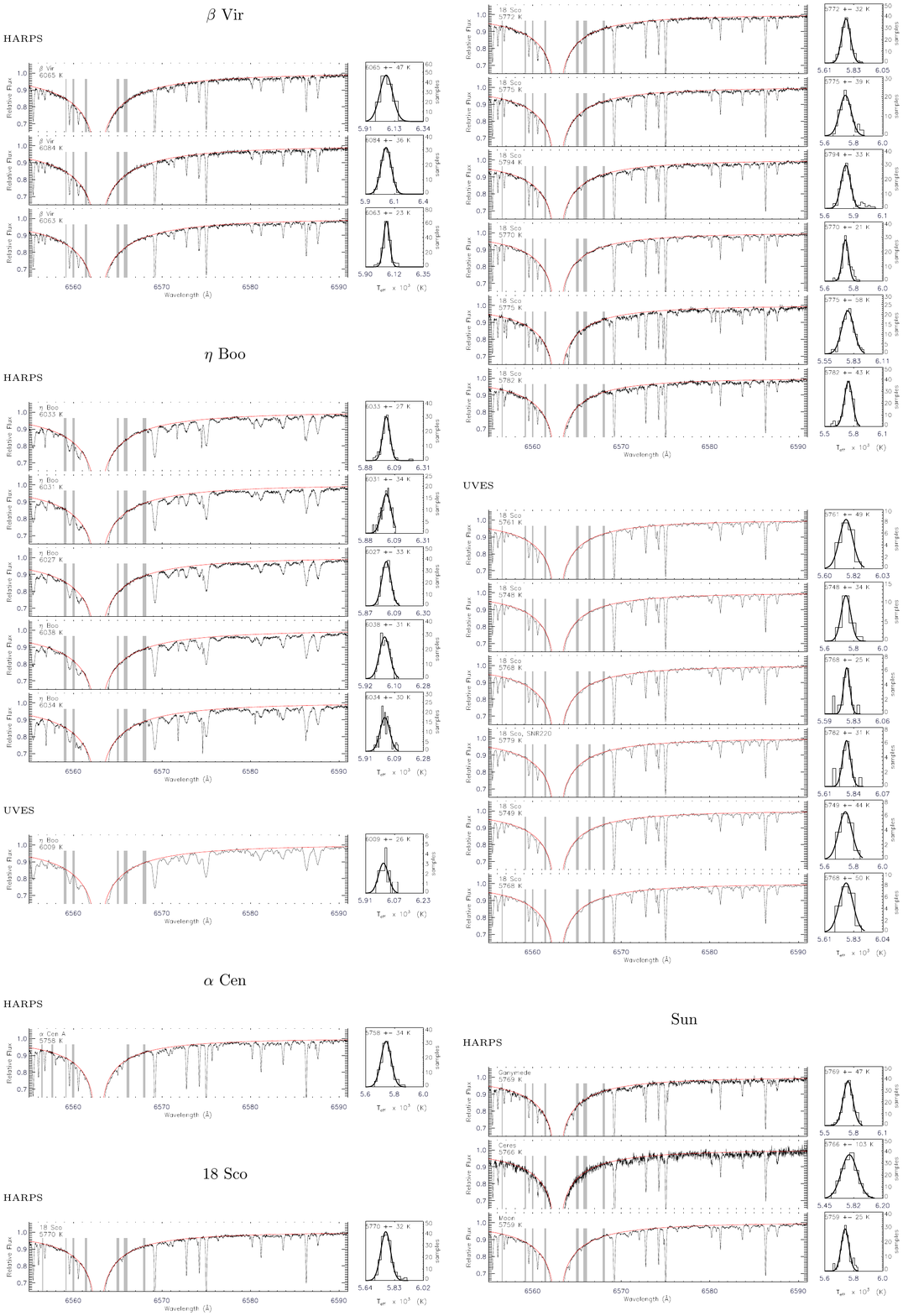}
\end{figure}
\begin{figure}[!ht]
    \centering
    \includegraphics[width=0.94\linewidth]{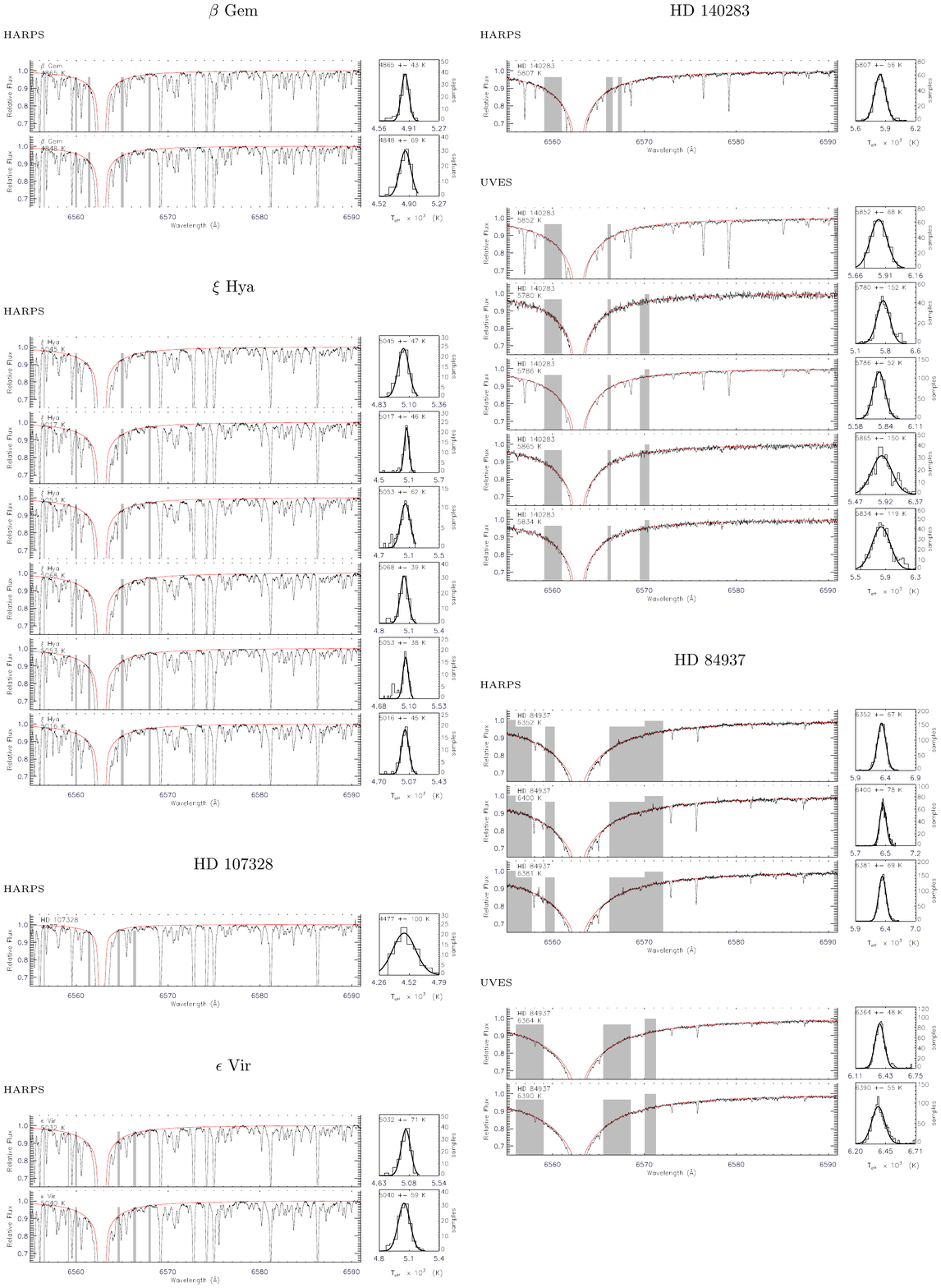}
\end{figure}
\begin{figure}[!ht]
    \centering
    \includegraphics[width=0.94\linewidth]{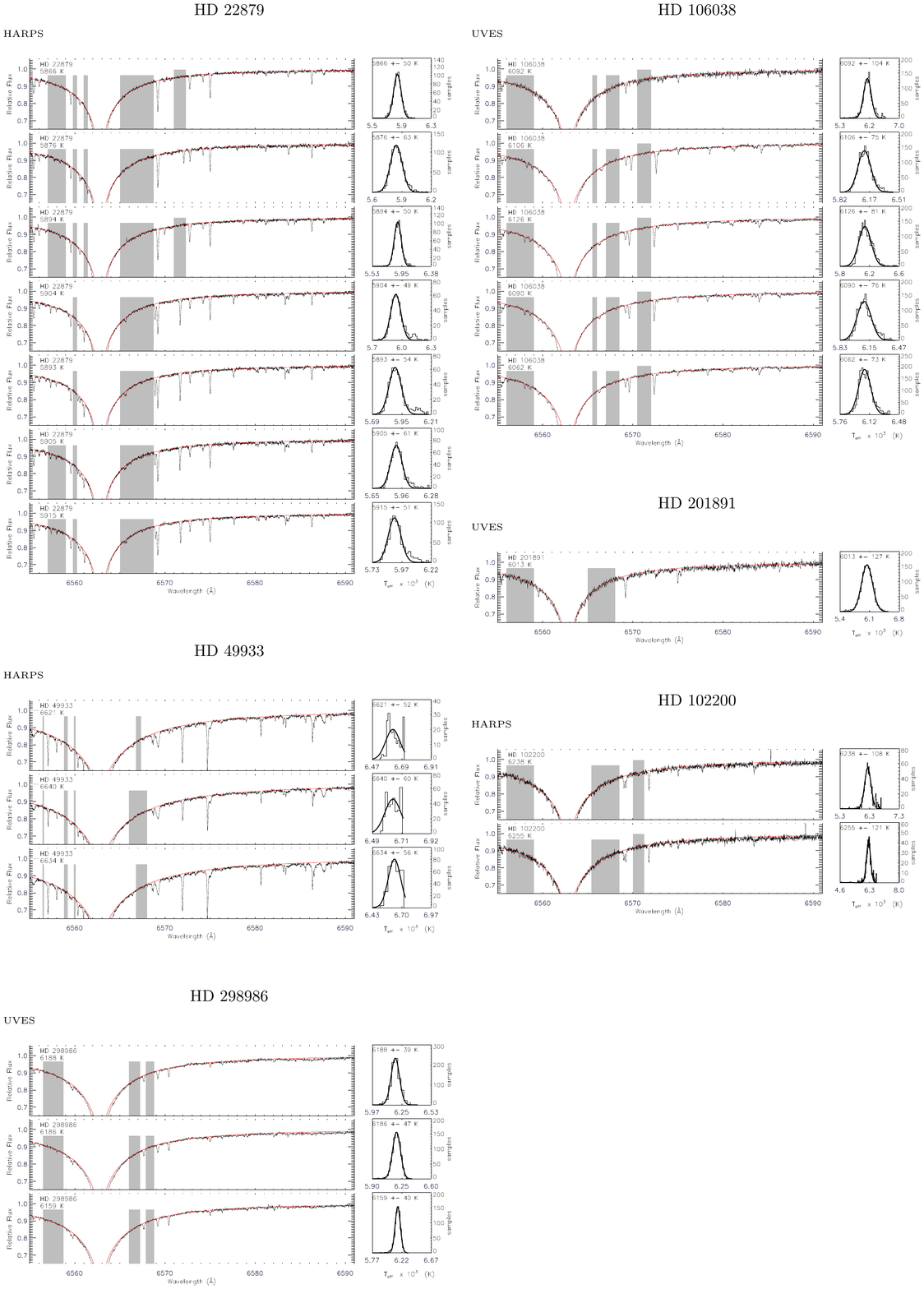}
\end{figure}
\end{appendix}

\end{document}